\documentclass[12pt, titlepage, openany]{amsbook}

\usepackage{amsmath, amsfonts, amssymb, amsthm}
\newcommand{\eg}{{e.g., }}

\usepackage[dvips]{graphicx}
\usepackage[inline]{enumitem}
\usepackage{listingsutf8, subfig, xcolor}
\usepackage{multirow}  % For tables with merged cells
\usepackage{xspace}    % For macros with automatic spacing
\usepackage{comment}   % For comment environments
\newcommand\YAMLcolonstyle{\color{red}\mdseries}
\newcommand\YAMLkeystyle{\color{black}\bfseries}
\newcommand\YAMLvaluestyle{\color{blue}\mdseries}

\makeatletter

\newcommand\language@yaml{yaml}

\expandafter\expandafter\expandafter\lstdefinelanguage
\expandafter{\language@yaml}
{
  keywords={true,false,null,y,n},
  keywordstyle=\color{darkgray}\bfseries,
  basicstyle=\ttfamily,                                 % assuming a key comes first
  sensitive=false,
  comment=[l]{\#},
  morecomment=[s]{/*}{*/},
  commentstyle=\color{purple}\ttfamily,
  stringstyle=\YAMLvaluestyle\ttfamily,
  moredelim=[l][\color{orange}]{\&},
  moredelim=[l][\color{magenta}]{*},
  moredelim=**[il][\YAMLcolonstyle{:}\YAMLvaluestyle]{:},   % switch to value style at :
  morestring=[b]',
  morestring=[b]",
  literate =    {---}{{\ProcessThreeDashes}}3
                {>}{{\textcolor{red}\textgreater}}1     
                {|}{{\textcolor{red}\textbar}}1 
                {\ -\ }{{\mdseries\ -\ }}3,
}

\lst@AddToHook{EveryLine}{\ifx\lst@language\language@yaml\YAMLkeystyle\fi}
\makeatother

\usepackage[all]{xy}

\usepackage{aliascnt} 
\usepackage[colorlinks, linktocpage, allcolors=black,breaklinks]{hyperref} 

\input{UNTdissertation.sty}

\newcommand{\name}{EdgeWeaver\xspace}
\newcommand{\namens}{EdgeWeaver}
\newcommand{\agent}{EdgeWeaver agent\xspace}
\newcommand{\agents}{EdgeWeaver agents\xspace}

\begin{document}

%%% These are the roman numbered pages -- the frontmatter.
\frontmatter

%%% Things for the title page ...
\title{Object Abstraction To Streamline\\Edge-Cloud-Native Application Development} 
\author{Pawissanutt Lertpongrujikorn} 
\degree{Dissertation Prepared for the Degree of\\ DOCTOR OF PHILOSOPHY}
\degreedate{December 2025}
\approved{
    Mohsen Amini Salehi, Major Professor\\
    Song Fu, Committee Member\\
    Tong Shu, Committee Member\\
    Hai Duc Nguyen, Committee Member\\
    Gergely Zaruba, Chair of the\\
    \hspace{36pt} Computer Science and Engineering Department\\
    Paul S. Krueger, Dean of the\\
    \hspace{36pt} College of Engineering\\
    Victor Prybutok, Dean of the\\
    \hspace{36pt} Toulouse Graduate School\\
  }

%%% This makes the title page ...
\maketitle

%%% The abstract page
%%%%%%%%%%%%%%%%%%%%%%%%%%%%%%%%%%%%%%
%%% Abstract
%%%%%%%%%%%%%%%%%%%%%%%%%%%%%%%%%%%%%%

\thispagestyle{empty}
\pagestyle{empty}
% \vspace*{44pt}

Lertpongrujikorn, Pawissanutt. \emph{Object Abstraction To Streamline Edge-Cloud-Native Application Development.} Doctor of Philosophy (Computer Science and
Engineering), December 2025, 181 pp., 8 tables, 51 figures, references, 127 titles.

Cloud computing has fundamentally transformed application development, yet a persistent gap remains between the serverless promise of simplified deployment and its practical realization. Current serverless platforms suffer from fundamental fragmentation: application logic resides in Function-as-a-Service (FaaS) runtimes, state management requires separate database services, and workflow orchestration demands additional coordination layers. This dissertation addresses this gap through empirical validation and technical innovation, establishing the Object-as-a-Service (OaaS) paradigm as a unified approach to cloud-native development. Grounded in empirical evidence from three complementary studies---an initial practitioner interview study (21 participants), a human study evaluating developer experience (39 participants), and NSF I-Corps National customer discovery (101 interviews across 86 organizations)---this work demonstrates that infrastructure complexity imposes substantial tax on developer productivity, with practitioners consistently prioritizing productivity, automation, and operational maintainability over cost optimization.

The dissertation makes five major contributions: (1) the OaaS paradigm unifies resource, state, and workflow management into a single object-oriented abstraction through the Oparaca prototype, demonstrating negligible performance overhead while achieving state-of-the-art scalability; (2) SLA-driven OaaS introduces declarative Non-functional Requirement (NFR) management, enabling developers to specify availability, throughput, consistency, and latency targets through high-level specifications; (3) OaaS-IoT with EdgeWeaver extends the paradigm across the edge-cloud continuum with SLA-driven placement and connectivity-aware invocation, achieving 31\% faster task completion, 44.5\% reduction in lines of code, and $10\times$ fewer configuration requirements compared to traditional FaaS approaches; (4) commercialization validation establishes a staged pathway targeting technology SMEs and startups as primary early adopters; and (5) the empirical methodology demonstrates the critical importance of grounding technical research in validated practitioner needs. By consolidating fragmented abstractions, automating performance optimization, and extending seamlessly to the edge, OaaS establishes a foundation for cloud-native platforms that truly deliver on the promise of hiding infrastructure complexity and empowering developers to focus on innovation.

\clearpage
\pagestyle{plain}
\setcounter{page}{2}

%%% The copyright page ... copyright.tex
\newpage
\phantom{Free space for short editorial comments.} \vglue -14pt
\vfill
\begin{center}
  Copyright 2025\\ by\\ Pawissanutt Lertpongrujikorn
\end{center}
\vfill
\newpage

%%% A dedication page (optional)
%%%%%%%%%%%%%%%%%%%%%%%%%%%%%%%%%%%%%%
%%% Dedication
%%%%%%%%%%%%%%%%%%%%%%%%%%%%%%%%%%%%%%

% \chapter*{DEDICATION}
% \addcontentsline{toc}{chapter}{DEDICATION}
\vspace*{3in}
\begin{center}
\textit{To my family,}

\textit{whose love sustained me through the challenges}

\textit{and celebrated with me in the triumphs.}
\end{center}

%%% An acknowledgments page (called acknowledgments.tex). This can be
%%% commented out if there are no acknowledgments.
%%%%%%%%%%%%%%%%%%%%%%%%%%%%%%%%%%%%%%
%%% Acknowledgments
%%%%%%%%%%%%%%%%%%%%%%%%%%%%%%%%%%%%%%

\chapter*{ACKNOWLEDGMENTS}
% \addcontentsline{toc}{chapter}{ACKNOWLEDGMENTS}

I express my deepest gratitude to my major professor, Dr. Mohsen Amini Salehi, for his unwavering support, insightful guidance, and patience throughout my doctoral journey. His expertise in distributed systems and cloud computing has been instrumental in shaping this research and my development as a researcher.

I am grateful to my committee members for their valuable feedback and constructive criticism, which significantly improved the quality of this work.

I would like to acknowledge the National Science Foundation for supporting this research and the I-Corps National program. I am particularly thankful to the 100+ industry professionals who participated in our empirical studies—their candid insights about cloud-native development challenges provided the empirical foundation that motivated and validated this research. I also extend my gratitude to the 39 college students who participated in the human study evaluating the OaaS platform; their dedication in completing tutorials, quizzes, and programming assignments provided crucial validation of the system's usability and developer experience.

I am grateful to Chameleon Cloud for providing the research infrastructure and to my collaborators, particularly Hai Duc Nguyen, for their significant contributions to this work.

I am deeply indebted to my family for their unconditional love and support. To my parents, who instilled in me a love of learning and perseverance—your sacrifices and encouragement made this achievement possible.

I am grateful to my labmates, colleagues, and friends for the community, perspective, and encouragement they provided throughout this journey.

Finally, I acknowledge the Institute of Electrical and Electronics Engineers (IEEE) and the Association for Computing Machinery (ACM) for permission to reprint previously published material. Portions of this dissertation were published by IEEE and ACM and are used with permission.

% \vspace{\baselineskip}

% Pawissanutt Lertpongrujikorn

% University of North Texas

% October 2025

%%% This makes the table of contents. 
\setcounter{tocdepth}{2}
\tableofcontents

%%% Uncomment the following to generate lists of tables or figures. 
\listoftables
\listoffigures

%%% Now for the actual dissertation -- the mainmatter.
\mainmatter

%%% This helps LaTeX to break a display across two pages in some
%%% circumstances. If it's commented out, the number of underful pages will
%%% probably go up (and the text will be distributed evenly on the
%%% underfilled pages, not pushed to the top).
\allowdisplaybreaks
  
%%% Include the chapters here.
%%%%%%%%%%%%%%%%%%%%%%%%%%%%%%%%%%%%%%
\chapter{Introduction}
\label{chapter1}
%%%%%%%%%%%%%%%%%%%%%%%%%%%%%%%%%%%%%%
The emergence of cloud technology has drastically transformed the application development process. With cloud infrastructure, provisioning can now be done in a few minutes instead of the weeks or months it used to take. Over the past decade, cloud services have replaced mundane tasks with software automation. The current state-of-the-art, serverless platform utilizes the function-as-a-service (FaaS) paradigm to enable developers to build applications by simply writing code in the form of a function and uploading it to the platform. The system then automates the process of building, deploying, and auto-scaling the application, making the overall development process more effortless and mitigating the burden for programmers and cloud solution architects. Major public cloud providers offer FaaS services (e.g., AWS Lambda~\cite{aws_lambda}, Google Cloud Function~\cite{gcloud_func}, Azure Function~\cite{azure_func}), and several open-source platforms for on-premise FaaS deployments are emerging (e.g., OpenFaaS~\cite{openfaas}, Knative~\cite{knative}). In the backend, the serverless platform hides the complexity of resource management and deploys the function seamlessly in a scalable manner. FaaS is proven to reduce development and operation costs via implementing scale-to-zero and charging the user in a truly pay-as-you-go manner. Thus, it aligns with modern software development paradigms, such as CI/CD and DevOps~\cite{bangera2018devops}.

\section{Empirical Motivation: Understanding Practitioner Challenges}

The rapid advancement of cloud computing has reshaped modern business strategies by enabling scalability, agility, and resilience~\cite{10913359}. However, as industries increasingly shift toward autonomous, real-time, and localized data processing in highly distributed systems, there is growing demand for computing models that also support low-latency responsiveness, localized decision-making, and flexible deployment~\cite{fatima2022production, modupe2024reviewing}. These capabilities are vital for several business domains—such as healthcare, IoT, finance, and robotics—where fast and adaptive software development and deployment are critical.

Despite significant advancements in cloud-native technologies, a persistent gap remains between the challenges acknowledged in academic research and those experienced by practitioners in real-world business domains. Scholarly efforts often overlook critical constraints such as steep learning curves, fragmented toolchains, and limited technical capacity—especially prevalent among small enterprises and domain-specific organizations. These oversights are partly due to academic reliance on idealized assumptions, including underestimated cost implications, simplified system models, and minimal consideration of organizational limitations. This disconnect is further widened by the rapid evolution of cloud platforms and a lack of sustained collaboration between academic researchers and industry stakeholders. As a result, many proposed cloud-native frameworks, despite their conceptual promises, fall short when applied in operational environments.

To help bridge this research-practice divide and ground the motivation for this dissertation, we present empirical findings from interviews with domain-specific professionals directly engaged in the design, development, and adoption of cloud-native systems. These interviews aim to illuminate issues that are often underrepresented in academic discourse—specifically, the practical pain points experienced by practitioners, the expectations they hold for emerging technologies, and the business factors that shape decisions around cloud adoption. By centering these real-world perspectives, we seek to surface latent challenges that hinder the practical uptake of academic innovations and advocate for a more grounded research agenda—one that not only prioritizes technical novelty but also accounts for the operational, organizational, and economic realities of modern software development ecosystems.

\subsection{Market Survey:  Hypothesis and Methodology}

Cloud computing emerged in the early 2000s alongside advancements in virtualization and web hosting technologies. Virtualization tools like VMware~\cite{vmware} enabled hardware abstraction, allowing multiple virtual machines to run on a single physical server, thus greatly improving resource utilization. By hiding infrastructure complexity, cloud platforms promised to reduce infrastructure management burdens and empower organizations, especially those with limited technical capacity, to develop and scale digital services more easily. This vision led to widespread adoption of cloud-native technologies such as container orchestration systems~\cite{k8s} and serverless computing~\cite{aws_lambda, gcloud_func}.

Despite their potential, modern cloud platforms have evolved into a highly complex and fragmented landscape. In fact, major commercial cloud providers (e.g., Amazon AWS, Google Cloud, and Microsoft Azure) have collectively introduced over \textbf{500} new services since 2006~\cite{borra2024comparison}. Many of these services demand deep technical expertise to configure, integrate, and operate effectively. Technologies like Kubernetes~\cite{k8s}, Function-as-a-Service (FaaS)~\cite{aws_lambda, gcloud_func}, CI/CD pipelines~\cite{github_actions}, and microservices architectures compound steep learning curves and require developers to master a large and ever-growing set of tools and best practices. As a result, fundamental routine tasks—such as deploying an application or configuring its performance—now require a high level of expertise in cloud-native principles and platform-specific knowledge~\cite{lertpongrujikorn2024streamlining}.

This escalating technical and conceptual complexity directly contradicts the cloud's foundational promise of streamlined IT and universal accessibility. For startups, small enterprises, and domain-focused organizations without dedicated technical teams, adopting cloud-native solutions can be prohibitively complicated. Organizations in these categories often face challenges such as talent shortages, prolonged development cycles, debugging and integration overhead, and constrained budgets.

This observation leads to the following hypothesis:

\vspace{1mm}
\noindent
\colorbox{red!10}{
\parbox{0.96\linewidth}{
\underline{\textbf{Hypothesis:}} \emph{The current real-world practice of cloud-native application development, deployment, and maintenance is complicated, requiring significant time and specialized expertise, which increases the overall costs of cloud software production.}
}}
\vspace{4mm}

This hypothesis stems from a thorough examination of existing cloud-native systems and industry reports. Through preliminary investigation, we identified a recurring pattern: domain-specific professionals often struggle with fragmented toolchains, complex deployment processes, and steep learning curves associated with distributed computing. 

To verify this hypothesis, we conducted interviews with \textbf{21 developers and industry professionals} across multiple business domains, asking questions focused on the pain points they faced, their expectations for solutions, and the obstacles to implementing them. The interviews covered diverse sectors, including agriculture, chips, cybersecurity, education, finance, healthcare, industrial IoT, social networks, and technology consulting, with participants split between technical and non-technical roles.

\subsection{Key Survey Findings: Pain Points}

During interviews, participants described a wide range of challenges encountered when adopting and scaling cloud-based solutions. These challenges span performance bottlenecks, DevOps limitations, and infrastructure trade-offs. Based on their responses, we classified the reported issues into six distinct categories, summarized below:

\paragraph{Operational Maintainability (57\% of participants):} The most frequently reported pain poin is, often linked toa  lack of dedicated technical personnel, limited cloud provider support, and the complexity of configuring and managing cloud services. This issue disproportionately affected non-technical organizations.

\paragraph{Development Complexity (48\% of participants):} Organizations reported significant difficulty in navigating fragmented toolchains and integrating multiple cloud services. Non-tech participants reported this issue at a rate \textbf{4× higher} than tech participants, indicating that technical expertise gaps substantially hinder cloud adoption.

\paragraph{Responsiveness (43\% of participants):} Performance and latency issues were common across both technical and non-technical domains, particularly for distributed and edge computing scenarios. Most participants reported that these issues remained partially unresolved.

\paragraph{Reliability and Availability (33\% of participants):} Concerns were particularly significant in organizations with IoT-centric and hybrid computing environments, especially those lacking dedicated DevOps or SRE teams. None reported full resolution of these issues, though some (57\%) implemented partial mitigations.

\paragraph{Security and Privacy (38\% of participants):} Highly reported by organizations handling sensitive data (finance, healthcare, education). Some addressed concerns by adopting hybrid or private cloud solutions. Remote sites with unreliable network connectivity faced additional challenges, particularly for authentication systems that must function during disconnections.

\paragraph{Cost (24\% of participants):} Appeared to be of less immediate concern, likely because many technical interviewees were not directly involved in budgeting decisions. However, cost pressures were indirectly evident through the resource constraints mentioned in other categories.

\vspace{2mm}
\noindent
\colorbox{blue!10}{
\parbox{0.96\linewidth}{
\underline{\textbf{Key Insight:}} \emph{Most reported pain points have remained unresolved, especially for non-tech organizations lacking technical expertise and dedicated support teams. Across all categories, non-technical participants consistently reported higher rates of challenges than their technical counterparts.
}}
}

\vspace{2mm}

\subsection{Key Survey Findings: Industry Expectations}

Participants articulated clear expectations for what cloud platforms or underlying technologies must provide to address their current pain points and better support future application development and deployment. We categorized these expectations into five feature groups:

\paragraph{Availability \& Reliability (43\% total; 67\% non-tech, 33\% tech):} High availability (minimum 99.9\%), quick support contact, and well-defined disaster recovery. Non-technical organizations particularly emphasized this expectation, as they typically lack in-house technical support and depend heavily on platform reliability.

\paragraph{Security \& Privacy (29\% total; 67\% non-tech, 33\% tech):} Strong and resilient security systems with data encryption, especially critical for sectors dealing with regulated or sensitive data.

\paragraph{Service Quality Assurance (33\% total; 57\% tech, 43\% non-tech):} High SLA offerings, rich configurability for performance optimization, system scalability, and capability. Tech participants prioritized this more highly, reflecting their reliance on underlying technology performance to deliver quality digital services.

\paragraph{Unified Maintainability (33\% total; 57\% tech, 43\% non-tech):} Support for hybrid cloud, automation (DevOps), and log aggregation. Tech interviewees emphasized streamlined deployment, updates, monitoring, and debugging to reduce operational overhead.

\paragraph{Programmability (38\% total; 50\% tech, 50\% non-tech):} Ease of development and user-friendly interfaces. While tech interviewees appreciated customization through APIs/SDKs, non-tech respondents focused on low-code capabilities that empower broader teams without deep technical skills.

\vspace{2mm}
\noindent
\colorbox{blue!10}{
\parbox{0.96\linewidth}{
\underline{\textbf{Key Insights:}} \emph{Tech participants prioritize QoS (availability, security, SLA) to deliver reliable digital services. Non-tech participants prioritize simplification (unified maintainability, programmability) due to limited technical expertise. The even distribution across categories indicates strong demand for comprehensive unified platforms.
}}
}

\vspace{2mm}

\subsection{Discussion: Survey Implications for this Dissertation}

Drawing from the empirical evidence, two major conclusions refine and expand our original hypothesis:

\paragraph{\textbf{Complexity Despite Technology Advancements:}} Although cloud technologies have continued to evolve, the interviews clearly show that cloud users, especially those in non-technical domains, still struggle with the complexity of real-world adoption. Instead of simplifying the development and deployment process, many modern cloud paradigms introduce new layers of abstraction that require even more specialized knowledge and operational overhead. These findings validate the original hypothesis that current cloud-native practices remain complicated, but they go further: the complexity is not simply due to immature technology but is often embedded in the very design of modern cloud platforms. This insight underscores the pressing need for a unified and streamlined approach to cloud infrastructure that can mitigate fragmentation and lower the technical barrier to adoption.

\paragraph{\textbf{Prioritizing Productivity and QoS over Cost and Migration:}} In contrast to common assumptions that practitioners primarily consider cost and migration concerns when adopting new technologies, our participants revealed a different reality. Across roles and domains, neither cost nor migration emerged as the primary barrier to adoption. Instead, their focus centered on expectations such as Quality of Service (QoS), simplified programmability, and improved accessibility. In fact, many participants expressed willingness to endure short-term migration challenges or higher upfront costs if the platform could offer meaningful improvements in service quality, operational simplicity, and development speed. This finding refines our original hypothesis by highlighting that while high cost is a consequence of complicated solution lifecycles, the root cause is deeper, stemming from productivity gaps and architectural complexity in today's cloud ecosystems.

\paragraph{\textbf{Implications for Cloud Platform Design:}} These empirical findings directly motivate the technical contributions of this dissertation. The survey reveals that practitioners need:
\begin{itemize}
    \item \textbf{Unified abstraction} to reduce fragmentation (operational maintainability, development complexity)
    \item \textbf{Quality-of-service guarantees} with intuitive interfaces (service quality assurance, availability/reliability)
    \item \textbf{Simplified programmability} that hides infrastructure complexity (programmability, ease of development)
    \item \textbf{Comprehensive platform support} across distributed environments (responsiveness, edge-cloud integration)
\end{itemize}

These practitioners needs point toward a fundamental rethinking of serverless platform design—one that unifies resource, state, and workflow management to reduce fragmentation; provides declarative QoS interfaces to guarantee service quality without deep expertise; and extends seamlessly across edge-cloud continua for distributed, responsive applications.

\section{Technical Motivation: Limitations of Current FaaS Platforms}

The empirical findings establish \textit{what} practitioners need from cloud platforms. To understand \textit{why} current solutions fall short and \textit{how} to address these needs, we must examine the technical limitations of existing serverless platforms. This section focuses on Function-as-a-Service (FaaS)—the current state-of-the-art serverless paradigm—and identifies how its fundamental design constraints directly manifest as the pain points reported by practitioners.

While FaaS represents significant progress toward allowing developers to focus on business logic by automating infrastructure management and scaling, it remains constrained by three fundamental limitations that directly correlate with the practitioner pain points identified earlier:

\paragraph{\textbf{Limitation 1}: Fragmented State Management:} The FaaS paradigm centers on \emph{stateless functions}, deliberately excluding \emph{state data} from its abstraction. In practice, most applications require persistent state—structured or unstructured—forcing developers to manually integrate external storage services (e.g., AWS S3~\cite{aws_s3}). This fragmentation manifests as the \textit{development complexity} pain point (48\% of practitioners): developers must write additional code to manage data persistence, coordinate access across functions, and handle consistency. For complex workflows, developers resort to using external storage as communication channels between functions, further compounding the integration burden and directly contributing to the \textit{operational maintainability} challenges (57\% of practitioners).

\paragraph{\textbf{Limitation 2}: Lack of Non-functional Requirement (NFR) Control:} FaaS platforms provide separate service abstractions for compute, databases, workflows, and messaging, preventing holistic application optimization. The platform operates with minimal knowledge of application semantics, while developers have limited ability to ``hint'' performance requirements or influence optimization decisions (e.g., data locality, caching policies). Non-functional requirements—such as Quality-of-Service (QoS) guarantees, consistency levels, and availability targets—cannot be declaratively specified within the FaaS abstraction. This architectural separation directly causes the \textit{service quality assurance} concerns reported by 33\% of practitioners—developers cannot declaratively specify performance targets, forcing them into trial-and-error tuning cycles that increase time-to-deployment and operational costs.

\paragraph{\textbf{Limitation 3}: Single-Cluster Deployment Model:} When applications must serve users across geographic regions or at the edge, meeting QoS requirements within a single cluster becomes infeasible. Developers must manually provision multiple clusters, partition application state and logic, and coordinate distributed execution—tasks that require deep expertise in distributed systems. Edge deployments (Fog Computing~\cite{costa2022orchestration}) compound this complexity with heterogeneous hardware, intermittent connectivity, and diverse protocol requirements. This limitation directly explains the \textit{responsiveness} pain point (43\% of practitioners) and \textit{reliability/availability} concerns in distributed environments, particularly for non-technical organizations lacking distributed systems expertise.

\section{Research Problems and Dissertation Approach}

The preceding analysis reveals a clear research agenda: current serverless platforms fail to address practitioner needs because of fundamental architectural constraints—fragmented state management, lack of NFR control, and single-cluster deployment models. These technical limitations directly cause the pain points experienced by practitioners. This dissertation addresses three interconnected research problems that systematically tackle these limitations:

\begin{itemize}
    \item \textbf{Unified Serverless Abstraction:} How can we design a serverless paradigm that abstracts resource, state, and workflow management into a cohesive programming model, eliminating fragmentation and allowing developers to focus solely on business logic?
    
    \item \textbf{Declarative NFR/SLA Automation:} How can we enable developers to specify non-functional requirements (quality-of-service, consistency, availability) through intuitive, high-level metrics that drive automated platform optimization without requiring deep cloud expertise?
    
    \item \textbf{Distributed Deployment:} How can we extend NFR-aware serverless systems to seamlessly deploy and optimize applications across geo-distributed clusters and edge-cloud continua while maintaining unified abstractions?
\end{itemize}

\begin{figure} [h]
    \centering
    \includegraphics[width=0.35\linewidth]{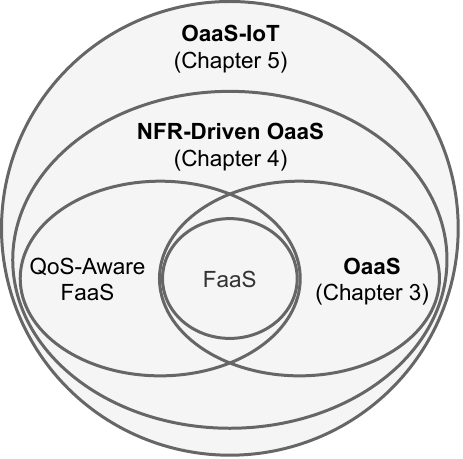}
    \caption{Three-stage evolution from FaaS to federated NFR-aware OaaS, progressively addressing fragmentation (Stage 1), NFR control (Stage 2), and distributed deployment (Stage 3).}
    \label{fig:ideal-platform}
\end{figure}

To address these problems, we propose a three-stage research approach illustrated in Figure~\ref{fig:ideal-platform}, progressively expanding serverless capabilities from unified abstraction (OaaS) to NFR-driven automation to federated edge-cloud deployment.

\subsection{Stage 1: Object-as-a-Service (OaaS) — Unified Abstraction}

\begin{figure}[h]
  \centering
  \subfloat[Function as a Service (FaaS)]{\includegraphics[width=0.48\textwidth]{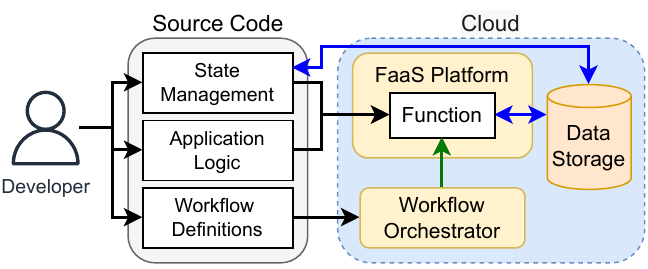}\label{fig:faas_cncpt}}
  \hfill
  \subfloat[Object as a Service (OaaS)]{\includegraphics[width=0.48\textwidth]{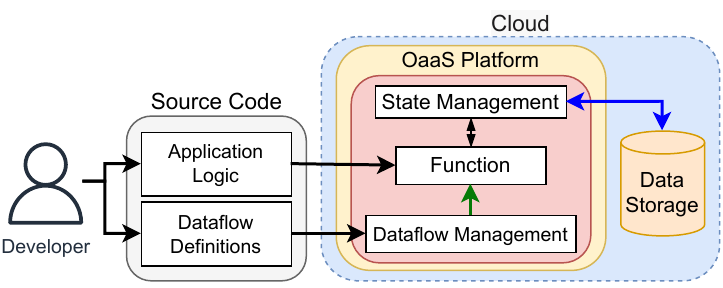}\label{fig:oaas_cncpt}}
  \caption{Architectural comparison: FaaS fragments state and workflow across external services, while OaaS unifies them within platform-managed objects.}
  \label{fig:intro}
  \vspace{-3mm}
\end{figure}

To address the fragmented state management limitation and the first research problem, we propose \textbf{Object-as-a-Service (OaaS)}~\cite{lertpongrujikorn2023object,lertpongrujikorn2024object}—a new serverless paradigm that borrows the notion of ``object'' from object-oriented programming (OOP) to unify resource, state, and workflow management within the platform abstraction. This directly tackles the \textit{development complexity} (48\%) and \textit{operational maintainability} (57\%) pain points by eliminating the need for developers to manually integrate fragmented services.

As illustrated in Figure~\ref{fig:intro}, OaaS fundamentally differs from FaaS by incorporating state management directly into the platform, making it transparent to developers. Where FaaS forces developers to coordinate separate storage services and manage data flow manually, OaaS provides:

\begin{itemize}
    \item \textbf{Platform-Managed State:} State is segregated from application code and managed by the platform as part of the object abstraction, eliminating manual storage integration.
    \item \textbf{Built-in Workflow Orchestration:} Dataflow semantics are native to the programming model—developers declare data flow between functions rather than implementing communication channels.
    \item \textbf{Unified Optimization Opportunities:} By encapsulating compute, state, and workflow within objects, the platform can apply holistic optimizations (data locality, caching, replication) without developer intervention.
\end{itemize}

Developers interact with OaaS through familiar \texttt{class} and \texttt{function} declarations, expressing application logic in an object-oriented style while the platform handles resource provisioning, state persistence, fault tolerance, and inter-function data navigation. This unified abstraction directly addresses practitioner expectations for \textit{simplified programmability} (38\%) and \textit{unified maintainability} (33\%).

\subsection{Stage 2: SLA-Driven OaaS — Declarative Performance Control}

\begin{figure*}[htb!]
    \centering
    \subfloat[Current FaaS lifecycle: Trial-and-error refinement cycles\label{fig:cloud-problems}]{%
        \includegraphics[width=0.80\textwidth, trim={10 50 10 50}, clip]{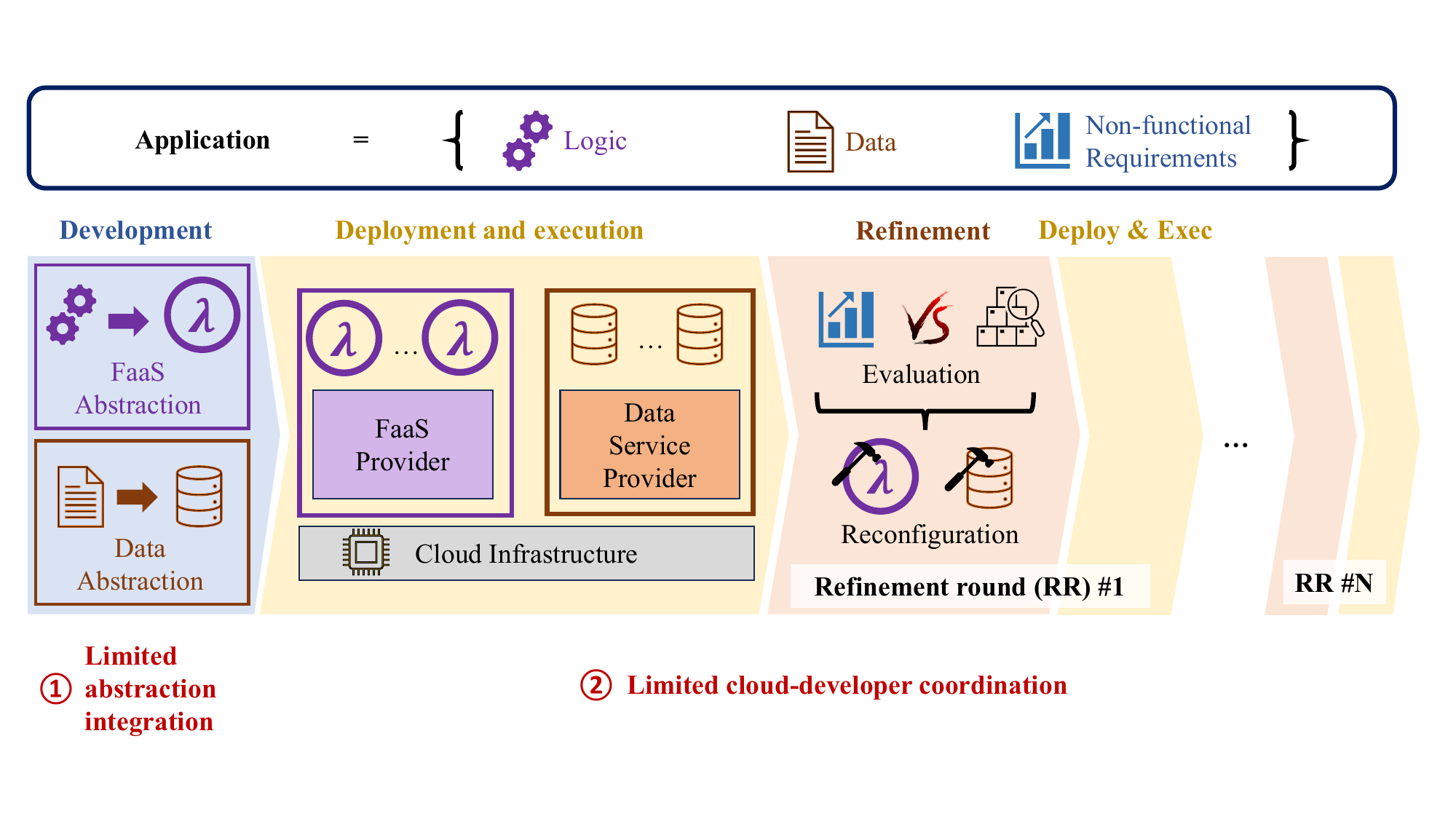}%
    }\\[1em]
    \subfloat[NFR-driven OaaS: Declarative requirements with automated refinement\label{fig:oaas-ideas}]{%
        \includegraphics[width=0.80\textwidth, trim={10 50 10 50}, clip]{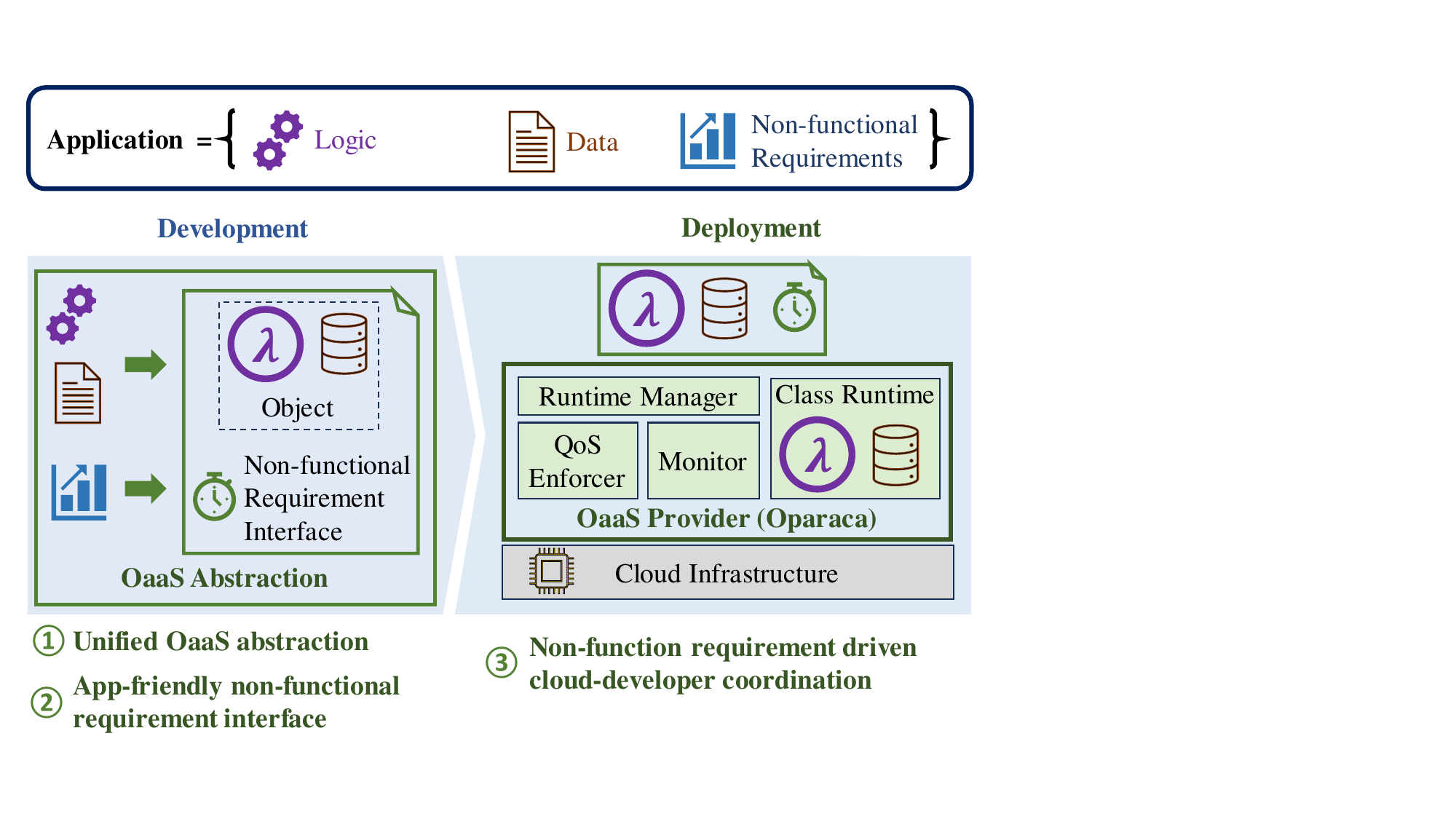}%
    }
    \caption{Comparison of development lifecycles: FaaS requires repeated manual refinement to meet NFR targets, while NFR-driven OaaS automates optimization through declarative specifications, directly addressing \textit{service quality assurance} expectations (33\% of practitioners).}
    \label{fig:problem-approach}
    % \vspace{-3mm}
\end{figure*}

With OaaS providing unified abstraction of resources, state, and workflows, we can now address the second research problem: enabling declarative NFR control. The key insight is that by unifying the programming model, the platform gains complete semantic knowledge of the application—enabling automated optimization across compute, storage, and workflow without cross-domain constraints. This directly tackles the \textit{service quality assurance} expectations (33\%) and reduces the trial-and-error refinement cycles that plague FaaS development (Figure~\ref{fig:cloud-problems}).

\textbf{NFR-Driven OaaS}~\cite{lertpongrujikorn2024streamlining} extends the base paradigm with two key mechanisms:
\begin{itemize}
    \item \textbf{Declarative NFR/SLA Interfaces:} Developers specify non-functional requirements—including performance targets (latency, throughput), consistency levels, and availability guarantees—alongside class definitions using intuitive, high-level metrics~\cite{batista2017qos}. The platform interprets these requirements to guide optimization decisions—or proactively rejects infeasible specifications, eliminating costly deployment failures.
    
    \item \textbf{Dynamic Class Runtimes:} Rather than sharing a monolithic runtime across all classes (which creates conflicting optimization goals), the platform dynamically provisions dedicated \textit{class runtimes} tailored to each class's NFR specifications. For example, cost-sensitive classes can disable expensive components (in-memory caches), while latency-critical classes receive aggressive caching and replication.
\end{itemize}

This approach creates \textit{self-contained, portable objects} where behavior and performance are determined by declarative specifications rather than manual tuning. Developers benefit from a programming model focused on business logic (addressing \textit{simplified programmability}, 38\%), while the platform guarantees performance across deployments—enabling seamless migration between providers without code changes.

\subsection{Stage 3: OaaS-IoT — Distributed Edge-Cloud Deployment}

The third research problem extends NFR-driven OaaS to geo-distributed environments, specifically the Edge-Cloud continuum, where applications must span heterogeneous, intermittently connected computing tiers. This addresses the \textit{responsiveness} pain point (43\%) and \textit{reliability/availability} concerns in distributed IoT deployments—from intelligent transportation and industrial automation to healthcare and smart cities. Current FaaS platforms require manual cluster management, state partitioning, and distributed coordination, creating prohibitive complexity for non-technical organizations.

\begin{figure}[th]
  \centering
  \includegraphics[width=0.9\textwidth]{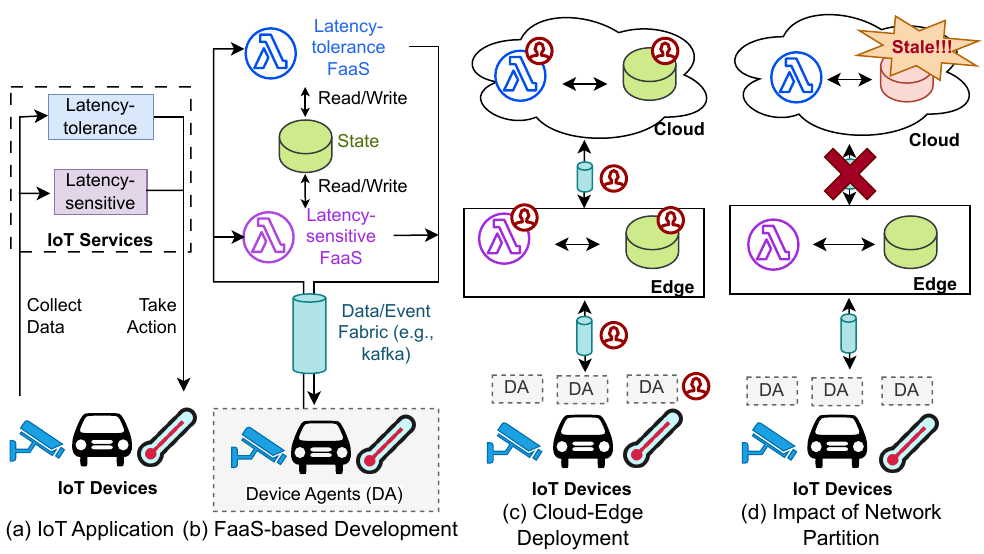}
  \caption{FaaS-based development and deployment challenges for IoT applications spanning edge-cloud continuum}
  \label{fig:faas-iot-challenges}
\end{figure}

The Edge-Cloud continuum introduces two fundamental challenges (Figure~\ref{fig:faas-iot-challenges}). First, \textit{amplified development complexity} arises from infrastructure heterogeneity (k3s at edge vs.~k8s in cloud), protocol diversity (MQTT, Kafka), and FaaS's lack of long-lived function support—essential for continuous IoT data processing. Component counts nearly double compared to cloud-only deployments~\cite{dritsas2025survey,patsch2024make}. Second, \textit{intermittent connectivity} between edge and cloud tiers creates network partitions where the CAP theorem forces unavoidable trade-offs between consistency and availability~\cite{brewer2012cap,esteves2024long}. Time-sensitive applications (industrial automation, healthcare) require availability during partitions, while others (financial transactions) demand consistency—yet current platforms lack mechanisms for declaratively specifying these application-specific requirements. Chapter~\ref{chapter2} provides detailed technical analysis of these Edge-Cloud challenges.

To systematically address these challenges, we propose \textbf{OaaS-IoT}—an extension of the OaaS paradigm across the Edge-Cloud continuum. We realize OaaS-IoT through the \textbf{EdgeWeaver} platform. OaaS-IoT's key contributions are:

\begin{itemize}
    \item \textbf{Distributed Objects:} Seamlessly span edge and cloud tiers, encapsulating application logic, state, IoT interactions, and communication patterns within unified class-based abstractions. Developers use the same object-oriented interface regardless of deployment location.
    
    \item \textbf{Declarative NFR/SLA Specification:} Extends NFR interfaces to include consistency levels, availability requirements under partitions, and geo-placement preferences. The EdgeWeaver platform interprets these specifications to guide function placement, state replication, and runtime adaptation.
    
    \item \textbf{Distributed Class Runtimes:} EdgeWeaver deploys class runtimes across Edge-Cloud tiers to hide infrastructure heterogeneity and enable location-transparent function calls. It leverages Raft and CRDTs for flexible consistency (eventual to strong), adaptive replication for availability during partitions, and rate-guarantee abstractions for performance targets.
\end{itemize}

This unified approach directly addresses all three practitioner pain points in distributed scenarios: \textit{development complexity} through unified abstractions, \textit{responsiveness} through edge placement, and \textit{reliability/availability} through declarative SLA enforcement—all without requiring distributed systems expertise.

\subsection{OaaS Positioning within the Cloud-Native Landscape}

Having presented OaaS's three-stage evolution—from unified abstraction to declarative NFR control to edge-cloud deployment—we now situate these contributions within the complete cloud-native development landscape. Figure~\ref{fig:architecture-landscape} organizes this landscape into five architectural layers, mapping where OaaS addresses critical gaps while honestly acknowledging limitations that remain for future research. Figure~\ref{fig:architecture-landscape} reveals how the three-stage OaaS approach addresses challenges across five architectural layers:

\begin{figure*}[ht]
    \centering
    \includegraphics[width=\linewidth]{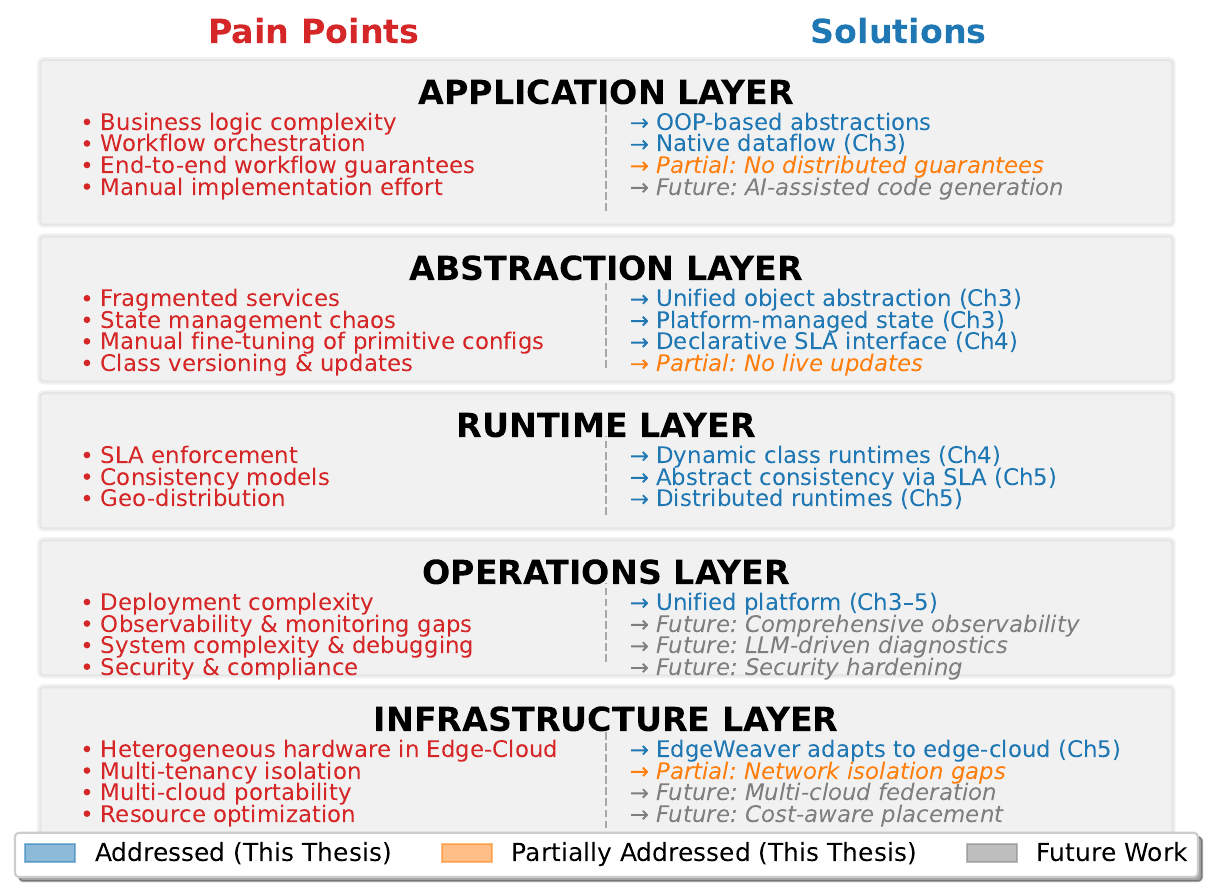}
    \caption{Architecture landscape showing five layers of cloud-native development challenges. Pain points and solutions map thesis contributions, partial solutions, and future work, contextualizing where OaaS addresses critical gaps while acknowledging limitations for future research.}
    \label{fig:architecture-landscape}
\end{figure*}

\paragraph{\textbf{Application Layer:}} OaaS's object-oriented abstractions and native dataflow (Chapter~\ref{chapter3}) address business logic complexity and workflow orchestration. End-to-end workflow guarantees across distributed components and AI-assisted code generation remain areas for future work.

\paragraph{\textbf{Abstraction Layer:}} Unified object abstraction (Chapter~\ref{chapter3}) consolidates fragmented services (compute, state, workflow) and eliminates state management chaos through platform-managed persistence. Declarative SLA interfaces (Chapter~\ref{chapter4}) replace manual fine-tuning with intuitive specifications—developers declare outcomes (availability, throughput, consistency) rather than mechanisms (cache sizes, replica counts). Class versioning and live updates remain partially addressed.

\paragraph{\textbf{Runtime Layer:}} Dynamic class runtimes (Chapter~\ref{chapter4}) enable SLA enforcement through automated resource allocation, while EdgeWeaver's distributed runtimes (Chapter~\ref{chapter5}) abstract consistency models and enable geo-distribution across edge-cloud tiers.

\paragraph{\textbf{Operations Layer:}} Observability, debugging, and security represent cross-cutting concerns spanning all layers, remaining largely unaddressed in current OaaS implementations. These operational challenges manifest differently at each layer—from application-level workflow debugging to runtime SLA compliance monitoring to infrastructure security auditing—yet require unified solutions for production readiness. Notably, \textit{deployment complexity}---stemming from service fragmentation and manual low-level configuration---is mitigated by OaaS's unified platform (Chapters~\ref{chapter3}--\ref{chapter5}), which consolidates compute, state, and workflow into cohesive objects and replaces primitive-level tuning with declarative SLAs. End-to-end release automation and full operational integration remain future work.

\paragraph{\textbf{Infrastructure Layer:}} EdgeWeaver (Chapter~\ref{chapter5}) demonstrates adaptation to heterogeneous hardware in edge-cloud environments through SLA-driven placement and connectivity-aware invocation. Multi-tenancy isolation, multi-cloud federation, and cost-aware placement optimization remain areas for future work.

This layered perspective reveals that while this dissertation makes substantial progress in the top three layers (Application, Abstraction, Runtime), significant opportunities remain for future research—particularly in operations tooling and infrastructure optimization. The honest assessment of partial coverage strengthens rather than diminishes the contribution by clearly delineating solved problems from open challenges, establishing concrete research directions addressed in Chapter~\ref{chapter7}.

\section{Contributions}

Grounded in empirical evidence from practitioner interviews that reveal persistent complexity in cloud-native development, this dissertation makes the following major contributions:

\begin{itemize}
    \item \textbf{Empirical Validation of Practitioner Challenges:} We present findings from interviews with 21 industry professionals across diverse domains, identifying six categories of pain points (operational maintainability, development complexity, responsiveness, reliability/availability, security/privacy, and cost) and five categories of expectations (availability/reliability, security/privacy, service quality assurance, unified maintainability, and programmability). These findings validate that current cloud-native practices remain complicated and costly, with non-technical organizations facing disproportionately higher challenges (4× higher development complexity rates).
    
    \item \textbf{Object-as-a-Service (OaaS) Paradigm:} We propose a new serverless paradigm that abstracts resource, state, and workflow management into the notion of an object, directly addressing the fragmentation and development complexity pain points identified by practitioners. OaaS unifies the programming model by incorporating state management and workflow orchestration into the platform, enabling developers to focus solely on business logic.
    
    \item \textbf{SLA-Driven Serverless System:} We present an NFR-aware serverless system that exploits the advantages of OaaS to drive performance optimization with intuitive high-level non-functional requirement specifications. This contribution directly addresses practitioner expectations for service quality assurance (33\% of participants) and simplified programmability (38\% of participants) by providing declarative NFR interfaces and automated optimization without requiring deep cloud expertise.
    
    \item \textbf{OaaS-IoT Paradigm and EdgeWeaver Platform:} We propose OaaS-IoT, an extension of the OaaS paradigm to the Edge-Cloud continuum, and realize it through the EdgeWeaver platform. This architecture scales QoS-aware serverless systems across geo-distributed edge-cloud environments with a unified deployment interface. This contribution addresses the responsiveness pain point (43\% of participants) and enables deployment across the computing continuum from cloud to edge.
    
    \item \textbf{Commercialization Validation:} We present insights from NSF I-Corps National customer discovery (101 interviews, 86 organizations) that corroborate and expand the initial survey findings, validating the real-world demand for simplified, unified cloud-native platforms and informing the path toward practical adoption of OaaS technologies.
\end{itemize}

\section{Dissertation Organization}

This dissertation is organized into seven chapters. Following this introduction, the remaining chapters are structured as follows:

\begin{itemize}
    
    \item \textbf{Chapter \ref{chapter2}: Background and Related Work} \\
    Presents technical background on serverless computing, stateful serverless systems, consistency models, and edge computing fundamentals. Synthesizes related work from multiple research domains including cloud-native platforms, QoS automation, and edge-cloud integration. This chapter provides the technical foundation necessary to understand the OaaS paradigm and its contributions.
    
    \item \textbf{Chapter \ref{chapter3}: Object-as-a-Service (OaaS) Core} \\
    Proposes the Object-as-a-Service paradigm, presenting the architecture, invocation model, consistency mechanisms, fault tolerance, and dataflow abstractions. Evaluates the OaaS implementation against state-of-the-art serverless platforms, demonstrating performance improvements and reduced development complexity. This chapter directly addresses the unified abstraction and simplified programmability needs identified in the empirical motivation. The OaaS implementation is available at \url{https://github.com/hpcclab/OaaS}. The results of this chapter have been published in the following papers:
    \begin{itemize}
        \item  \textbf{P. Lertpongrujikorn} and M. Amini Salehi, ``Object as a Service: Simplifying Cloud-Native Development through Serverless Object Abstraction,'' \textit{IEEE Transactions on Computers}, accepted in Oct. 2025, In Press.
        \item \textbf{P. Lertpongrujikorn} and M. Amini Salehi, ``Object as a Service (OaaS): Enabling Object Abstraction in Serverless Clouds,'' \textit{In Proceedings of the 16th IEEE International Conference on Cloud Computing (IEEE CLOUD '23)}, Chicago, IL, USA, 2023, pp. 238-248.
    \end{itemize}
    
    \item \textbf{Chapter \ref{chapter4}: SLA-Driven OaaS: Declarative Performance Control} \\
    Presents an NFR-driven serverless system that exploits OaaS to enable performance optimization through intuitive, high-level non-functional requirement specifications. Introduces declarative NFR interfaces, dynamic class runtimes, and automated optimization mechanisms. Demonstrates how declarative NFR specifications address the service quality assurance expectations identified by practitioners. The SLA-driven OaaS implementation is available at \url{https://github.com/hpcclab/OaaS}. The results of this chapter have been published in the following paper:
    \begin{itemize}
        \item \textbf{P. Lertpongrujikorn}, H. D. Nguyen, and M. Amini Salehi, ``Streamlining Cloud-Native Application Development and Deployment with Robust Encapsulation,'' \textit{Proceedings of the ACM Symposium on Cloud Computing (SoCC '24)}, Redmond, WA, USA, 2024.
    \end{itemize}
   
    \item \textbf{Chapter \ref{chapter5}: OaaS-IoT: Extending Object as a Service to the Edge--Cloud Continuum} \\
    Proposes OaaS-IoT, an extension of the OaaS paradigm to geo-distributed edge-cloud environments, and its realization through the EdgeWeaver platform. Addresses edge placement strategies, connectivity-aware invocation, state synchronization across the computing continuum, and geo-location-aware performance optimization. This chapter directly tackles the responsiveness and distributed deployment challenges identified in the empirical study. The EdgeWeaver/OaaS-IoT implementation is available at \url{https://github.com/hpcclab/OaaS-IoT}. The results of this chapter have been submitted to the following venue:
    \begin{itemize}
        \item \textbf{P. Lertpongrujikorn}, H. D. Nguyen, and M. Amini Salehi, ``EdgeWeaver: Seamless Edge-Cloud Integration for Stateful Serverless Applications,'' \textit{40th IEEE International Parallel and Distributed Processing Symposium (IPDPS '26)}, 2026 (under review).
    \end{itemize}
    
    \item \textbf{Chapter \ref{chapter6}: Exploration and Analysis of the OaaS Productization} \\
    Presents customer discovery findings from NSF I-Corps National Summer 2025 program (101 interviews, 86 organizations). Documents methodology, synthesizes industry-specific pain points and desired outcomes, and validates the real-world demand for OaaS-style unified platforms. Discusses value proposition, adoption risks, and paths to commercialization. This chapter corroborates and expands the initial empirical motivation with $5\times$ more data points. The results of this chapter have been submitted to the following venue:
    \begin{itemize}
        \item \textbf{P. Lertpongrujikorn}, H. D. Nguyen, J. Kwon, and M. Amini Salehi, ``Exploring Challenges in Developing Edge-Cloud-Native Applications Across Multiple Business Domains,'' \textit{Cloud-Network Convergence}, 2026 (under review).
    \end{itemize}
    
    \item \textbf{Chapter \ref{chapter7}: Conclusion and Future Directions} \\
    Summarizes the contributions of this dissertation, discusses lessons learned and limitations, and outlines future research directions, including LLM-driven self-evolving objects and multi-cloud object federation. Integrates insights from both empirical studies to propose a research agenda grounded in practitioner needs and operational realities.
    
\end{itemize}
%%%%%%%%%%%%%%%%%%%%%%%%%%%%%%%%%%%%%%
\chapter{Background and Literature Study}
\label{chapter2}
%%%%%%%%%%%%%%%%%%%%%%%%%%%%%%%%%%%%%%

\section{Towards the Next Generation of Cloud Computing Systems}
As established in Chapter~\ref{chapter1}, empirical evidence from practitioners reveals persistent complexity in cloud-native development, with fragmented toolchains, inadequate NFR control, and limited support for distributed deployment identified as major barriers to adoption~\cite{lertpongrujikorn2024streamlining}. These practical challenges stem from fundamental architectural constraints in current serverless platforms—particularly the Function-as-a-Service (FaaS) paradigm~\cite{aws_lambda,azure_func,gcloud_func}—which force developers to manually coordinate separate services for compute, state management, and workflow orchestration.

Cloud platforms are evolving beyond VM- and container-centric abstractions toward managed, intent-driven services that emphasize elasticity, programmability, and developer productivity. This evolution manifests as: (i) event-driven and dataflow-centric programming models~\cite{yu2023following}; (ii) deeper disaggregation of compute, storage, and control planes~\cite{agache2020firecracker}; and (iii) automation that optimizes for application-level objectives (e.g., latency percentiles, availability, and cost)~\cite{tariq2020sequoia,zhou2022aquatope}. The three-stage approach presented in this dissertation---OaaS core (Chapter~\ref{chapter3}), NFR-driven OaaS (Chapter~\ref{chapter4}), and OaaS-IoT with EdgeWeaver (Chapter~\ref{chapter5})---builds on this trajectory by systematically addressing the practitioner pain points through unified abstraction, declarative NFR control, and seamless edge-cloud integration.

This chapter surveys the technical landscape that both motivates and contextualizes our contributions. We examine state-of-the-art research in stateful serverless computing, QoS-aware systems, and edge-cloud platforms to identify specific gaps that the OaaS paradigm addresses. Where Chapter~\ref{chapter1} establishes \emph{what} practitioners need and \emph{why} current platforms fail, this chapter explores \emph{how} existing research has attempted to solve similar problems and where opportunities remain for the unified, NFR-driven, and distributed approach proposed in this dissertation.

\section{Serverless Computing and Function-as-a-Service (FaaS)}
Serverless computing popularized the separation of application code from infrastructure management. In its prevalent \emph{Function-as-a-Service} (FaaS) form, developers package functions that are triggered by events and scale on demand. Major providers offer managed FaaS (AWS Lambda~\cite{aws_lambda}, Azure Functions~\cite{azure_func}, Google Cloud Functions~\cite{gcloud_func}) while open-source alternatives (OpenFaaS~\cite{openfaas}, OpenWhisk~\cite{openwhisk}, Fission~\cite{fission}) enable self-hosting. As compositions grow, managing triggers and intermediate data becomes complex, motivating the use of complementary \emph{workflow orchestrators} such as AWS Step Functions~\cite{aws-sf} and Azure Durable Functions~\cite{azure_df}.

Beyond FaaS, Container-as-a-Service (CaaS)~\cite{caas} exposes container-level control with serverless-like operations. Kubernetes~\cite{k8s} and Knative~\cite{knative} bring autoscaling, scale-to-zero, and eventing to containerized services. While these offerings reduce operational burden, their \emph{stateless-by-default} model complicates the design of applications that require affinity, low-latency state access, or cross-invocation consistency.

Empirical studies revealed key characteristics and constraints of production FaaS. Wang et al. measured cold starts, resource variability, and isolation overheads across providers~\cite{wang18peeking}, while Shahrad et al. characterized large-scale provider workloads and highlighted opportunities for optimization~\cite{shadrad20serverless}. Subsequent work explored microVMs~\cite{agache2020firecracker}, function caching and prewarming~\cite{fuerst2021faascache,du2020catalyzer}, and provider mechanisms such as provisioned concurrency~\cite{aws-lambda-provisioned-concurrency} to mitigate tail latencies. Beyond control-flow orchestration, recent work advocates data-centric orchestration where the movement and placement of data guide function execution~\cite{yu2023following}. Platforms like Nightcore~\cite{jia2021nightcore} and systems research on low-latency isolates aim to better support interactive microservices under the serverless model.

\section{Stateful Serverless}
\begin{figure}[tbp]
  \centering
  \includegraphics[width=\textwidth]{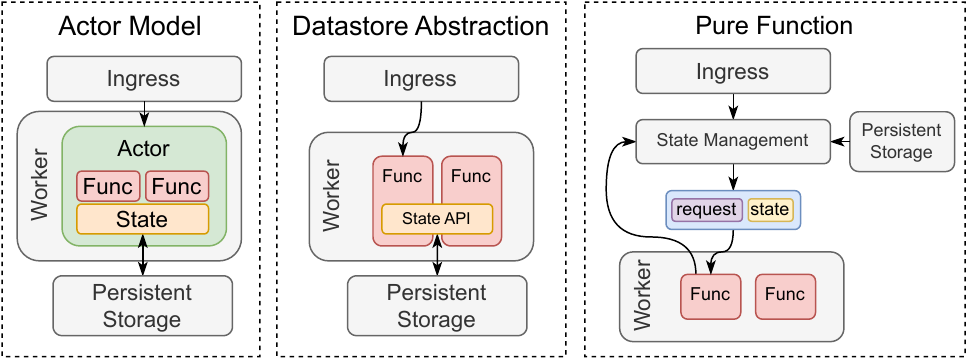}
  \caption{\small{Comparison of three models for stateful serverless.}}
  \label{fig:stateful_serverless}
\end{figure}

Recent work extends serverless beyond statelessness to support \emph{stateful} functions and objects (e.g., \cite{jia2021boki,beldi,cepless}). These approaches primarily aim to address the stateless nature of the FaaS model, where the burden of managing application data, access control, and function workflows is often shifted to the developer through separate cloud services. As summarized in Figure~\ref{fig:stateful_serverless}, three design patterns emerge based on state placement and access: (1) \textbf{actor model}, (2) \textbf{datastore abstraction}, and (3) \textbf{pure function}.

\subsection{Actor Model.} In the actor model, the platform fetches the state from persistent storage and places (i.e., caches) it inside a worker instance, then dispatches the request to where the state resides to achieve data locality. This favors message-driven designs but can make maintainability difficult for bulky unstructured data, as the platform needs to balance each node's computing and storage aspects. The deployment granularity is an actor that contains multiple functions sharing the same state. 

The foundational platform in this space is Orleans~\cite{bykov2011orleans}, which introduces ``virtual actors''. Its influence is evident in modern platforms such as Azure Entity Functions~\cite{azure_enfunc}, which are part of Azure Durable Functions. While effective for certain use cases, this model can tightly couple state and compute within a single language and environment. The actor model approach has been popular in programming languages and object-oriented programming because it spurs asynchronous messaging across actors and lends itself to distributed deployments. In contrast, OaaS (Chapter~\ref{chapter3}) manages the object abstraction at the platform level, allowing functions implemented in different languages or runtimes to operate on an object, thereby offering greater flexibility.

\subsection{Datastore Abstraction}
The datastore abstraction is a hybrid approach where the platform provides an API for the function to access data on demand. Like pure functions, it relaxes the need for state and function co-location, but can utilize caching to preserve data locality. Several systems utilize this pattern. Cloudburst~\cite{cloudburst} uses a shared distributed key-value database, while Boki~\cite{jia2021boki} provides API access to a distributed logging system. This log-based approach for consistency differs from OaaS's mechanisms (Chapter~\ref{chapter3}), which use a combination of fail-safe state transitions for unstructured data and localized locking to handle race conditions. Beldi~\cite{beldi}, on the other hand, provides the database and transaction API to the state. While providing direct transactional APIs offers strong guarantees, OaaS abstracts these concerns from developers through its fail-safe versioning scheme and guarantees of exactly-once execution, aiming for a higher-level resilience model.

FAASM~\cite{faasm} optimizes the function-state interaction by using WebAssembly~\cite{wasm}. Although this enables multiple functions to share memory and achieve data locality, it requires compiling code to WebAssembly, which can limit library compatibility. Crucial~\cite{crucial} also explores shared-memory abstractions. OaaS's hybrid model avoids these limitations by combining the pure function and datastore abstraction approaches, offering broader language support.

\subsection{Pure Function}
In the pure function approach, state is placed on another system and transferred to the worker instance upon invocation, appearing as part of the function's input arguments. This disaggregates state management and computation for system design simplicity, but can compromise data locality. Apache Flink Stateful Function (StateFun)~\cite{statefun} is a representative solution based on this approach. OaaS (Chapter~\ref{chapter3}) combines the pure function approach for structured data with a lazy-fetch mechanism for unstructured data, practically employing both the pure function and datastore abstraction approaches.

\subsection{Additional encapsulation patterns}
Beyond the three primary models, additional patterns have emerged. Unified deployment units such as proclets~\cite{ruan2023nu} and recent actor surveys~\cite{spenger2024survey} further explore co-location of methods with state. Storage-resident execution such as Shredder~\cite{zhang2019narrowing} and Apiary~\cite{kraft2022apiary} pushes computation into storage engines, inverting the traditional separation. OaaS (Chapter~\ref{chapter3}) spans these approaches: it provides an object API for encapsulation while retaining datastore and pure-function modes when appropriate, offering developers flexibility in choosing the right abstraction for their use case.

\subsection{Process abstractions.} While OaaS raises the level of abstraction, a complementary line of work reduces abstraction to expose OS/process-like control for performance. Process-as-a-Service (PaaS) introduces a cloud process abstraction to unify elasticity with stateful execution~\cite{copik2024process}. For long-running services, Nu proposes migratable logical processes to achieve microsecond-scale resource fungibility~\cite{ruan2023nu}. These efforts align with OaaS in spirit—structuring control and state—but differ in the locus of programmability and who manages adaptation.

\subsection{Serverless workflow and dataflow.} Beyond state management for individual functions, composing them into complex applications is another significant challenge. Commercial orchestrators such as AWS Step Functions~\cite{aws-sf} and Azure Durable Functions~\cite{azure_df} can mitigate the burden of chaining events and managing control-flow; however, developers are often still required to navigate data between steps manually. The research community has proposed more integrated solutions. DataFlower~\cite{li2023dataflower} introduces a dataflow paradigm specifically for serverless workflow orchestration to optimize data movement between functions. Netherite~\cite{azure_df} focuses on the efficient execution of these workflows through techniques like partitioned state management and collocated execution. While these systems significantly advance workflow performance and orchestration, OaaS (Chapter~\ref{chapter3}) approaches the problem from a different angle by integrating workflow directly into its core object abstraction. OaaS's dataflow functions allow developers to declaratively define a workflow as a directed acyclic graph (DAG) based on data dependencies rather than task dependencies. The platform transparently manages parallelism, data navigation, and state consistency, abstracting these complexities from the developer.

\subsection{Consistency and fault tolerance.} A central theme across these systems, and a core focus of Chapter~\ref{chapter3}, is the consistency model offered for reads/writes across invocations and failures. Techniques range from at-least-once execution with idempotence, to transactional updates via logging and versioning (e.g., Beldi~\cite{beldi}), to object-centric consistency with conflict resolution via CRDTs and virtual actors~\cite{bykov2011orleans,shapiro2011conflict}. The design of OaaS adopts dataflow-aware state management and structured recovery to balance availability and correctness without imposing monolithic transactions.

\section{QoS-Aware Serverless Computing}
While serverless abstracts infrastructure, developers still need to meet \emph{non-functional} requirements such as tail latency, availability, throughput, and cost. Emerging systems expose interfaces to declare QoS intents and automate the choice of runtime configurations (e.g., memory size, concurrency, placement, batching) and control policies (e.g., timeouts, retries, prewarming). Chapter~\ref{chapter4} formalizes this as a contract between the developer and the platform and introduces a dynamic class runtime that materializes objects according to the declared QoS. An optimization loop closes the gap between observed and target metrics with minimal developer intervention, harmonizing with the consistency and fault-tolerance mechanisms in Chapter~\ref{chapter3}.

\subsection{Non-functional requirements enforcement and prior art.} A large body of work targets latency and isolation symptoms of today\textquotesingle s platforms. Cold-start mitigation spans ahead-of-time preparation and sandbox reuse (e.g., Catalyzer~\cite{du2020catalyzer}, Nightcore~\cite{jia2021nightcore}, and recent analyses~\cite{ebrahimi2024cold}); lightweight isolation via microVMs improves performance and multi-tenancy~\cite{agache2020firecracker}. Providers expose pre-allocation (e.g., provisioned concurrency~\cite{aws-lambda-provisioned-concurrency}) to trade cost for predictability, while real-time serverless extends the API with rate guarantees~\cite{nguyen2019real}. Beyond single functions, Sequoia~\cite{tariq2020sequoia} proposes QoS-aware scheduling, Aquatope~\cite{zhou2022aquatope} reasons about uncertainty at workflow scope, and Astrea~\cite{jarachanthan2022astrea} automates analytic job configuration. Data-centric orchestration~\cite{yu2023following} further reduces coordination overhead by aligning placement and movement with data dependencies. These efforts motivate OaaS\textquotesingle s declarative QoS interface and adaptive runtime.

\section{Serverless Computing in Edge--Cloud Settings}
Edge-to-cloud applications introduce intermittent connectivity, constrained resources, and locality-sensitive state access. Chapter~\ref{chapter5} (EdgeWeaver) specializes OaaS for IoT by adding connectivity-aware invocation, edge affinity/placement policies, and state synchronization across edge and cloud. The design must reconcile eventual disconnections and updates with bounded staleness or conflict-resolution strategies while preserving the "serverless" developer experience. This positions objects as a natural unit for mobility, caching, and coordination across the continuum.

Research on FaaS across the continuum considers resource provisioning and placement~\cite{ascigil2021resource,wang2021lass}, QoS-aware function delivery~\cite{yu2023faasdeliver}, and consistency models adapted to geo-distribution~\cite{shapiro2011conflict}. Edge-oriented orchestration emphasizes data proximity and network variability; our EdgeWeaver design builds on these insights to enable connectivity-aware execution and reconciliation.

Two fundamental challenge threads recur in edge--cloud IoT deployments. First, \emph{complexity and heterogeneity}: edge FaaS often runs on constrained platforms and differs from cloud-side stacks, leading to fragmented designs and duplicated concerns across tiers. Diverse devices and middleware add to configuration and lifecycle overheads, especially across multiple sites~\cite{noaman2022challenges,carvalho2021edge,aslanpour2024load}. Second, \emph{intermittent connectivity and consistency}: partitions and variable links are common at the edge; systems must balance consistency and availability in light of CAP~\cite{brewer2012cap,lee2021quantifying}, choosing application-appropriate reconciliation or bounded staleness strategies without blocking urgent actions~\cite{nguyen2019real}. The following subsections examine these challenges in detail, providing the technical foundation for EdgeWeaver's design (Chapter~\ref{chapter5}).

\subsection{Amplified Development Complexity in Edge-Cloud Deployments}
The ``function" abstraction in FaaS focuses solely on application logic, leaving data management and communication to developers. This results in fragmented implementations requiring complex interactions among FaaS invocations, databases, and orchestrators. Edge deployments compound this fragmentation: developers must manually integrate not only cloud services but also IoT devices, edge servers, communication protocols (MQTT, Kafka), and heterogeneous FaaS implementations (k3s at edge vs.~k8s in cloud)~\cite{dritsas2025survey}. 

Critically, FaaS's stateless nature \emph{lacks built-in support for long-lived functions}—essential for IoT scenarios involving frequent or continuous data processing~\cite{patsch2024make}. Traditional FaaS platforms are designed around ephemeral, event-triggered executions that complete within seconds or minutes. However, IoT applications often require persistent connections to device streams, continuous monitoring of sensor data, or long-running analytics pipelines. This mismatch forces developers to orchestrate intricate dataflows across many short-lived functions and shared-state services, dramatically increasing system complexity and introducing coordination overhead.

Furthermore, large-scale IoT solutions span diverse technologies, application models, and communication protocols~\cite{noaman2022challenges,dave2024data}. While IoT middleware provides abstraction layers (e.g., device agents managed by fleet managers) for easier administration~\cite{miquel2025middleware}, these layers add management overhead. Even with a single edge node, the number of components developers must implement and manage nearly doubles compared to cloud-only architectures. With multiple edge sites, this complexity escalates as each node may differ in software stacks (different versions of Kubernetes or container runtimes), capacity (CPU, memory, storage constraints), network conditions (bandwidth, latency, jitter), and policies (security requirements, data sovereignty rules)~\cite{carvalho2021edge,aslanpour2024load}. Developers must account for these variations when designing, testing, and deploying applications, leading to fragile configurations that break when assumptions about the underlying infrastructure change.

\subsection{CAP Theorem and Application-Specific Trade-offs}
The Edge-Cloud continuum often suffers from \emph{intermittent connectivity} between tiers, caused by network congestion, physical obstructions (buildings, terrain), power limitations, or fluctuating bandwidth~\cite{esteves2024long}. Lost or delayed connections (\emph{network partitions}) disrupt data flow, causing the cloud's view of application state to become stale while edge nodes continue processing locally. In time-sensitive applications like industrial automation or healthcare, these inconsistencies have serious consequences: unsynchronized production line data may result in incorrect machine behavior, and outdated patient information could lead to delayed or dangerous clinical decisions~\cite{abbas2024iomt}. The CAP theorem~\cite{brewer2012cap,lee2021quantifying} formally shows that distributed systems cannot simultaneously guarantee \emph{consistency}, \emph{availability}, and \emph{partition tolerance} under network partitions. Edge-Cloud IoT deployments must balance these trade-offs based on application-specific QoS needs: time-sensitive applications (industrial automation, emergency response) prioritize availability, accepting temporary inconsistency, while financial transactions demand strong consistency even if operations are delayed.

Current platforms lack mechanisms for developers to declaratively specify these NFR trade-offs. For example, an intelligent transportation system detecting an accident should immediately dispatch an ambulance rather than waiting for cloud acknowledgment~\cite{nguyen2019real}, yet developers must implement custom distributed protocols (two-phase commit, vector clocks, operational transformation) from scratch, requiring deep expertise in distributed systems. EdgeWeaver addresses these challenges by providing connectivity-aware execution, object-centric state synchronization, and policy-driven placement that allow developers to declaratively specify consistency and availability requirements without implementing low-level coordination protocols.

\section{Positioning of this Dissertation}
The literature reviewed in this chapter reveals both the promise and limitations of current serverless computing paradigms. While FaaS has democratized cloud application deployment, its stateless-by-default model burdens developers with state management complexities. Stateful serverless systems have emerged to address this gap, yet they often impose trade-offs: actor models tightly couple state and compute, pure function approaches sacrifice data locality, and datastore abstractions require explicit state navigation. Furthermore, existing systems typically focus on either functional correctness \emph{or} performance optimization, rarely integrating both concerns within a unified abstraction. Finally, the extension of serverless to edge--cloud environments remains fragmented, with most solutions addressing edge deployment as an afterthought rather than a first-class design consideration.

This dissertation addresses these limitations by advancing serverless computing along three complementary dimensions. First, we introduce \emph{Object as a Service} (OaaS), a paradigm that unifies stateful computing, dataflow orchestration, and fault tolerance within a single object-oriented abstraction (Chapter~\ref{chapter3}). Unlike prior actor-based or pure-function approaches, OaaS manages object abstractions at the platform level, supporting both structured and unstructured data through fail-safe state transitions and exactly-once execution semantics. The platform transparently handles consistency, recovery, and dataflow dependencies, freeing developers from low-level coordination concerns.

Second, we extend OaaS with a declarative interface for non-functional requirements, enabling developers to specify quality-of-service (QoS) objectives---such as latency bounds, throughput targets, and cost constraints---as first-class contracts (Chapter~\ref{chapter4}). A dynamic class runtime materializes objects according to these declarations, continuously optimizing resource configurations and execution strategies through feedback-driven adaptation. This approach harmonizes functional and non-functional concerns within the same abstraction, avoiding the fragmentation between application logic and performance tuning prevalent in current systems.

Third, we specialize OaaS for the edge--cloud continuum through OaaS-IoT and its realization in the EdgeWeaver platform (Chapter~\ref{chapter5}). Rather than treating edge deployment as a variant of cloud execution, EdgeWeaver provides connectivity-aware invocation, edge-affinity placement policies, and object-centric state synchronization that explicitly account for intermittent connectivity, resource constraints, and geo-distribution. This positions objects as natural units for mobility, caching, and coordination across heterogeneous edge and cloud tiers.

Together, these contributions establish a cohesive framework that raises the level of abstraction in serverless computing while maintaining the flexibility, performance, and reliability required by modern distributed applications. The following chapters detail the design, implementation, and evaluation of each component, demonstrating how this integrated approach addresses the challenges identified in the literature while opening new directions for cloud application development.

\section{Summary}
\label{sec:ch2-summary}

This chapter surveyed serverless computing research across stateful systems, QoS-aware platforms, and edge-cloud deployments, revealing fundamental limitations that motivate the Object-as-a-Service paradigm. While stateful serverless approaches—actor models, pure functions, and datastore abstractions—advance beyond stateless FaaS, they fragment state management, workflow orchestration, and quality-of-service control across competing abstractions. QoS-aware systems target specific symptoms like cold starts and provisioned concurrency but lack integrated mechanisms for declarative NFR specifications and automated adaptation. Edge-cloud extensions introduce additional complexity from intermittent connectivity and resource heterogeneity, yet current platforms lack declarative mechanisms for application-specific consistency-availability trade-offs mandated by the CAP theorem.

This dissertation addresses these gaps through three integrated contributions: OaaS (Chapter~\ref{chapter3}) unifies stateful computing, dataflow orchestration, and fault tolerance within a platform-managed object abstraction; declarative NFR interfaces (Chapter~\ref{chapter4}) enable developers to specify QoS objectives with dynamic runtime optimization; and EdgeWeaver (Chapter~\ref{chapter5}) specializes OaaS for edge-cloud continuum through connectivity-aware execution and object-centric state synchronization. Together, these contributions establish a cohesive framework advancing serverless computing beyond current fragmentation while preserving developer simplicity.

%%%%%%%%%%%%%%%%%%%%%%%%%%%%%%%%%%%%%%
\chapter{Object as a Service (OaaS): Enabling Object Abstraction in Serverless Clouds\protect\footnotemark}
\label{chapter3}
%%%%%%%%%%%%%%%%%%%%%%%%%%%%%%%%%%%%%%

\footnotetext{This chapter is based on and includes material from the following publications: (1) P.~Lertpongrujikorn and M.~Amini Salehi, ``Object as a Service: Simplifying Cloud-Native Development through Serverless Object Abstraction,'' \textit{IEEE Transactions on Computers}, accepted Oct.~2025, In Press; and (2) P.~Lertpongrujikorn and M.~Amini Salehi, ``Object as a Service (OaaS): Enabling Object Abstraction in Serverless Clouds,'' in \textit{Proceedings of the 16th IEEE International Conference on Cloud Computing (IEEE CLOUD '23)}, Chicago, IL, USA, 2023, pp.~238--248. Reprinted and adapted here with permission from IEEE.}

\section{Overview}
\label{ch3-sec:overview}

The empirical findings resulted from our market survey, presented in Chapter~\ref{chapter1}, revealed that development complexity and operational maintainability constitute the most pressing challenges in cloud-native adoption, particularly for organizations with limited technical expertise. Current serverless platforms, despite their promise of simplified deployment, suffer from fundamental fragmentation: application logic resides in Function-as-a-Service (FaaS) runtimes, state management requires separate database services, and workflow orchestration demands additional coordination layers. This fragmentation forces developers to master multiple abstractions, manually integrate disparate services, and navigate complex inter-service communication patterns—directly contradicting the serverless vision of infrastructure abstraction.

This chapter introduces Object as a Service (OaaS), a unified paradigm that consolidates resource, state, and workflow management into a single object-oriented abstraction. By borrowing established object-oriented programming (OOP) concepts—classes, methods, attributes, inheritance, and polymorphism—OaaS enables developers to define entire applications through familiar constructs while the platform automatically handles deployment, scaling, and state management. We present Oparaca, an open-source prototype implementation that demonstrates how this unified abstraction streamlines cloud-native development without sacrificing performance or scalability. The chapter is organized as follows: Section~\ref{sec:oaas} defines the OaaS paradigm and its conceptual model; Section~\ref{sec:oprc} describes the Oparaca architecture, including class deployment, function invocation, and data management strategies; Section~\ref{sec:discus} discusses security, multi-tenancy, and cold start considerations; and Section~\ref{sec:evltn} evaluates Oparaca through experiments demonstrating that OaaS imposes negligible overhead while significantly simplifying development workflows.

\section{Object as a Service (OaaS) Paradigm}\label{sec:oaas}

\subsection{Conceptual Modeling of OaaS}
To realize OaaS, \textit{\underline{first}}, we need to establish the notion of \emph{cloud object} as an entity that possesses a \emph{state} (i.e., data) and is associated with one or more \emph{functions}. We empower objects to support both structured (\eg JSON records) and unstructured (\eg video) forms of state. Upon calling an object's function, OaaS creates a task that can safely take action on the state. 

\textit{\underline{Second}}, OaaS provides the \emph{class} semantic as a framework to develop objects. Inspired by OOP, the developer has to define a set of functions and states within the class. Then, an arbitrary number of objects---that is bound to the functions and states declared in that class---can be instantiated. To improve cloud software reusability and maintainability, we enable class \textit{inheritance} for cloud functions and states from other classes, plus the ability to \textit{override} any derived function. 

\textit{\underline{Third}}, OaaS offers built-in \textit{access control} to provide the ability to declare the ``scope of accessibility'' for a state or function. Importantly, when defining a set of classes, the developer can declare it within a single package that includes the access modifier to prevent unauthorized access from other packages. This is particularly useful when cloud application developers utilize imported third-party packages.

\textit{\underline{Fourth}}, OaaS enables higher-level abstractions by allowing cloud objects to be nested, where a high-level object references lower-level objects. Functions can use these references to fetch inputs or invoke dataflow functions (called \textit{macro functions}) that chain operations across lower-level objects. Unlike traditional FaaS workflows \cite{aws-sf}, macro functions determine execution flow based on dataflow rather than task (i.e., function call) dependency. Developers only define the data flow, while OaaS manages parallelism, data navigation, and state consistency transparently.

\subsection{Developing Classes in OaaS}

In OaaS, developers define one or more classes within a package using configuration languages like YAML or JSON. The package definition contains the class section and the function section. The functions section defines the configuration and deployment details of each function. The class section defines the object's structure, which includes the state and function it links to.

\begin{figure} [t]
  \centering
\includegraphics[width=0.7\linewidth]{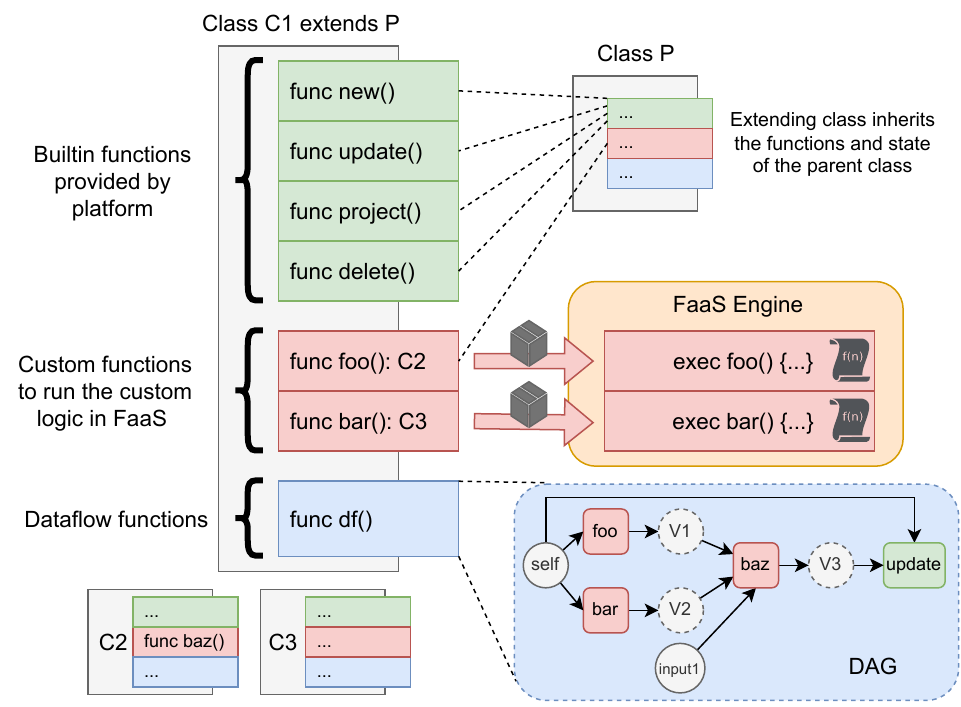}
  \caption{Different types of functions supported by OaaS.}
  \label{ch3-fig:class_representation}
\end{figure}

As shown in Figure~\ref{ch3-fig:class_representation}, OaaS supports three function types. First, \textit{built-in} functions that are provided by the platform. These functions could be the standard functions such as \texttt{CRUD} (create, read, update, and delete), which are the common data manipulation operations. The platform manages the execution of these functions without intervention from the developer. Second, \textit{custom} (a.k.a. \textit{task}) functions that are developed by developers (OaaS users) to provide their business logic. To handle the invocation of these functions, OaaS employs existing FaaS engines in its underlying layers to exploit their auto-scaling and scale-to-zero capabilities. Third, \textit{dataflow} (macro) functions are defined as a DAG representing the chain of invocations to objects.

As an example of package definition, Listing~\ref{lst:cls_exp} represents a declaration example for a package that includes one class called \texttt{video} that has a state named \texttt{mp4} (Line 6), \textit{built-in} function named \texttt{new} (Line 9), and \textit{custom} function named \texttt{transcode} (Line 1). The state \texttt{mp4} refers to video data that is unstructured data. The class has a public \textit{custom} function called \texttt{transcode}. The definitions of the \textit{custom} function are declared in Lines 15---17. The \texttt{type} of a function (Line 16) can be a \texttt{task} (or a \texttt{macro}, as noted earlier). This function creates another object instance of type \texttt{video} as an output. Line 17 declares the container image URI for executing function code.

\begin{minipage}{0.9\linewidth}
 \linespread{0.8}
\begin{lstlisting}[
    language=yaml, 
    label={lst:cls_exp}, 
    caption=\small{An example simplified script that declares \texttt{multimedia} package with a \texttt{video} class, and a \texttt{transcode} function for it in the \texttt{YAML} format. }
]
name: multimedia
classes:
  - name: video
    stateSpec:
      keySpecs:
        - name: mp4
          access: PUBLIC     
    functions:
      - function: new
        access: PUBLIC
      - function: transcode
        access: PUBLIC
        outputCls: .video
functions:
  - name: transcode
    type: TASK
    image: transcode-py:latest
    ...
\end{lstlisting}
% \vspace{-1mm}
\end{minipage}

\section{Oparaca: A Platform for the OaaS Paradigm}
\label{sec:oprc}

\subsection{Design Goals}
\label{sec:oprc_goals}

The Oparaca platform is designed with its foundational goal of providing object abstraction with two additional design goals: \textit{backward compatibility} and \textit{extensibility}. \textit{\underline{first}}, while OaaS simplifies cloud-native application development, it is not always a replacement for FaaS; thus, Oparaca supports stateless FaaS and direct data access to storage systems. \textit{\underline{Second}}, for extensibility, Oparaca decouples the control plane from the execution plane, allowing the execution plane to operate independently via standardized APIs. This platform-agnostic design accommodates various execution planes optimized for specific use cases, such as latency-constrained function calls~\cite{singhvi2021atoll} or access to hardware accelerators~\cite{yang2022infless}.

\subsection{Overview of the Oparaca Architecture}
\label{sec:oprc_arch}

\begin{figure}
  \centering
  \includegraphics[width=0.85\linewidth]{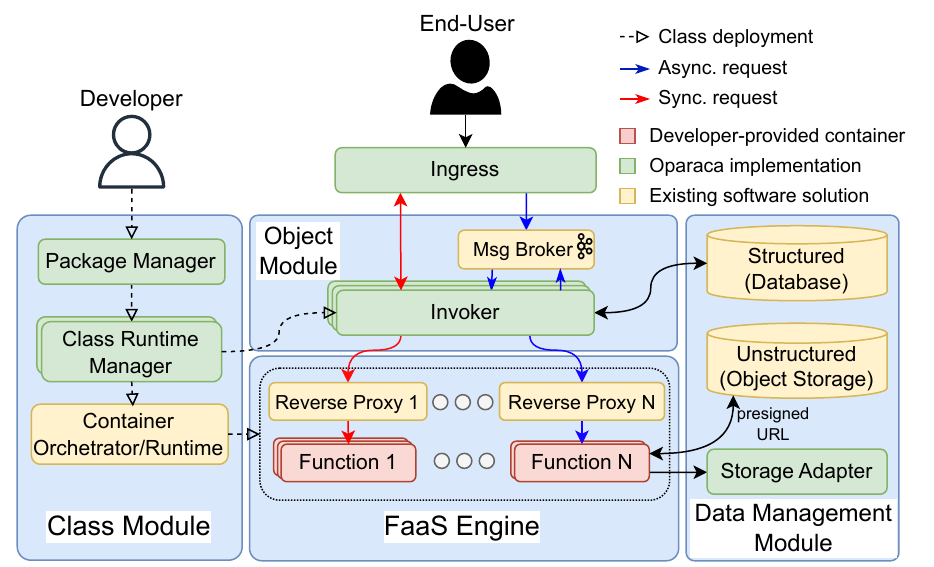}
  \caption{A bird-eye view of the Oparaca architecture. Dashed lines show actions of the developer defining classes and objects, and solid lines show actions using objects and invoking functions. % either synchronously (red arrows) or asynchronously (blue ones).
  }
  \label{ch3-fig:oaas_architecture}
\end{figure}

The Oparaca platform is designed based on multiple self-contained microservices that communicate within a serverless system. Figure \ref{ch3-fig:oaas_architecture} provides a birds-eye view of the Oparaca architecture that is composed of five modules: 
\begin{itemize} [leftmargin=*]
    \item \textbf{Class Module} serves as the interface for developers to create and manage classes and their functions.
    \item \textbf{Object Module} serves as the cornerstone of Oparaca that has two main objectives: (a) providing the ``object access interface'' for the user application to access an object(s); and (b) offering the object abstraction while transparently handling function invocation and state manipulation.
    \item \textbf{FaaS Engine} is the underlying execution engine of Oparaca, which can be any existing FaaS system (e.g., Knative).
    \item \textbf{Data Management Module} is to manage object data persistence via employing database (e.g., document database) and object storage (e.g., S3-compatible storage). To bind these storages to the functions, the Invoker abstracts data access for structured data, while the Storage Adapter is employed to handle access to unstructured data in the object storage 
    \item \textbf{Ingress Module} whose purpose is to provide a single end-point for the user application.
\end{itemize}

Details of these modules, their interactions, and how they fulfill the consistency and fault-tolerance objectives (described in Section~\ref{ch3-sec:overview}) are elaborated in the following subsections.
 
 \subsection{Class Module}
\label{sec:oprc_dep}
To define a new class and its functions in Oparaca, the developer defines them as a package definition and registers it to the \emph{Package Manager}, shown in Figure~\ref{ch3-fig:oaas_architecture}. Upon successful package validation by the Package Manager, the \textit{Class Runtime Manager} (termed \textit{CRM} for brevity) performs the class registration process that includes two operations: 

\noindent(\textbf{a}) Informing the Object Module about the new/updated class. Upon receiving a class registration, the Object Module creates a handler instance to be prepared for handling object invocation. We elaborate on this process in Section \ref{sec:oprc_inv}. 

\noindent(\textbf{b}) Registering the custom functions of the new class in the FaaS engine for future invocation. Recall that we aim to make Oparaca agnostic from the underlying FaaS engine. We design Oparaca to host a dedicated CRM for each FaaS engine. Accordingly, a new FaaS engine can be integrated into Oparaca by simply plugging its dedicated CRM into the system. When a function registration event occurs, the corresponding CRM processes this event by translating the function configuration into the specific format for that engine (e.g., Knative) and forwards it. Consequently, the underlying FaaS Engine creates the actual function runtime to be invoked by the Object Module.

\subsection{Object Module and FaaS Engine}
\label{sec:oprc_inv}

\begin{figure}[tbp]
  \centering
  \includegraphics[width=1\linewidth]{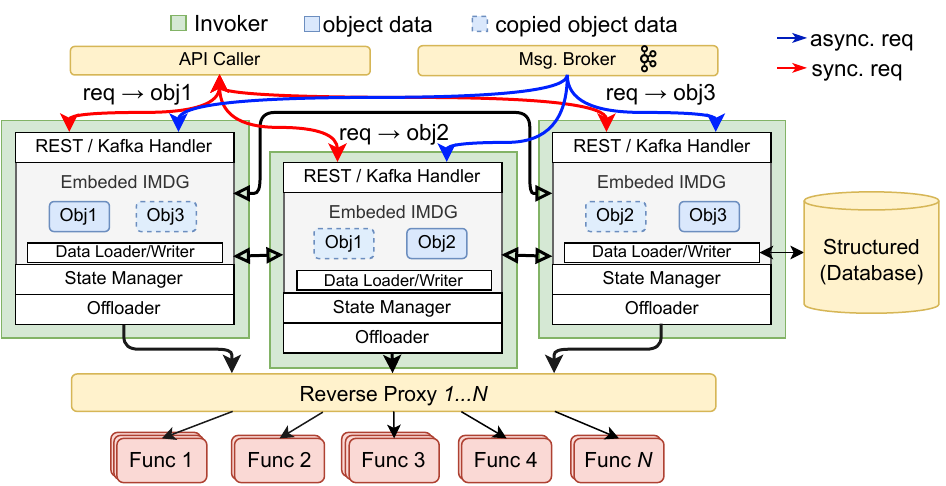}
  \caption{The cluster of Invokers replicates and distributes object data across the cluster via IMDG with consistent hashing. The invoker offloads the invocation to a corresponding FaaS function.
  }
  \label{ch3-fig:invoker}
\end{figure}

Recall that OaaS needs to support three types of functions: built-in, custom, and dataflow. Unlike built-in functions and dataflow functions that can be executed without the direct need of the FaaS engine, custom functions need to execute the developer-provided code on the FaaS engine. Thus, Oparaca requires a mechanism to utilize the FaaS engine to execute the custom function code while allowing it to access the object state transparently and with the minimum data transfer overhead. Needless to say, this mechanism also maintains the separation between the Object Module and the FaaS engine. 

To fulfill the above expectations, we design the object invocation mechanism in the Object Module by distinguishing between structured and unstructured states and managing it so that the data access overhead is minimized. We develop a hybrid approach that leverages the ``\emph{pure function}'' technique for structured data access and the ``\emph{datastore abstraction}'' technique for unstructured data access. The rationale of this design choice is that the unstructured state (i.e., BLOB) is usually large and expensive to transfer; hence, to maintain efficiency, the FaaS engine should retrieve the state directly from the object storage (e.g., S3) in a lazy, on-demand manner. This differs from the structured state, for which we include the state as an input argument to maintain a clear separation between the Object Module and the FaaS engine and let the FaaS engine maintain its statelessness.

In the Oparaca architecture (Figure~\ref{ch3-fig:oaas_architecture}), the mechanism for handling invocation and state management is managed by the \textit{Invoker} component. In particular, to offload the object invocation to the FaaS engine, Invoker bundles the request and the related structured object data as a ``\textit{task}'', as described in the next part, and passes it to the associated FaaS engine for execution.

\paragraph{\textbf{Task Generation in the Invoker}}
\label{subsubsec:task_gen}
Upon receiving a function call, the Invoker bundles the invocation request and associated object data into the task and offloads it to be executed on the FaaS engine. To further reduce the data transfer overhead of providing the object abstraction in the task generation process, we design Invokers to maintain the object data (i.e., state and metadata) in a distributed hash table~\cite{hassanzadeh2021dht}, thereby reducing the cost of data transfer in a scalable manner. As shown in Figure \ref{ch3-fig:invoker}, we equip each Invoker instance with an embedded in-memory data grid (IMDG)~\cite{zhang2015memory}. IMDG partitions the entire data space into multiple segments and distributes them across Invoker instances. The Invoker with IMDG determines the segment for a given object by consistent hashing of the object ID and assigns the object data to the selected segment. Similarly, to retrieve the object data, IMDG determines the owner of the data and then fetches it from the owner of the segment in one hop.

\paragraph{\textbf{Unstructured Data Accessing}}
\label{subsubsec:unstruc_access}
To minimize the overhead of accessing unstructured data, Oparaca allows function code to access the unstructured data on-demand and directly through a \textit{presigned URL} and \textit{redirection} mechanism. The presigned URL is the specific HTTP URL that includes the digital signature in query parameters to grant permission for anyone with this URL to access the specific data without the secret token. When a function needs to access the unstructured data, it sends an HTTP request to the storage adapter to receive the redirection response that points to the presigned URL of specific state data. Then, the function code can fetch the content directly from object storage via the given presigned URL. In addition to minimizing the overhead, using the presigned URL is important in protecting the function container from unauthorized access to other objects' data by analyzing their URL patterns.

\paragraph{\textbf{Task Completion}}
\label{subsubsec:task_completion}
After the FaaS engine completes the task, it sends the task completion data to the Invoker to update the state. If the function reports a failed task, the state remains unchanged. Otherwise, the Invoker updates the object data in IMDG and then writes it to the persistent database immediately or asynchronously.  If an invocation involves both structured and unstructured states, we use pure and datastore techniques together, which can potentially lead to ``state inconsistency challenges''. We address this challenge in section~\ref{sec:oprc_consistency}.

% Importantly, in the circumstances where an invocation involves both structured and unstructured states, we utilize pure and datastore techniques together. In this case, the invocation encompasses two different state persistence of Oparaca (\ie Object storage and Structured database in Figure~\ref{ch3-fig:oaas_architecture}) at the same time. However, this invocation type can potentially make the system prone to the ``state inconsistency challenge'' when a failure happens after the unstructured state is updated and before the structured state is changed. We address this challenge in section~\ref{sec:oprc_consisitency}.

\paragraph{\textbf{Synchronous and Asynchronous Invocation}}
As mentioned in Section~\ref{sec:oprc_goals} and shown in Figure~\ref{ch3-fig:invoker}, we designed Oparaca to offer synchronous and asynchronous function invocations. In synchronous mode, the function is executed immediately upon invocation and returns the result to the caller. Meanwhile, in asynchronous mode, the invocation ID is provided to the caller as a reference so they can check the invocation result later. The request is placed into the message broker to be reliably processed at a later time. To accommodate both modes, the Invoker utilizes the handler instance to accept the invocation request for either the REST API (synchronous) or the message broker (asynchronous). Subsequently, the handler instance forwards the request to be processed in the same way by the other part of the Invoker.

\begin{figure} [tbp]
  \centering
  \includegraphics[width=0.8\linewidth]{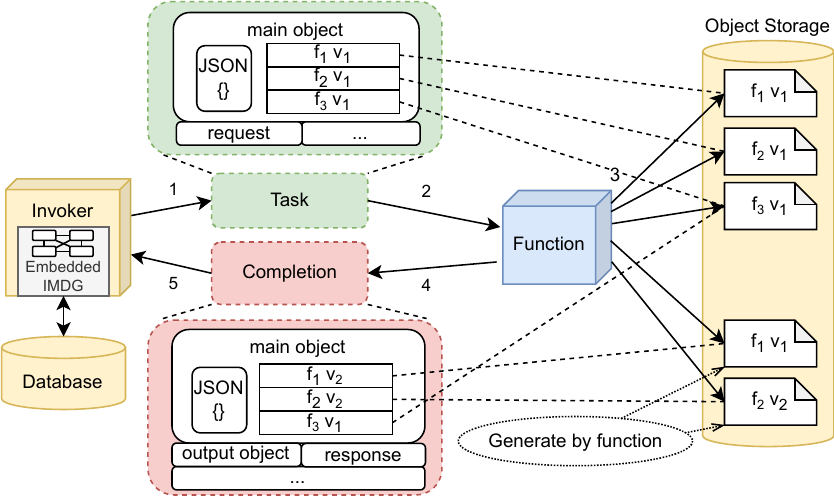}
  \caption{The process of offloading invocation task into the function runtime. Invoker bundles the request input and object state into a task and offloads it to the function to be executed. With \emph{fail-safe state transition}, when the function needs to update the file in object storage, it creates a new file and updates the corresponding version ID via the returning completion message. }
  \label{ch3-fig:invocation}
\end{figure}

\subsection{Ingress Module}
\label{sec:oprc_serve}

To provide the end user with a single access point, we position the Ingress Module in front of the cluster of Invokers. Additionally, to minimize data movement, the Ingress Module is designed to be aware of the object data distribution through consistent hashing of DHT. This allows the Ingress Module to correctly forward the object invocation request to the Invoker that owns the primary object data. As a result, the designated Invoker is able to access the data in its memory.

\subsection{Resilience Measures of Oparaca}
\label{sec:oprc_consistency}
Oparaca is prone to the data inconsistency problem that stems from both \textit{failure} and \textit{race} conditions. In this section, we describe the internal mechanisms of Oparaca designed to make it resilient against these conditions.

\paragraph{\textbf{Resilience against failure}}
% \textbf{Resilience against failure.}
Data inconsistency from failure can happen if the system stops while performing multiple update operations, causing some of the update operations to be incompletely executed. The pure function model, used for structured data in Oparaca, is inherently immune to this problem because a function returns the modified state to the platform only when its execution is complete. Nonetheless, the datastore abstraction used for the unstructured data in Oparaca is still prone to the data inconsistency problem between the structured database and object storage.

Maintaining data consistency across two data storages implies guaranteeing both storages are either successfully updated or fail for the same invocation. Otherwise (i.e., if only one of them succeeds), it leads to data inconsistency. To overcome this problem, we develop the \emph{fail-safe state transition} mechanism that disregards the data update in the object storage if Invoker fails to update the structured part of the object data in the structured database. For that purpose, the mechanism uses a two-phase versioning scheme to keep track of the unstructured data. As shown in Figure \ref{ch3-fig:invocation}, in the first phase, the mechanism creates a version ID for each file (unstructured data) and keeps them as structured data (metadata of object data) to track the current version of the file. In the second phase, which occurs upon function completion, Invoker changes all version IDs associated with the updated files (unstructured data) and then writes them to IMDG and the structured database. 

For example, consider object $o_1$ that has file $f_1$ with the version ID $v_1$. Upon function invocation, $f_1$ is updated and written to the object storage with version ID $v_2$. After the execution, the Invoker must change the version ID from $v_1$ to $v_2$ and commit the new structured object data. If any operation fails within this process, the next invocation still loads $o_1$ with version ID $v_1$, as if the previous invocation never happened. In the last step, when the invocation is complete, the Invoker purges the old and unused versions of data.

\paragraph{\textbf{Resilience against race condition}}
% \textbf{Resilience against race condition.}
Race conditions in Oparaca can occur when multiple invocations modify the same object data simultaneously, resulting in potential data inconsistency. One way to prevent this issue is by using database transactions; however, this method lacks abstraction as it allows direct function code access to the database and is tightly dependent on the type of database. 
An alternative approach to avoiding race conditions is the cluster-wide pessimistic locking mechanism to synchronize the locking state for all invokers. Nevertheless, this approach necessitates additional network communication to coordinate the locking state, which can lead to scalability issues. Alternatively, we develop an improved version of this mechanism, called ``localized locking,'' which relies on consistent hashing to direct the invocation request to the invoker that owns the primary copy of the targeted object data. Each invoker will only need to lock the object locally without additional network communication, making it more scalable than the cluster-wide version. Additionally, our localized locking approach guarantees that requests to the same object are executed in the arrival order, which is necessary in certain use cases where order matters, such as seat reservations. This is difficult to achieve with cluster-wide locking.

\paragraph{\textbf{Failure recovery in Oparaca}}
\label{sec:oprc_fault}
To further establish resilience against failures, Oparaca is equipped with a mechanism to self-recover from the failure. Broadly speaking, a function invocation failure can be recovered by simply retrying the invocation. However, this approach can cause data incorrectness owing to the execution of the function more than once. The retrying approach could be undesirable for synchronous invocations because the failure can be handled on the client side. For asynchronous invocations, however, we need to guarantee that any invocation is only executed \textit{exactly once}.  

To achieve the \textit{exactly-once} guarantee, we have to prevent three sources of the problem that are: (a) losing messages, (b) duplicating messages, and (c) processing messages more than once. Message brokers with stable storage (e.g., Kafka~\cite{kreps2011kafka}) have features that can be leveraged to address these problems. To solve the first problem, upon failure occurrence, the Invoker can detect and reprocess the incomplete request using an offset number that is automatically generated by the message broker. The offset number is the auto-incremental number based on the message's arrival order and can be used to track the message's position in the queue. The second problem of producing duplicated request messages can be resolved using the message broker's ``idempotent producer'' feature. 

However, the message broker cannot completely address the third problem. That is, the Invoker can process the same invocation request more than once when the message broker has not acknowledged the completed one before the system failure occurs. We prevent this problem by tracking the offset number of the last processed request and adding it to each object metadata. In this manner, before processing an invocation request, Invoker checks the offset number of the target object to see if it is lower than the offset number of an incoming request. When the condition is met, the Invoker can detect that it has not been processed and perform the normal operation. Otherwise, it must be skipped to avoid reprocessing.

\subsection{Dataflow Abstraction in Oparaca}
\label{sec:oprc_dataflow}

To offer a high-level abstraction to declare a workflow, Oparaca provides the dataflow abstraction as a built-in feature that enables developers to declaratively define the invocation steps as a directed acyclic graph (DAG) in a domain-specific language (DSL) with YAML format. In every step, the developer can declare the output of each invocation as a temporary variable within the workflow. Then, the next invocation can use the temporary variables from previous steps as the input or target to call the function. Upon registering a dataflow function by the developer, Invoker constructs the DAG by having the invocation step as the edge and the objects as nodes.

Upon calling the dataflow function, one of the Invokers takes on the role of orchestrator, similar to the orchestrator pattern~\cite{richardson2018microservices} in microservices.  It breaks down the dataflow into multiple lower-level invocations and forwards them based on the topological order of DAG. Using consistent hashing, the invoker can determine the address of the target object and send the request directly to another Invoker that holds the target object. When each step is completed, the orchestrator keeps track of the intermediate dataflow state to transparently operate the data exchange between invocation steps. With the orchestrator pattern, the dataflow control logic is centralized into a single invoker, simplifying the management, monitoring, and error-handling implementation. 

When using the orchestrator pattern, the exact-once guarantee may be compromised because the object data is stored separately from the dataflow state. If the guarantee is needed, Oparaca allows flagging all invocation steps as immutable. Upon handling the dataflow request, Oparaca can generate the output ID in advance for each step, making each step of dataflow execution idempotent and safely re-executable.

%Typically, the behavior of failure recovery required by workflow varies based on the use case. In a short-running dataflow, we need to make the state transitions atomic across the workflow (if the failure happens, the state should not be changed). However, long-running dataflows do not prefer to guarantee the atomicity of the state across the workflow but still want to keep the result (output) of the task before the failure happens and retry only the failed task because retrying the whole chain of the dataflow can be expensive. Supporting both of those schemes is desirable, but different methods are needed to achieve them.

%For an atomicity-preserved dataflow, Oparaca can execute the entire graph in one pass by keeping all the changed states in memory and then, persisting them to the database when the whole graph is complete. This approach does not conflict with the recovery scheme because if the Invoker that executing the dataflow fails and the state is lost, the whole graph needs to be re-executed from the start anyway. For non-atomic dataflow, Oparaca checkpoints the state after each step of dataflow is executed and then, passes the message for executing the next step through message queue, as a reliable channel.

\section{Discussion of other Concerns}\label{sec:discus}

\subsection{Security}
Certain security measures can be implemented in Oparaca to strengthen it against potential attacks. 
The \emph{first} measure is to reduce the attacking surface by limiting the necessary inbound traffic to the function container. As the function container is only accessed by the Invoker, the traffic policy can be configured to block inbound traffic except from the Invoker. The \emph{second} measure is to avoid reusing secret tokens. To prevent the function container from accessing out-of-context data via analyzing the URL path, we use the presigned URL mechanism for object storage. Thus, object storage in Oparaca is more secure than in FaaS, where the same secret key is used for every request. To secure the storage adapter, we can make the Invoker generate a unique secret token for each task, and every request for the storage adapter must be authenticated via the secret token. 

\subsection{Multi-tenancy}

The primary concern of multi-tenancy is ensuring data and resource isolation. The fundamental idea is to prevent sharing classes and functions among tenants. Since custom functions are offloaded and executed in a FaaS engine that provides strong isolation---with no shared functions---the execution environment is effectively contained within the FaaS engine. 
Regarding the Invoker and data management module, it is possible to share these components, as the data is stored separately in each class. However, depending on the billing model and isolation requirements, we can enhance security and resource isolation by separating these components for each tenant.

\subsection{Cold Start}
The developer functions and the Oparaca components can benefit from scale-to-zero to reduce the cost when there is no usage. However, this has the side effect of causing more cold starts. Since Oparaca components are shared across functions, we can effectively keep it warm to eliminate the additional cold start impact. In such a case, the cold start performance entirely depends on the underlying serverless execution engine.

\section{Performance Evaluation}
\label{sec:evltn}

% \captionsetup[subfloat]{captionskip=-0.1mm, farskip=-2mm}

\begin{figure*}[ht]
    \centering
    \subfloat[Video transcoding (sync)]{\includegraphics[width=0.48\linewidth]{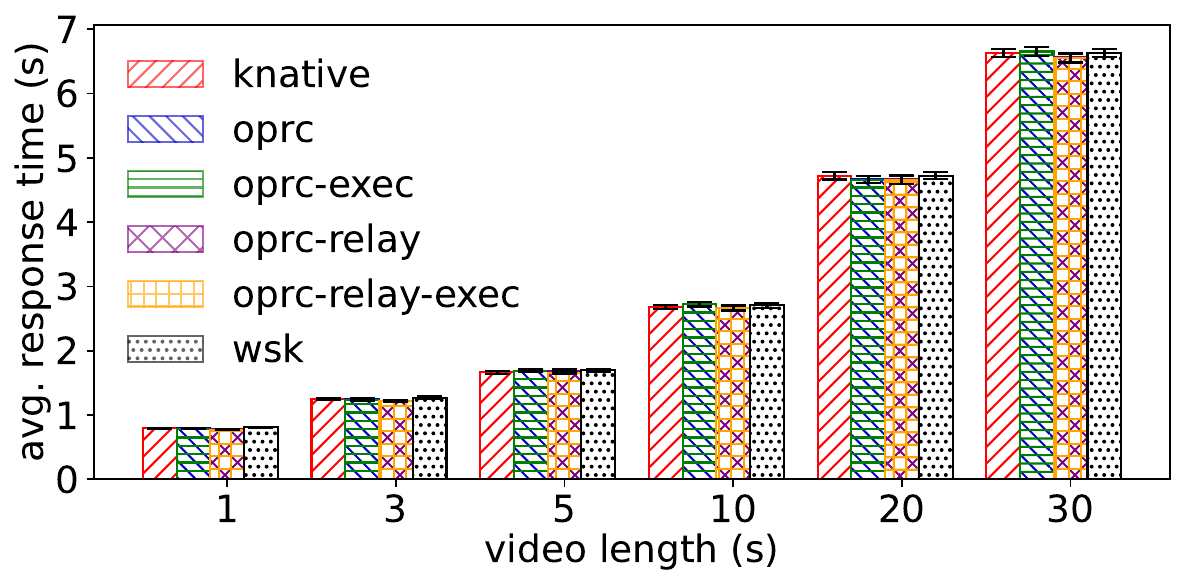}\label{ch3-fig:evlt:state_video_sync}}
    \hfill
    \subfloat[Video transcoding (async)]{\includegraphics[width=0.48\linewidth]{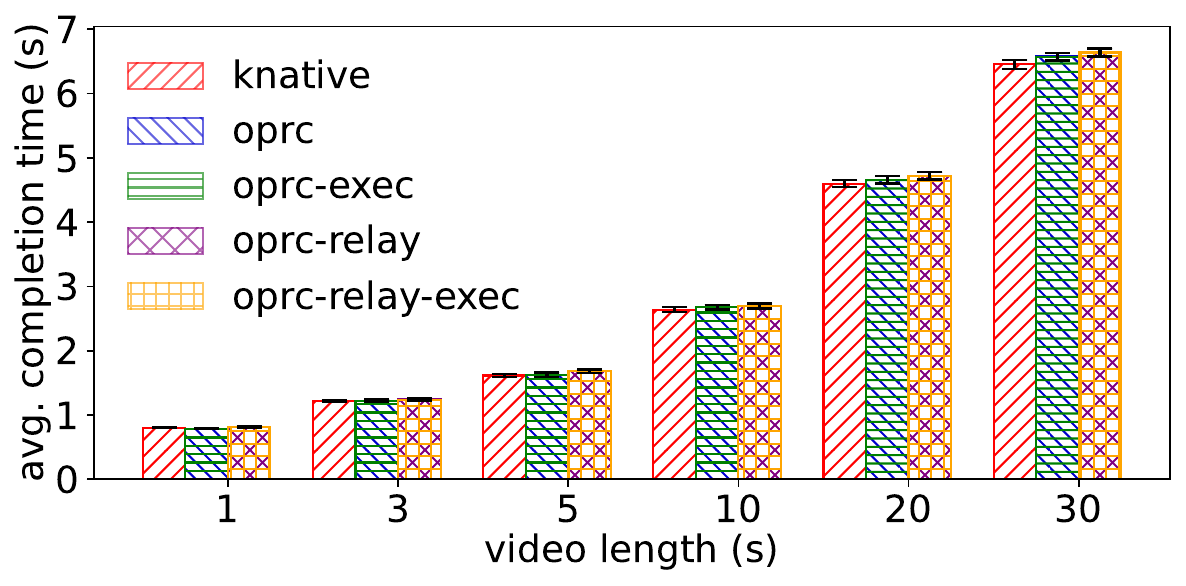}\label{ch3-fig:evlt:state_video_async}}

    \medskip

    \subfloat[Text concatenation (sync)]{\includegraphics[width=0.48\linewidth]{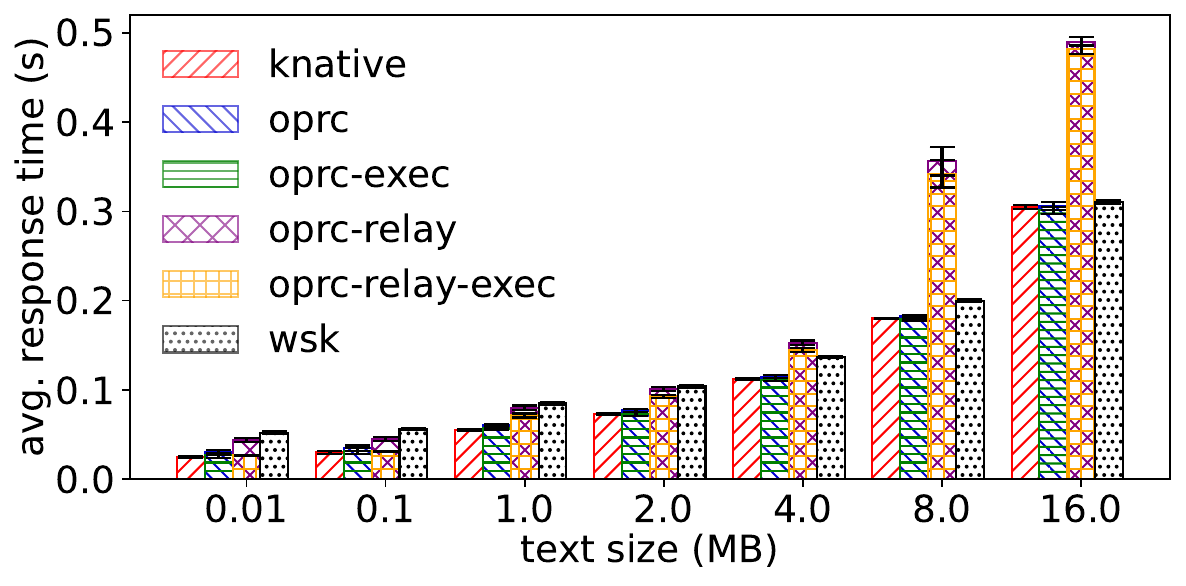}\label{ch3-fig:evlt:state_concat_sync}}
    \hfill
    \subfloat[Text concatenation (async)]{\includegraphics[width=0.48\linewidth]{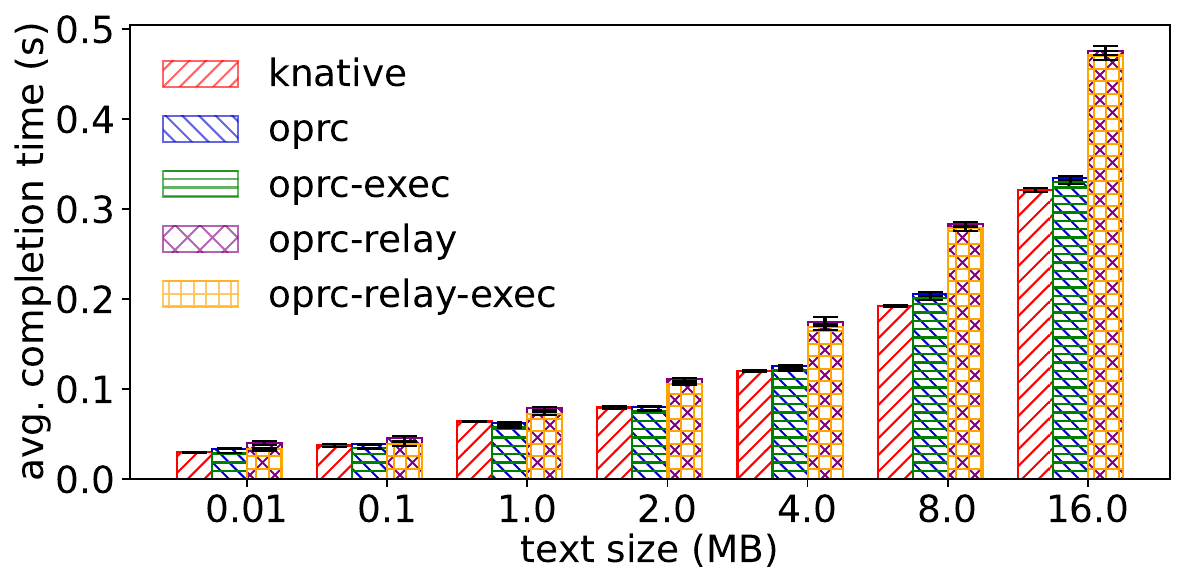}\label{ch3-fig:evlt:state_concat_async}}

    \medskip

    \subfloat[JSON update (sync)]{\includegraphics[width=0.48\linewidth]{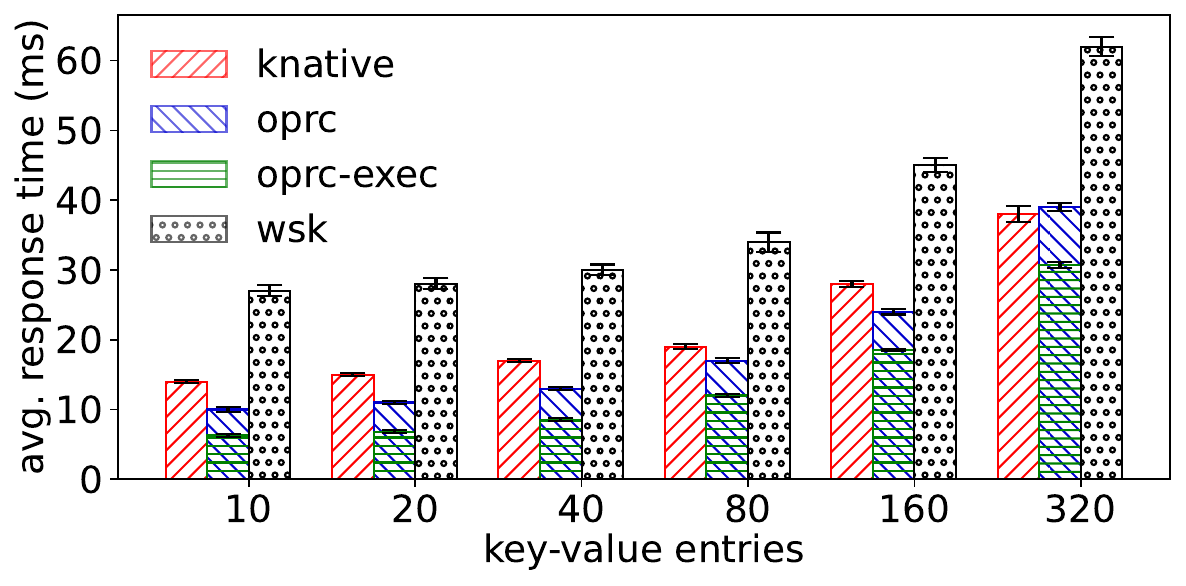}\label{ch3-fig:evlt:state_json_sync}}  
    \hfill
    \subfloat[JSON update (async)]{\includegraphics[width=0.48\linewidth]{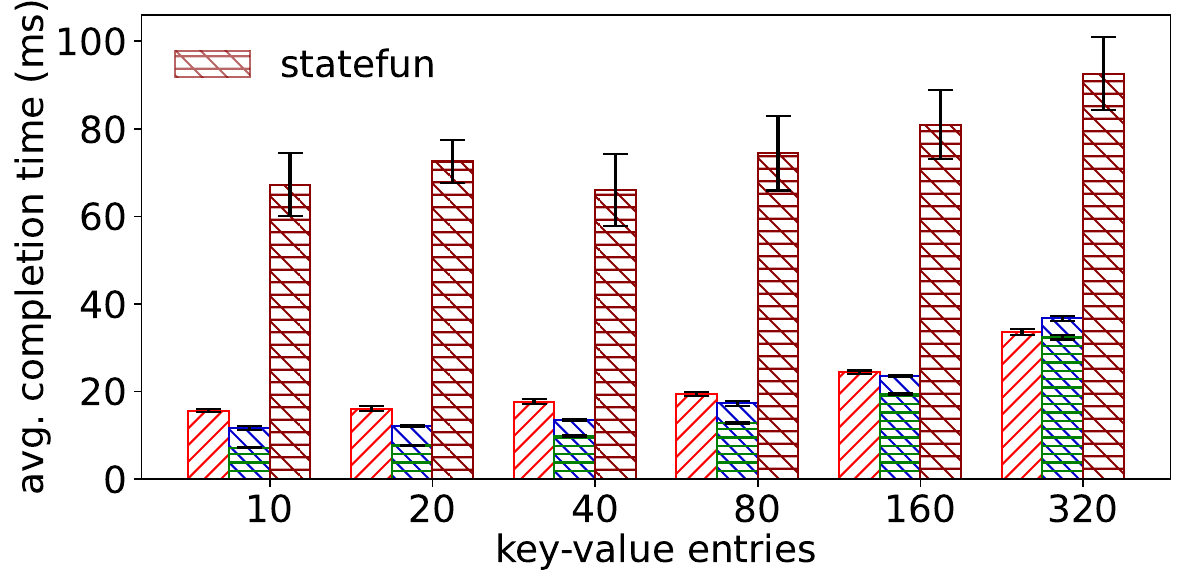}\label{ch3-fig:evlt:state_json_async}}

    \caption{
    The average execution time of functions for objects with various state sizes in synchronous and asynchronous invocations. Two versions of Oparaca are examined: the full version and the version without URL-redirection (\textit{oprc-relay}). We also capture the time used by the internal Knative in both Oparaca versions and show them with the suffix \emph{-exec} and plot them in the same bar as their Oparaca version.
    }
    \label{ch3-fig:evlt:state}
\end{figure*}

\subsection{Experimental Setup}

We deploy the Oparaca platform on 4 machines of Chameleon Cloud \cite{chameleon_cloud}, each with 2 sockets of 24-Core Intel(R) Xeon(R) Gold 6240R CPU processors that collectively have 192 cores, 768 GB memory, and SSD SATA storage. We set up the Kubernetes cluster, which includes 15 VMs with 16 vCPUs and 32 GB of memory. We made another 2 VMs for the S3-compatible storage (Minio~\cite{minio}) for unstructured data and ArangoDB (\cite{arangodb}) for structured data. Oparaca is implemented using Java with Infinispan~\cite{infinispan} for IMDG.  The source code is available at \url{https://github.com/hpcclab/OaaS}.

\paragraph{\textbf{Baselines.}}
We configure Apache Flink Stateful Function (StateFun) \cite{statefun}, OpenWhisk \cite{openwhisk}, and Knative \cite{knative} to serve as the baselines. Unlike Oparaca and OpenWhisk, which focus on API calls and event handling, StateFun is an open-source stateful serverless system focusing on stream processing. Because StateFun does not manage the function worker instances out of the box, we configure Knative to complement it. OpenWhisk and Knative are popular open-source stateless FaaS platforms that we use to represent the state management done by the developer. %However, because OpenWhisk cannot execute the function concurrently on each container instance, its performance is marginally worse than the other baseline in some experiments, and we decided to exclude it. 
 
We used Gatling\cite{gatling} for load generation and implemented three applications to serve as the workload. First is the video transcoding function, which utilizes FFmpeg~\cite{ffmpeg}, a CPU-intensive application. The second is a text concatenation function that concatenates the content of a text file (state) with an input string. This function represents a highly IO-intensive workload. Third is the JSON update function, which uses only structured data in JSON and is used to insert key-value pairs into the JSON state data randomly. The remaining workload characteristics are specific to each experiment and are explained in the respective sections. All three functions are implemented in the Python language. 
% For asynchronous invocation, since that system responds immediately after taking the request, we record the completion time in a database to be extracted when all requests are done.
%As StateFun does not support unstructured data as the state, we exclude it for the video transcoding and text concatenation functions. StateFun only executes the function call asynchronously. Hence, instead of reading it from Gatling, we calculate the execution time by subtracting the difference between request and response timestamp in Kafka topics. 

\subsection{Analyzing the Imposed Overhead of Oparaca}
\label{sec:evltn:ovh}
The abstractions provided by Oparaca are not free of charge and introduce some time overhead to the applications using these abstractions. In this experiment, our aim is to measure this overhead and see how the efficient design of Oparaca can mitigate this overhead. The latency of a function call is the metric that represents the overhead. We mainly study two sources of the overhead: (a) The \emph{state data size} that highlights the overhead of OaaS in dealing with the data, and (b) The \emph{concurrency of function calls} that highlights the overhead of the Oparaca system itself.

%We examine three types of objects: (i) An object with a one-second-long video file (105 KB with resolution 1920$\times$1080) as its state and a \texttt{transcoding} function, which exhibits a compute-intensive behavior; (ii) An object with a text file (1 MB) as its state and a function that \texttt{concatenates} the state with its input string (8 Bytes) argument. Because the processing time is only a fraction of the data loading time, we consider it as data-intensive; (iii) An object with structured (JSON) data as the state and a JSON \texttt{update} function that doubles a set of persisted random key-value pairs.

\begin{figure*}[ht]
    \centering
    \subfloat[Video transcoding function (sync)]{\includegraphics[width=0.48\textwidth]{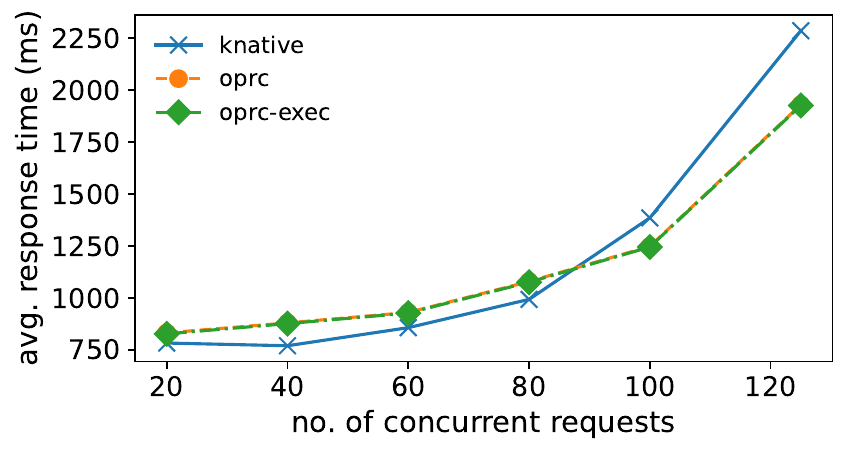}\label{ch3-fig:evlt:ccr_video_sync}}
    \hfill
    \subfloat[Video transcoding function (async)]{\includegraphics[width=0.48\textwidth]{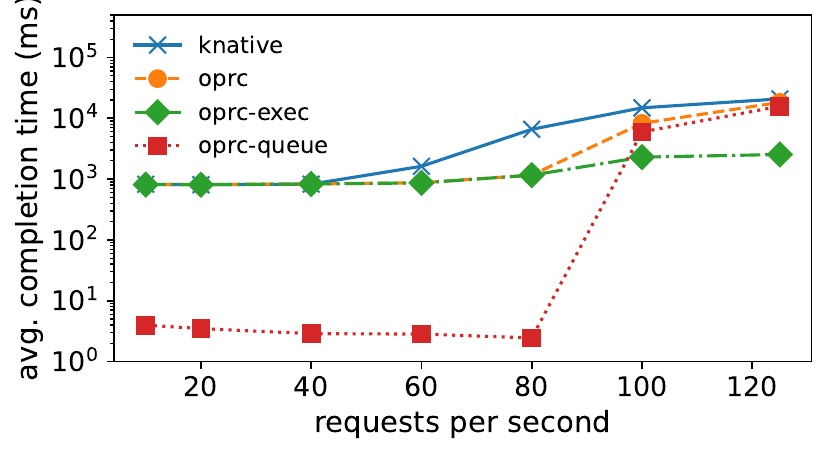}\label{ch3-fig:evlt:ccr_video_async}}
    
    \medskip
    
    \subfloat[Text concatenation function (sync)]{\includegraphics[width=0.48\textwidth]{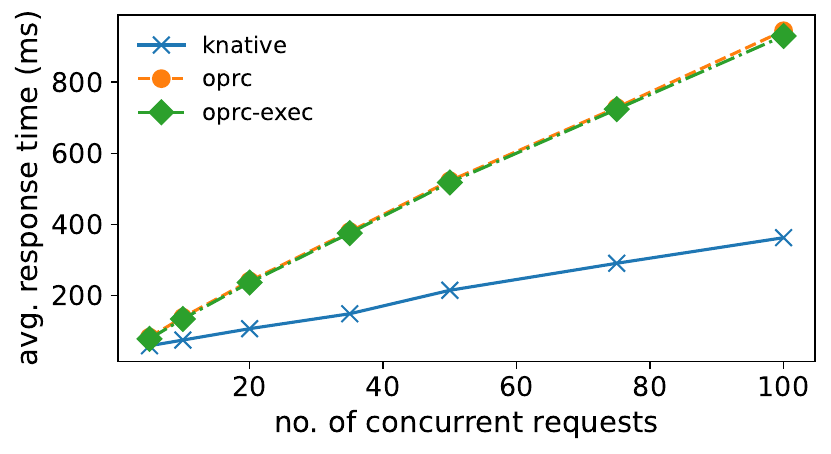}\label{ch3-fig:evlt:ccr_concat_sync}}
    \hfill
    \subfloat[Text concatenation function (async)]{\includegraphics[width=0.48\textwidth]{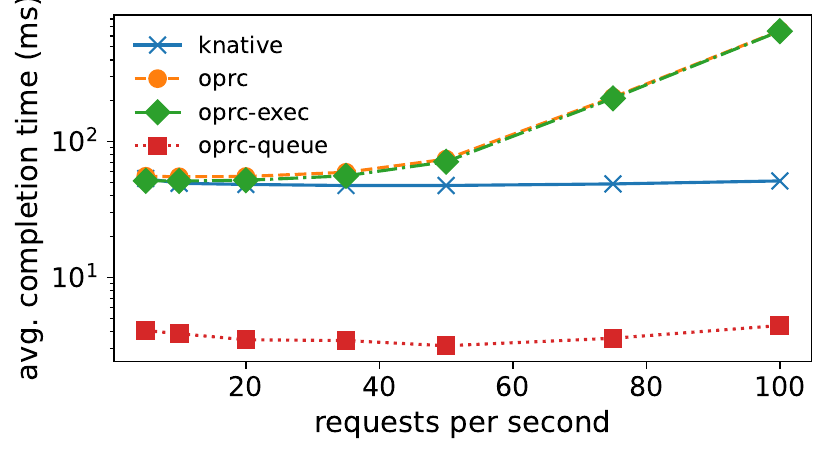}\label{ch3-fig:evlt:ccr_concat_async}}

    \medskip

    \subfloat[JSON update function (sync)]{\includegraphics[width=0.48\textwidth]{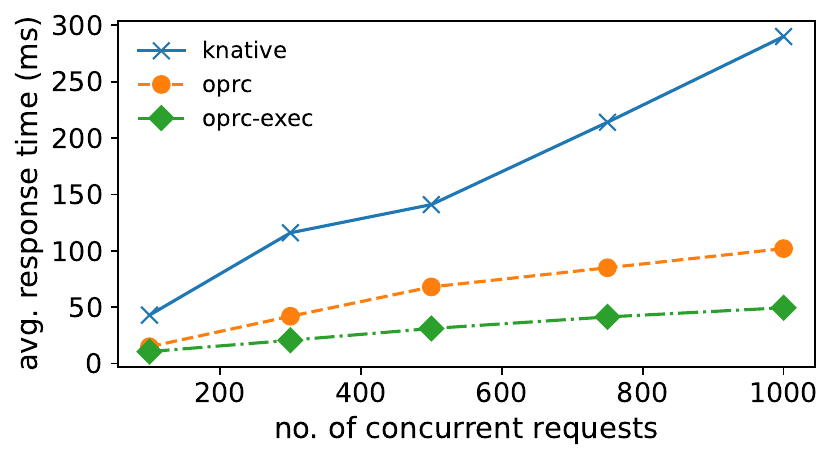}\label{ch3-fig:evlt:ccr_json_sync}}
    \hfill
    \subfloat[JSON update function (async)]{\includegraphics[width=0.48\textwidth]{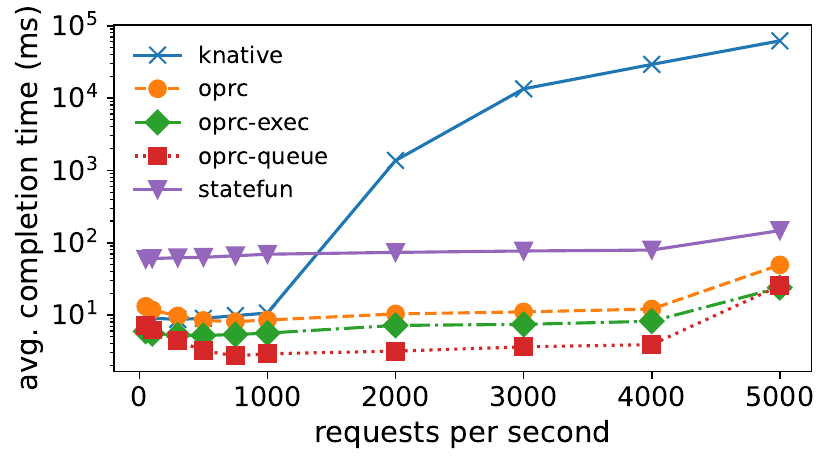}\label{ch3-fig:evlt:ccr_json_async}}
  
  \caption{The average completion time of functions upon varying the rate of incoming requests in synchronous and asynchronous invocations. \textit{oprc-queue} is the queuing time that requests stay within the message queue}
  \label{ch3-fig:evlt:ccr}
\end{figure*}

\paragraph{\textbf{The impact of changing the state size}} As shown in Figure~\ref{ch3-fig:evlt:state}, to generate objects with various state sizes, we increased the input video length from 1---30 seconds. To remove the impact of video content on the result, the longer videos were generated by concatenating the same 1-second video. Similarly, the text files are from 0.01---16 MB. For the JSON object, the key and value sizes are 10 and 40 bytes, respectively, and the number of key-value pairs varies from 10---320 pairs. 
To concentrate only on the overhead of data access and avoid other sources of overheads, we configure Gatling to assign only one task at a time and set it to repeat this operation 100 times. To analyze the improvements offered by the URL redirection, we examine two versions of Oparaca: the full version (expressed as \emph{oprc}) and without URL redirection (expressed as \emph{oprc-relay}). The horizontal axes represent different state sizes for video, text, and JSON, respectively, and the vertical axes represent the average response/completion time (latency).

In Figure \ref{ch3-fig:evlt:state}, the average task execution time increases for larger state sizes. For the video transcoding function, all of the platforms perform with similar latency, which is expected because of the compute-intensive nature of the video transcoding that dominates the completion time. In a text concatenation function, however, Knative performs slightly better than Oparaca because of the overhead of unstructured state access by the redirection of the presigned URL. However, if we compare Oparaca with another version that uses a relay mechanism to provide the state abstraction, it performs much lower than its alternative with an average of 30\% lower response time. Lastly, we can see all the described trends happen similarly for synchronous and asynchronous request types.

In the JSON update function (Figures~\ref{ch3-fig:evlt:state_json_sync} and \ref{ch3-fig:evlt:state_json_async}), Oparaca can perform with lower latency than Knative because the function does not need to fetch the object data from the database because of the pure function semantic. Nevertheless, Knative can catch Oparaca by increasing the key-value entries to 320. The reason is that the gain from eliminating the database connection is surpassed by the overhead of moving the data to the function code for larger records. OpenWhisk and Knative have the same pattern because both of them are FaaS, but OpenWhisk performs significantly worse. In Figure~\ref{ch3-fig:evlt:state_json_async}, the Statefun shares the same pattern with Oparaca with the consistent gap because it also relies on local storage to keep the function state without the need to fetch the data from the database. We also observe that Statefun performance degraded compared to our initial results \cite{lertpongrujikorn2023object}. This is because the storage hardware being used for the experiment has a lower through, which impacts the performance of Statefun.

\begin{figure*}[ht]
    \centering
    \subfloat[Speedup results by horizontal scaling]{
    \includegraphics[width=0.49\linewidth]{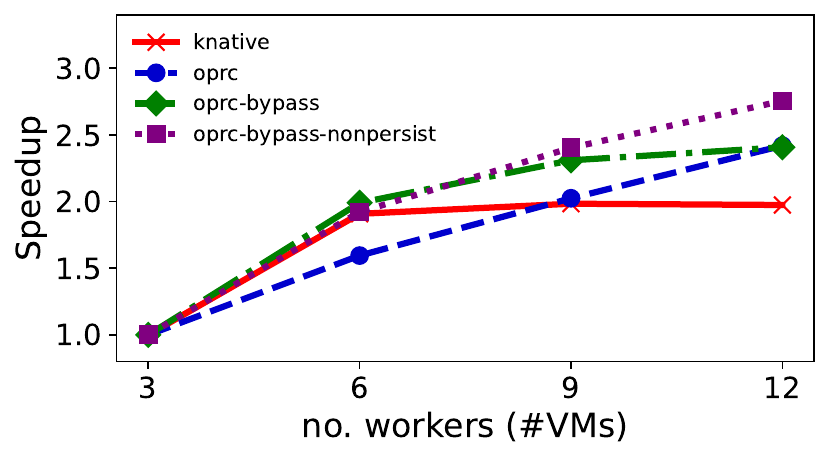}\label{ch3-fig:evlt:spd_json}
    }
    % \hfill
    \subfloat[Throughput results from horizontal scaling]{
    \includegraphics[width=0.49\linewidth]{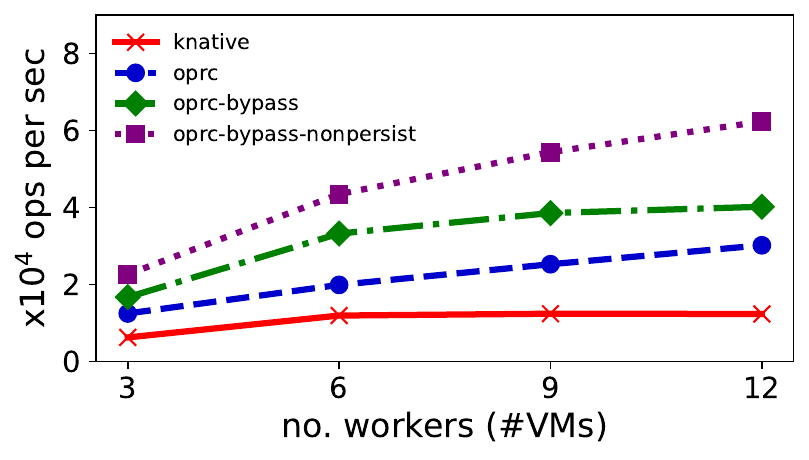}\label{ch3-fig:evlt:rps_json}
    }
    \caption{Evaluating the scalability of the OaaS platform against other baselines.}
    \label{ch3-fig:evlt_scl_up}
\end{figure*}

\begin{figure*}[ht]
    \centering
    \includegraphics[width=0.5\linewidth]{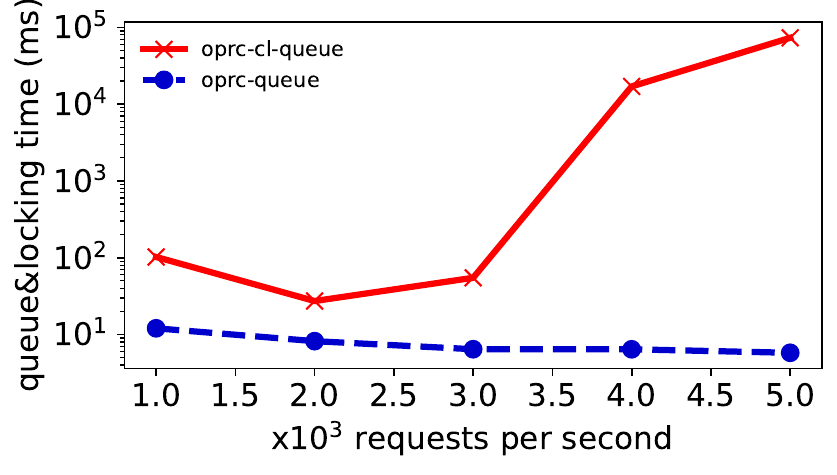}
    \caption{Evaluating the performance of localized locking compared to cluster-wide locking}
    \label{ch3-fig:evlt:lock}
\end{figure*}

\paragraph{\textbf{The impact of concurrent function invocations}} on the Oparaca overhead is shown in Figures \ref{ch3-fig:evlt:ccr}. In synchronous invocation, we increase the number of concurrent invocations of the same function (horizontal axes), whereas, for asynchronous invocation, concurrency depends on the system implementation which cannot be forced directly; thereby, we use the request arrival rate to increase the concurrency of invocations. To remove the impact of any randomness, We disabled the auto-scaling and limited the number of worker instances to 6. We also exclude OpenWhisk from this section because the Python runtime in OpenWhisk does not support container-level concurrency. 

For the transcoding function (Figures \ref{ch3-fig:evlt:ccr_video_sync} and \ref{ch3-fig:evlt:ccr_video_async}), at the low concurrency levels ($<$ 80 invocations), Oparaca has average response times higher than Knative, but for the higher concurrency levels, the response time of Knative grows faster than Oparaca due to computing resource limitations. Oparaca doesn't need to fetch video file metadata, giving it an edge at high concurrency. In the concatenation function (Figures \ref{ch3-fig:evlt:ccr_concat_sync} and \ref{ch3-fig:evlt:ccr_concat_async}), however, this phenomenon does not happen. The difference is that text concatenation is IO-intensive and desires high network bandwidth. The overhead of unstructured data access overwhelms the performance gain from eliminating structured data fetching.  

For the JSON update function (Figures \ref{ch3-fig:evlt:ccr_json_sync} and \ref{ch3-fig:evlt:ccr_json_async}), Oparaca can effectively reduce the latency by eliminating the need to fetch from the database. In Figure~\ref{ch3-fig:evlt:ccr_json_async}, because Statefun also shares this invocation scheme and, therefore, offers less completion time than Knative. However, since it relies on local storage to keep the state, while Oparaca uses the memory, Statefun's completion time is higher than Oparaca's. %We observe a high completion time as Knative's resource gets saturated when there are more than 2000 requests per second. 

% Another aspect that we can see in asynchronous invocation with transcoding and JSON update functions is that when the request rate is higher than the system throughput, Oparaca will keep requests in the queue instead of loading them to overwhelm the execution module. As a result, the execution time does not much increase, but the queue time increases instead. This is good for the execution module because it can reduce the peak of memory usage and the chance of system failure.

 In sum, Oparaca improves performance by eliminating database fetching but adds overhead by accessing unstructured data for secure state abstraction. Depending on the workload, this can either improve or impair object function invocation performance. The overhead may outweigh I/O-intensive workloads, but Oparaca can improve latency by up to $2.27\times$ compared to Knative for workloads without unstructured data.

\vspace{2mm}
\noindent
\colorbox{blue!10}{
\parbox{\linewidth}{
\underline{\textbf{Takeaway}:} \emph{Object abstraction can be provided with an insignificant latency overhead for objects with only a structured state. The main object overhead occurs as a result of securing unstructured data access.}
}}
\vspace{2mm}

\subsection{Scalability of the Oparaca Platform}

\label{sec:evltn:scl}

To study the scalability, we scale out the Kubernetes workers from 3---12 VMs, each with 16 vCPU cores (in total 48---192 vCPUs). We measured throughput and speedup metrics, focusing on the JSON update function, which does not rely on slow object storage, which becomes the bottleneck of this experiment. We measure the throughput by continually increasing the concurrency until the throughput stops growing (Figure \ref{ch3-fig:evlt:rps_json}). We assume three VMs as the base speedup=1, and the speedup of other cluster size is calculated with respect to the base value. Moreover, we add two other versions of Oparaca: first is \textit{oprc-bypass} that uses a standard Kubernetes deployment as its underlying function execution instead of Knative; Second is \textit{oprc-bypass-nonpersist} that does not persist the object data to the database to measure if Oparaca is not bottleneck by the database write operation. 

According to Figure~\ref{ch3-fig:evlt:spd_json}, the speedup of Knative plateaus after reaching 6 VMs. We realized that this plateau is attributed to the database write operation throughput bottleneck. Conversely, Oparaca exhibits the potential for higher speedup enhancement due to its reliance on the distributed in-memory hash table to consolidate data for batch write operations. This approach can boost maximum throughput by up to $3.27\times$ when comparing \textit{oprc-bypass} with \textit{knative}. 

Figure \ref{ch3-fig:evlt:rps_json} shows that \textit{oprc-bypass} yields a higher throughput over the baseline Oparaca. This is because Oparaca sends task data through the Knative internal proxy to offload the task to Knative. While this setup allows for scale-to-zero functionality, bypassing these components leads to even higher throughput. Furthermore, by disabling the database writing operation, which is the bottleneck, \textit{oprc-bypass-nonpersist} can achieve even higher throughput. Although there isn't linear scalability due to the limitations of the database write performance, Oparaca significantly improves maximum throughput compared to traditional FaaS systems.

\vspace{2mm}
\noindent
\colorbox{blue!10}{
\parbox{\linewidth}{
\underline{\textbf{Takeaway}}: \emph{In addition to offering a higher-level abstraction, Oparaca can improve the throughput and response time of its underlying Knative engine via reducing database operations, thereby, mitigating its bottleneck.
}
}
}
\vspace{2mm}

\subsection{Performance of Localized Locking}

To analyze effectiveness of localized locking, we created a variation of Oparaca, called \textit{oprc-cl}, that has cluster-wide locking and evaluated it using a cluster of 12 Invokers while increasing the request arrival rate to measure the locking overhead. To generate requests involving the locking mechanism, we created multiple requests targeting the same object. %Since the localized locking involves Kafka, we have to measure the time of invocation requests passing through Kafka. 
From Figure~\ref{ch3-fig:evlt:lock}, the overhead of localized locking remains mostly constant, while the overhead of cluster-wide locking rises for higher request rates. %The overhead decreases as arrival rates increase because the load generator and Invoker can be delayed longer with a higher system load, reducing the necessary locking time. 
The cluster-wide version does not exhibit this behavior, as the network communication overhead limits the throughput and hinders high invocation rates.

\vspace{-4mm}
\subsection{Case Study: Development Efficiency Using OaaS}
\label{sec:evltn:dev}

\begin{figure}[tbp]
  \centering
  \includegraphics[width=0.8\linewidth]{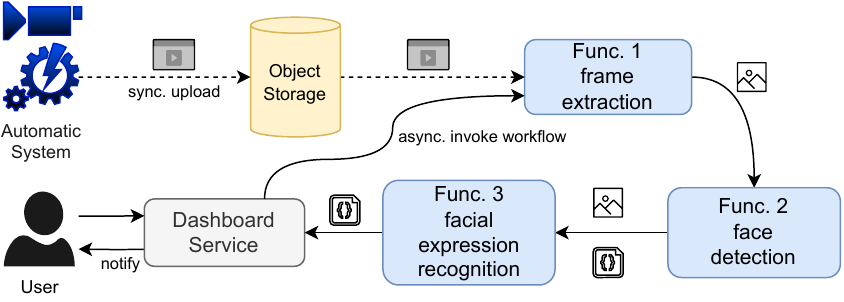}
  \caption{The use case of developing a face expression recognition workflow for an input video.}
  \label{ch3-fig:evltn:scn}
\end{figure}

In this part, we provide a real-world use case of object development using OaaS and its FaaS counterpart and then demonstrate how OaaS makes the development process of cloud-native serverless applications easier and faster.

\paragraph{\textbf{\underline{Case Study \# 1. expression detection system.}}} This case study is a video processing workflow that performs face detection and facial expression recognition. Figure \ref{ch3-fig:evltn:scn} shows the automatic system uploads the video file to the object storage to be processed by the workflow of functions. 
The workflow includes: \texttt{Func\_1} to split the video into multiple image segments; \texttt{Func\_2} to detect the face on each sample image frame; and \texttt{Func\_3} to perform facial expression recognition on the detected face image and generate a \texttt{JSON} format label. These functions must persist their output object so that the next function in the workflow can consume it.
 
\vspace{1mm}
\noindent\textbf{FaaS implementation.} The developer must implement the following steps: (a) Configuring cloud-based object storage and managing access tokens. (b) Implementing business logic to respond to trigger events. (c) Manage data within the functions that involve three steps: allocating the storage addresses, authenticating access to the object storage, and performing fetch and upload operations to the allocated addresses. 
Upon implementing these functions, the developer has to connect them as a workflow via a function orchestrator service. Finally, the dashboard service invokes the workflow upon receiving a user request and collects the results. 
	
\vspace{1mm}
\noindent\textbf{OaaS implementation.} The developer defines three classes, namely \texttt{Video}, \texttt{Image}, and \texttt{Expression}, in the form of the three following classes: 
(a) \texttt{Video} class with \texttt{frame\_extract()} functions; and a macro function, \texttt{df\_exp\_recog(detect\_interval)}, that includes the whole dataflow of function calls, with the requested sampling period as its input, and an \texttt{expression\_data} object as the output.
(b) \texttt{Image} class with the \texttt{face\_detect()} and \texttt{exp\_recognize()} functions. 
(c) \texttt{Expression} class that does not require any function. 
The dashboard service calls the \texttt{wf\_exp\_recognize(detect\_interval)} dataflow function directly using the object access interface and receives the \texttt{Expression} object as the output.
We note that the developer does not need to be involved in the data locating and authentication steps when developing the class functions because of the abstraction that OaaS provides.

\begin{figure}[t]
    \centering
    \subfloat[FaaS-based]{
        \includegraphics[width=0.8\columnwidth]{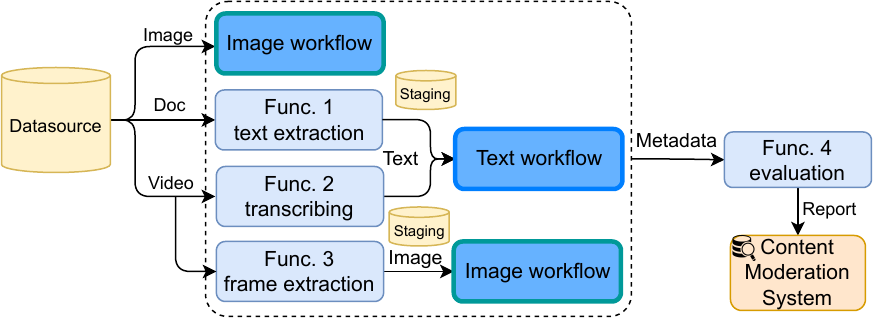}
        \label{ch3-fig:evlt:case-study2-faas}
    }
    \vspace{3mm}
    \subfloat[OaaS-based]{
    \includegraphics[width=0.8\columnwidth]{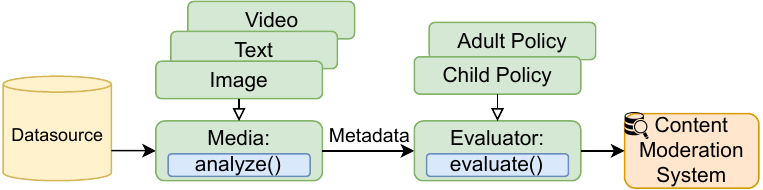}
        \label{ch3-fig:evlt:case-study2-oaas}
    }
    \caption{The automatic content moderation system.}
    \label{ch3-fig:evlt:case-study2}
\end{figure}

\paragraph{\textbf{\underline{Case Study \# 2. content moderation system.}}} Moderating the content at scale in various formats, including image, document, and video. We first present how the application is deployed in FaaS, the limitations of this approach, and how OaaS can resolve those limitations.

\vspace{1mm}
\noindent\textbf{FaaS implementation.} To simplify complex multimedia processing workflows~\cite{aws-cms}, we abstract the workflow to extract the metadata from the image files as \texttt{Image workflow}, and the workflow to extract metadata from the text files as \texttt{Text workflow} as shown in Figure~\ref{ch3-fig:evlt:case-study2-faas}. Before using both workflows, the content must be pre-processed to extract raw data via using the pertinent FaaS functions: (a) \texttt{text extraction} to extract text from the document. (b) \texttt{transcribing} to extract text from the video. (c) \texttt{frame extraction} to sample image frame from the video. After feeding the data into both workflows to extract the metadata, we use the \texttt{evaluation} function to generate a report to the content moderation system.

The FaaS implementation has three main drawbacks: \textbf{(A)} developers must explicitly manage application state and data using separate storage services that increase the complexity. \textbf{(B)} even though the common workflow can be reused, the intermediate data management is not abstracted. That is, if the developer needs to separate/change the staging storage, it must be done manually. \textbf{(C)} functionalities may require numerous and heterogeneous FaaS deployments---e.g., requiring a separate workflow for each content type, where the Video format requires a more complicated workflow than the other types. These drawbacks complicate development, deployment, and management as the application evolves to handle various document types and integrates more functionalities and options (e.g., using multiple evaluation services instead of one).

\vspace{1mm}
\noindent\textbf{OaaS implementation.} To demonstrate the efficacy of OaaS, we transform the given FaaS-based solution into OaaS with minimal effort to resolve the aforementioned drawbacks.
\begin{itemize}[leftmargin=*, noitemsep, topsep=0.5pt]
\item\textbf{Workflow Construct.}
We encapsulate related FaaS functions and states into classes representing two key functionalities: \texttt{Media} to extract the metadata and \texttt{Evaluator} to evaluate metadata and report to the content moderation system. The two classes form the critical path of the application processing pipeline, as shown in Figure \ref{ch3-fig:evlt:case-study2-oaas}.

\item\textbf{Object Encapsulation.} We use inheritance and polymorphism to enhance software reuse by encapsulating FaaS functions and states in classes derived from two base classes. This approach hides the need for storage services behind object abstraction and allows their implementation to be managed in the cloud. It simplifies development, as developers only build the processing pipeline once in the base classes. They can then focus on specific functionalities in the derived classes, avoiding \textbf{redundant} implementation when adding new content types or evaluator services.
\end{itemize}

\vspace{2mm}
\noindent
\colorbox{blue!10}{
\parbox{\linewidth}{
\textbf{\underline{Takeaway}:} \emph{The OaaS paradigm aggregates the state storage and the function workflow in the object abstraction and enables cloud-native dataflow programming. Thus, developers are relieved from the burden of state management, learning the internal mechanics of the functions and pipelining them. 
}
}
}
\vspace{2mm}

\section{Summary}\label{sec:conclsn}

In this research, we presented the OaaS paradigm, which aims to simplify state and workflow management for cloud-native applications. Our prototype, Oparaca, supports both structured and unstructured data types, ensuring consistency through fail-safe state transitions. It also provides secure and low-overhead management for unstructured data using presigned URLs and redirection mechanisms. 
For structured data, it employs the pure function scheme to transparently manage application data to the developer code and using DHT and consistent hashing to scalably cache the object data and improve the data locality. To make the Oparaca fault-tolerant, we developed the \textit{exactly-once} and \textit{localized locking} schemes. To support cloud-native workflow, Oparaca enables declarative dataflow abstraction that hides the concurrency and synchronization concerns from the developer's perspective. The evaluation results demonstrate that Oparaca streamlines cloud-native programming and is ideal for use cases that require persisting the state or defining a workflow. Oparaca offers scalability with negligible overhead, particularly for compute-intensive tasks. With the core OaaS abstraction and Oparaca architecture in place, Chapter~\ref{chapter4} builds on these foundations with declarative non-functional requirement management, and Chapter~\ref{chapter5} extends OaaS across the edge--cloud continuum.

\chapter{SLA-Driven OaaS: Declarative Performance Control for Cloud-Native Applications\protect\footnotemark}
\label{chapter4}

\footnotetext{This chapter is based on: P.~Lertpongrujikorn, H.~D.~Nguyen, and M.~Amini Salehi, ``Streamlining Cloud-Native Application Development and Deployment with Robust Encapsulation,'' in \textit{Proceedings of the ACM Symposium on Cloud Computing (SoCC '24)}, Redmond, WA, USA, 2024. Reprinted and adapted here with permission from ACM.}

\section{Overview}
\label{ch4-sec:overview}

While Chapter~\ref{chapter3} demonstrated how the Object as a Service paradigm unifies application logic and state management, a critical limitation remains: \textit{developers still lack systematic control over non-functional requirements such as performance, availability, and consistency}. The empirical studies in Chapter~\ref{chapter1} revealed that practitioners prioritize service quality assurance and unified maintainability, expecting platforms to provide high SLA offerings and rich configurability for performance optimization without requiring deep infrastructure expertise. Current serverless platforms force developers into iterative trial-and-error cycles—deploying applications, measuring performance, manually adjusting configurations, and redeploying—a process that contradicts both productivity goals and the serverless promise of automated resource management.

This chapter extends OaaS with a declarative Non-functional Requirement (NFR) interface that enables developers to specify desired outcomes—availability targets, throughput requirements, consistency guarantees, and latency constraints—through high-level, measurable specifications. The platform then automatically configures and adapts deployments to meet these requirements, eliminating manual tuning while establishing a symbiotic relationship between developers and cloud providers. We present the enhanced Oparaca system that realizes NFR-driven OaaS, demonstrating how declarative specifications can be systematically translated into runtime enforcement mechanisms. The chapter is organized as follows: Section~\ref{sec:approach} presents the design rationale for unifying abstraction and NFR control; Section~\ref{sec:architecture} describes the Oparaca architecture for NFR enforcement, including SLA specification interfaces, class runtime management, and automated optimization strategies; and Section~\ref{sec:evaluation} evaluates NFR enforcement demonstrating that Oparaca automatically satisfies declared throughput, latency, and availability requirements while reducing deployment complexity and eliminating manual tuning cycles.

\section{Design Rationale: Unifying Abstraction and NFR Control in OaaS}
\label{sec:approach}
To establish an agile and cost-efficient application delivery, the limitations of current FaaS platforms presented in Chapter~\ref{chapter1} (specifically, fragmented state management and lack of NFR control) must be properly addressed. In this section, we propose solutions for each challenge and then combine them to form a novel approach for FaaS-based application development and deployment. 

\begin{table*}[ht]
    \centering
    \small
    \begin{tabular}{|p{2.7cm}|p{2.2cm}|p{1.3cm}|p{8cm}|}
        \hline
        \textbf{Name} & \textbf{Value Type} & \textbf{Unit} & \textbf{Definition}\\ \hline \hline
        \multicolumn{4}{|l|}{\textit{NFR Specifications}}\\ \hline
        \textbf{Throughput} & \textbf{Integer} & \textbf{Rps} & \textbf{Minimum number of invocations guaranteed to be executed per second}\\ \hline
        \textbf{Availability} & \textbf{Real} & \textbf{Percent} & \textbf{The percentage of time an object/function must be available for service}\\ \hline
        \textbf{Locality} & \textbf{\{Local, None\}} & \textbf{N/A} & \textbf{How function invocations are dispatched with respect to object state location}\\ \hline\hline
        \multicolumn{4}{|l|}{\textit{Deployment Constraints}}\\ \hline 
        \textbf{Persistent} & \textbf{Yes/No} & \textbf{N/A} & \textbf{Should the data associated with the object be persistent}\\ \hline
        \textbf{Runtime Req.} & \textbf{Dict} & \textbf{N/A} & \textbf{Specific object runtime configuration (e.g., choice of FaaS engine)}\\ \hline
        Budget & Integer & Credit & Object deployment and operation budget; all costs must not exceed this value\\ \hline
        Consistency & Enumerate & N/A & Object consistency model: eventual, sequential, linearization, or none\\ \hline
        Jurisdiction & Enumerate & N/A & Candidate places to deploy an object\\ \hline
        Data Encryption & Enumerate & N/A & Specify or disable the encryption algorithm for the stored data\\ \hline
    \end{tabular}
    \caption{Potential Non-functional requirements and constraints. Those with bold font are currently supported by Oparaca.}
    \label{tab:non-functional-requirements}
\end{table*}

\subsection{Unified OaaS Abstraction} 

We extend the FaaS abstraction, called Object as a Service (OaaS), that borrows the object-oriented programming (OOP) concepts to unify application logic and data within a single abstraction. Specifically, each application is defined as a collection of cloud objects where its data (a.k.a. state) is modeled as ``attributes'' with supported data types in current cloud data abstraction, and its logic is modeled as methods realized by serverless functions.
In this manner, OaaS abstraction alone is sufficient for the entire application development phase---eliminating the need for multiple distinct abstractions and the complexities of effectively gluing them.

OaaS also offers the notions of abstract class, inheritance, and polymorphism to establish software reuse across cloud objects, thereby reducing redundancy and enhancing development productivity at the FaaS workflow level. This is in contrast to traditional FaaS, which typically limits software reuse to the function or invocation level (e.g., through shared libraries).
Beyond these, OaaS transformation unlocks new opportunities for deployment optimizations that were previously difficult or impossible.
This is because the object abstraction provides richer information for optimization and grants the cloud greater control over the deployment to exploit them. For example, OaaS lets application data and logic be encapsulated and managed together under the object abstraction. Thus, OaaS can easily find the data associated with each method and proactively distribute them across the cloud database instances that are close to the deployed method, thereby minimizing the data transmission overhead.

\subsection{Non-functional Requirement Interface}

Within the OaaS abstraction, we develop a non-functional requirement interface that lets the developer express their non-functional requirements in a human-friendly manner. Through the interface, developers can declare their non-functional requirements for a whole object or even for a specific part (attribute or method) of it. The requirements are defined as high-level and measurable metrics either in the form of NFR specifications (e.g., availability and throughput) or deployment constraints (e.g., budget and jurisdiction). During the deployment, the cloud provider takes these non-functional requirements as input to its internal services and adjusts their operations to meet the requirements. The benefits are three-fold: 
\begin{itemize}[leftmargin=*, itemsep=0pt, topsep=2pt]
    \item \textit{Productivity}: applications no longer need to consider low-level resource configuration for non-functional requirements. This relieves the burden of performance optimization from their deployment process, thus improving productivity.
    \item \textit{Portability}: as long as the cloud provider supports OaaS, the application can rely on the object abstraction to maintain its functionality, meet its NFR and constraint expectations (via the non-functional requirement interface), and comfortably deploy across scenarios with minimal changes. 
    \item \textit{Cloud-application symbiosis}: since applications use cloud resources for execution, the common sense is that the cloud should fulfill the non-functional requirement, as it has sufficient knowledge and privilege on the underlying infrastructure. With the non-functional requirement interface, however, the cloud does not take this responsibility alone. Here, the interface acts as a ``glue'' to make a symbiosis between the cloud and the application developer.Specifically, the requirements declared through the interface are valuable guidelines for cloud service providers to know which optimization they should follow so as not to impact the applications negatively. On the other hand, the interface is a useful means of communication that lets the developer actually configure for performance and quality, as opposed to going through multiple rounds of playing a ``trial-and-error'' game with the cloud providers to meet the desired outcomes.
\end{itemize}

% \begin{figure}
%     \centering
%     \includegraphics[scale=0.5]{Figures/OaaS_SoCC_2024/oaas-ideas.png}
%     \caption{Cloud application life cycle using the new approach leveraging the concept of OaaS and Non-functional Requirement Declaration Interface.}
%     \label{fig:oaas-ideas}
% \end{figure}

% \subsection{Requirements Driven Cloud-Developer Coordination}
\subsection{Simplified, Refinement-Free Deployment}

Based on the ideas above, as shown in Figure \ref{fig:oaas-ideas}, we propose a novel paradigm to develop and deploy cloud applications. In this paradigm, cloud applications are modeled as a set of objects, each can be developed and deployed independently. An object can possess deployment constraints and NFR specifications declared through the non-functional requirement interface. The object is deployed and managed on the cloud by means of the OaaS abstraction. 
Specifically, an OaaS-based platform (we call it Oparaca and introduce it in Section \ref{sec:architecture}) receives the object deployment packages from the developer,
deploys them on the cloud, and also automatically configures and monitors their resource allocation to meet the defined non-functional requirements.

The proposed paradigm greatly simplifies the process of delivering cloud-native applications. Instead of having multiple logic/data deployments with multiple rounds of development-deployment-evaluation that are subjected to many uncertainties caused by the cloud's shared environment and uncooperative abstraction realization, the application now needs to deal with only one type of abstraction. Moreover, with the non-functional requirements serving as the driving force for the underlying OaaS orchestration, no re-deployment or re-configuration is needed to meet the desired non-functional requirements.

\section{Oparaca: Realizing the NFR-Driven OaaS}
\label{sec:architecture}

In this part, we first describe the design goals of Oparaca---an open-source platform realizing the ideas of the OaaS paradigm with NFR enforcement. Then, we introduce new concepts and interfaces needed for this realization, and finally, we discuss its development details.

\subsection{Design Goals and Requirements}
\label{sec:architecture-goals-requirements}

We use Oparaca as a proof of concept to (1) illustrate how OaaS can reshape cloud application deployment, making it more productive and cost-effective; and (2) highlight how OaaS unlocks new opportunities for a more efficient, collaborative application deployment optimization. To achieve these objectives, we outline the following requirements and try to ensure Oparaca meets them throughout the entire design and implementation process.

\begin{enumerate}[leftmargin=*, noitemsep, topsep=1pt]
    % \item Encapsulate computation, data, and non-functional requirements into a single deployment entity.
    \item \textit{Simplicity}: Extend the concept of \textit{object} in OOP to a service abstraction that allows application developers to encapsulate their application logic, data, and non-functional requirements into a single deployment entity.
    % \item Simple (high-level abstraction) interface allows users with no experience to deploy high-quality services at low cost.
    \item \textit{Declaratory}: Provide a simple, human-friendly interface for non-functional requirements that allows developers to express and achieve desired non-functional requirements with minimum configuration/deployment effort.
    % \item Flexible class runtime management system to satisfy different non-functional requirements.
    \item \textit{Efficiency}: Oparaca can enforce application requirements at comparable cost versus state-of-the-art solutions. 
    % \item Enable performance optimization automation to meet the non-functional requirement even if workload or environment changes.
    \item \textit{Portability}: Oparaca allows applications to deliver proper functionality with desired NFRs anytime, anywhere.
\end{enumerate}

Oparaca is implemented in Java and comprises approximately 20,000 lines of code. The platform offers a YAML-based OaaS API for defining objects and their non-functional requirements. Oparaca operates with FaaS functions at the container level using Knative and Kubernetes, and it provides a supported SDK for working with Python. The source code is available at \url{https://github.com/hpcclab/OaaS}.

\subsection{OaaS Abstraction Interface}
\label{sec:archtiecture-interface}

To fulfill the first two requirements (i.e., simplicity and declaratory), we provide a deployment interface for OaaS to help developers define the entities of their cloud-native application and non-functional requirements akin to OOP concepts. To that end, the cloud-native application is built on the foundation of \textit{classes}. Each class defines the structure of independent executable objects that are responsible for carrying out one or multiple functionalities. Upon deployment, Oparaca allocates appropriate cloud resources to realize the corresponding objects of the class and manage them to handle workloads. Moreover, Oparaca supports \textit{inheritance} and \textit{polymorphism} for its classes.

Within each class, we can define \textit{methods} and \textit{attributes} to encapsulate the application logic and state (that can be in the form of structured or unstructured data, i.e., BLOB), respectively. For structured state data, Oparaca allows the developer to keep the data as a JSON-based document, similar to the document database \cite{carvalho2022nosql}. For unstructured data, however, object storage is employed to store them.
We model each method as a serverless function\footnote{we use the term function and method interchangeably in this paper}. Oparaca shares object states among methods of the same object following the OOP encapsulation principles.
% \textcolor{red}{For structured data, we design Oparaca to utilize the pure function approach that bundles the input argument and the object state as a standalone task to be executed in the developer code within a FaaS engine. With this approach, the function can behave as a stateless operation that returns the modified state as the output back to the Oparaca platform to update the state. For unstructured data, however, Oparaca provides the API that allows the function code to call to the object storage directly to avoid unnecessary data transmission overhead.} 

In Oparaca, application NFRs are declared through the \textit{non-functional requirement interface}. The interface allows the developer to associate a class or its methods with one or a set of requirements that the cloud provider has to meet once objects of the assigned class or methods are deployed successfully. Table \ref{tab:non-functional-requirements} shows the list of NFR specifications and constraints currently supported by Oparaca. Non-functional requirement declarations are treated as properties of classes or methods, so they are enforced according to the OOP inheritance principles. If a method and its class have conflicting requirements, then the method-level requirement prevails.

\begin{figure} [t]
    \centering
    \includegraphics[width=0.8\linewidth]{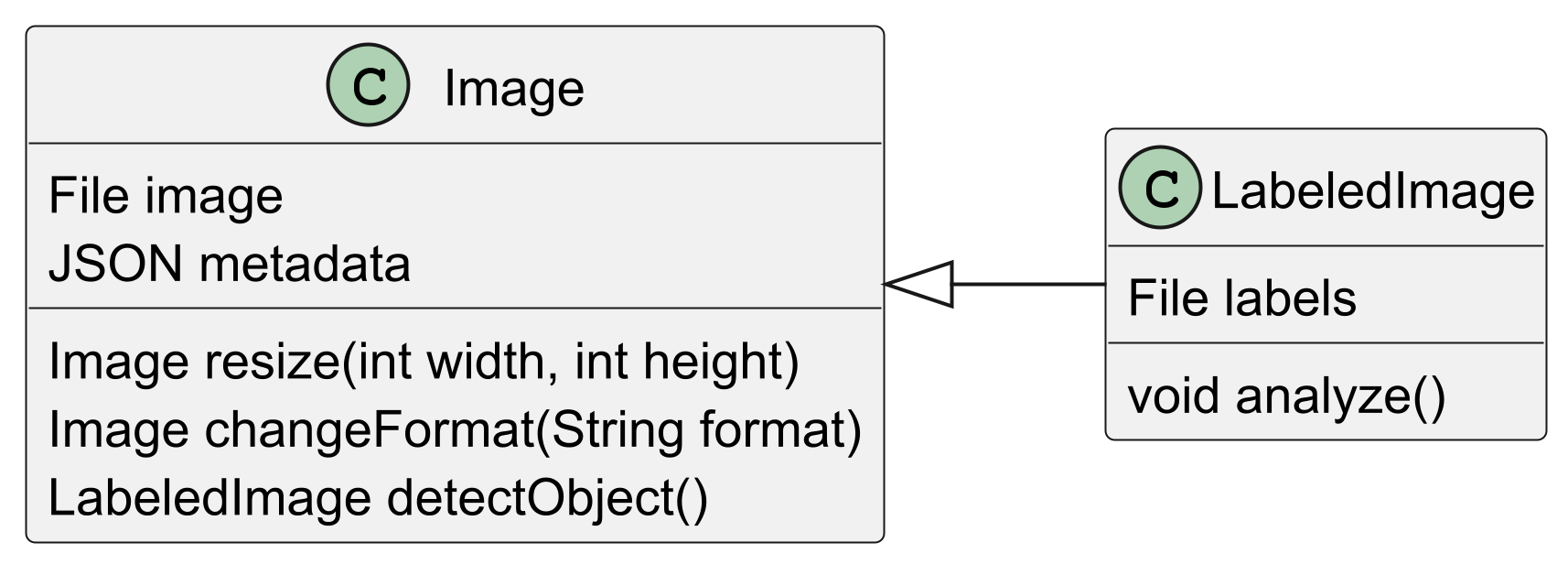}
    \caption{Class diagram for the image processing example. The developer can translate the class diagram directly to cloud deployment in Listing \ref{lst:image-yaml} through OaaS abstraction.}
    \label{fig:oaas-example}
\end{figure}

\begin{figure}[t]
\begin{minipage}{0.95\linewidth}
 \linespread{0.8}
\begin{lstlisting}[
    language=yaml,
    caption=OaaS Deployment for Image Processing,
    label=lst:image-yaml
]
classes:
  - name: Image
    qos:
        availability: 99.9
    constraint:
        persistent: true
    keySpecs:
      - name: image #File Image;
    functions: 
      - name: resize
        qos:
            throughput: 100  #rps
        #container image
        image: img/resize
      - name: changeFormat
        image: img/change-format
      - name: detectObject
        qos:
            throughput: 100
        image: img/detect-object
  - name: LabelledImage
    parent: Image
    keySpecs:
      - name: labels #File labels;
    functions: 
      - name: analyze
        qos:
            throughput: 50
\end{lstlisting}
\end{minipage}
\end{figure}

\begin{figure} [t]
    \centering
    \includegraphics[width=0.7\linewidth]{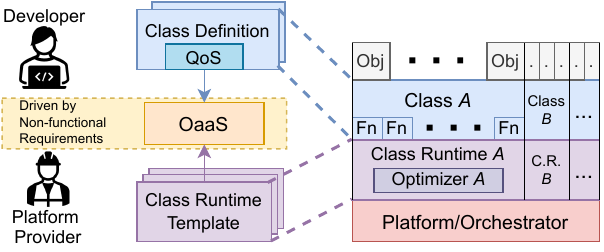}
    \caption{Realizing objects with class runtime and template: OaaS maintains templates customized for various deployment scenarios. For a specific class, Oparaca uses one of its predefined templates to create a class runtime to manage the deployed classes optimally.}
    \label{fig:class-runtime}
    % \vspace{-3mm}
\end{figure}

Figure \ref{fig:oaas-example} shows the class diagram of an example application providing image processing functionalities, such as resizing and changing the format. A developer can translate the diagram directly to OaaS classes. Specifically, OaaS allows images to be wrapped inside the \texttt{Image} class abstract where the image itself can be defined as a single unstructured file and its metadata is structured data. %Operations on the image are defined as serverless functions. For instance,
The \texttt{resize} function receives width and height as its inputs and produces a new image object as its output. The \texttt{changeFormat} function receives the new format name as input and produces a new image as the output object. The developer can add a new class \texttt{LabelledImage} for the image that can have the label data of image content. This class extends the \texttt{Image} class with the additional \texttt{labels} data and \texttt{analyze} function. The \texttt{Image} class also has a \texttt{detectObject} function to perform object detection to create the \texttt{labels} data and create the \texttt{LabelledImage} object as an output. The \texttt{analyze} function is to perform further analysis to label data. Oparaca currently supports the OaaS Abstraction Interface in \texttt{YAML} format. The class declaration of the example is in Listing~\ref{lst:image-yaml}.

Based on inheritance, in this example, the \texttt{LabelledImage} class inherits the non-function parameters from \texttt{Image} class (i.e., availability=99.9). The \texttt{resize} and \texttt{changeFormat} functions that the class \texttt{LabelledImage} inherit also maintain the non-functional parameter from class \texttt{Image}. Also, note that the \texttt{detectObject} function should inherit the class throughput requirements (150 rps); however, since its own NFR specification also includes throughput, Oparaca overrides its throughput (50 rps) instead.

\subsection{Object Realization}
\label{sec:architecture-realization}

\paragraph{\textbf{Class Runtime and Template}}
\label{sec:architecture-realization-cr}

Oparaca uses \textit{class runtime} to deploy and manage objects derived from user-defined classes (Figure~\ref{fig:class-runtime}).
% The class runtime is a set of components that execute the class logic and maintain its state.
To meet the third requirement (i.e., efficiency), the class runtime must be optimized to fulfill the non-functional requirements within a reasonable cost and overhead. However, given the non-functional requirements that Oparaca supports, there is a vast diversity of possible non-functional requirement combinations that need different specializations to satisfy. 
% must be flexible enough to fulfill the non-functional requirements of different classes.
Thus, it is impractical to have a single design for the class runtime that can efficiently satisfy all of the requirements.
% It also eliminates the possibility of providing specialized runtime for specific use cases.
% Therefore, the class runtime must be dynamic, extendable, and dedicated to a specific combination of non-functional requirements.
% This way, each class runtime can be optimized and scaled independently. To simplify matters, each class runtime should be dedicated to a single class.  

% To resolve the issue, Oparaca assigns a dedicated class runtime to each class. The class runtime is derived from a \textit{class runtime template},
To resolve the problem, Oparaca introduces \textit{class runtime template}, which provides a configurable class runtime design optimized for a specific set of requirement combinations. 
% and then optimized by the OaaS Runtime Manager (see Section \ref{sec:oaas-architecture}) to meet the requirements of the assigned class. Each class runtime template
Oparaca maintains a list of different templates to support as many requirement combinations as possible.
When deploying a class, Oparaca will choose from the list the most suitable template to realize the class requirement and then follow the template design to create a dedicated class runtime for this class. 
% Each is a design of a class runtime that is dynamic, configurable, and extendable enough to satisfy as many non-functional requirements as possible.
This approach allows Oparaca to satisfy both portability and efficiency design requirements.

In terms of portability, the class runtime template enables Oparaca to have freedom and flexibility in realizing objects. Instead of seeking a one-size-fits-all object realization mechanism, Oparaca decomposes the object realization into a set of sub-problems, each one aiming to find the optimal solution (i.e., class runtime template) for a specific infrastructure setting and requirement combinations. The approach makes Oparaca's implementation modular and flexible. One can upgrade existing solutions, extend the implementation to include new non-functional requirements, or even adjust for new infrastructure by adding/modifying templates without worrying about compatibility issues.
% define templates based on off-the-shelf solutions that already work well with some specific infrastructure and requirement combinations. 

In terms of efficiency, Oparaca can use off-the-shelf solutions to implement its class runtime templates. This allows Oparaca to take advantage of a vast diversity of existing state-of-the-art solutions, which have been proven to be efficient in practice, to reliably enforce non-functional requirements at minimum time, cost, and effort. Further, since class runtime templates are configurable, depending on specific object deployment scenarios, the class runtime derived from the template can be customized for further efficiency. 
Oparaca also allows platform provider to customize the template configurations, selection conditions, and priority for their operation objective (e.g., resource utilization).

\begin{figure}
    \centering
    \includegraphics[width=\linewidth]{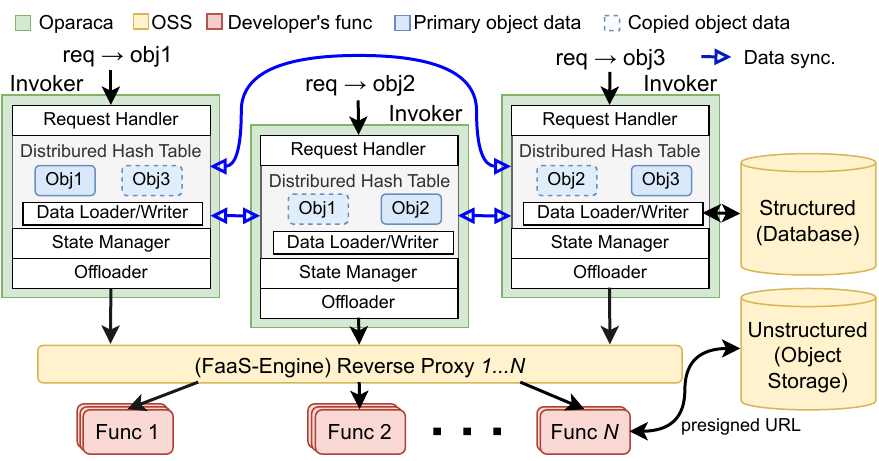}
    \caption{LTAG (Latency, Throughput, and Availability Guarantee): An example of a class runtime template designed for enforcing class latency, throughput, and availability requirements (OSS: Open-source software).}
    \label{fig:classruntime}
    % \vspace{-2mm}
\end{figure}

\paragraph{\textbf{Class Runtime Example}}
\label{sec:architecture-realization-ltag}

Figure \ref{fig:classruntime} shows LTAG (Latency, Throughput, and Availability Guarantee)---a class runtime template that Oparaca currently uses to enforce class latency, throughput, and availability requirements. Each class runtime derived from the template has three modules: \textit{invoker}, \textit{FaaS engine}, and \textit{data storages}. The invoker is responsible for handling all of the object-related operations. For each operation, the invoker finds its corresponding function and offloads the operation to that function managed by the FaaS engine. LTAG can maintain the object state in both unstructured and structured databases.

In the offloading mechanism, the invoker utilizes the pure function approach that bundles the invocation request and the object attributes as a standalone task within a FaaS engine. Each invocation takes the object attributes as input, modifies them, and then returns the updated attributes as the output to the invoker. The invoker maintains an internal in-memory distributed hash table (DHT) \cite{hassanzadeh2021dht} to keep the object data (i.e., attributes and metadata) for reducing database access operation, thereby speeding up the object invocation.

\paragraph{\textbf{Throughput Enforcement}}
OaaS currently supports throughput enforcement by allowing applications to specify a guaranteed invocation rate $A$ per  FaaS function \cite{nguyen2019real}. Oparaca ensures that sufficient resources are available so that at least one invocation can start immediately (i.e., without cold-start delays) every $\frac{1}{A}$ seconds. LTAG customizes the Invokers and FaaS engine based on Real-time Serverless \cite{nguyen2019real, nguyen2023storm} to estimate and periodically adjust resource allocation for each class and its functions, ensuring they can handle operation requests up to the specified rate guarantee.

\paragraph{\textbf{Latency Enforcement}}
Recent work on latency enforcement aims to minimize end-to-end latency in a best-effort manner \cite{jin2023ditto, lei2023chitu, zhang2023online, zhang2021caerus, lin2020modeling}, giving no guarantee to construct/realize non-functional requirements. Besides, other efforts try to keep latency within a specific target deadline \cite{szalay2022real, ascigil2021resource, verma2024lease, moghimi2023parrotfish, bhasi2022cypress}, but this is extremely difficult from the cloud provider's perspective due to the highly dynamic and unpredictable nature of invocation logic \cite{shadrad20serverless, joosen2023does, eismann2021state}, data size \cite{bhasi2022cypress, eismann2020predicting, mvondo2021ofc}, and communication requirements \cite{yu2023following}.
Thus, to enforce the latency in a feasible and controllable way, OaaS offers guarantees to minimize the system overhead of invocation executions, focusing on cold-start and communication, enabling the developers to optimize their functionality execution time barely based on improving their codes. The developer can address cold-start via throughput enforcement, as described above. For communication, OaaS provides a \textit{locality} guarantee, allowing developers to specify the location for invocation dispatch. This can be either (i) \textit{local}: attributes are read and written as if they are in the same FaaS container executing the function logic, and (ii) \textit{none}: no locality restriction.

LTAG enforces the \textit{local} guarantee by exploiting the class function-attribute relationships. Specifically, Oparaca uses consistent hashing, maintained by invokers, to track object data locations and route invocation requests to the corresponding place.

\paragraph{\textbf{Availability Enforcement}}
OaaS provides availability enforcement as a reliability guarantee, defining the percentage of time that an object (or its methods) are available for invocation execution.
LTAG enforces availability through replication. Specifically, given an object with availability requirement $A$ (e.g., 99.99\%), we enforce $A$ by creating $N$ replicas of the object with $N$ is defined according to Meroufel and Belalem \cite{meroufel2013managing} as follows. 
\begin{equation}
    A = 1 - (1 - P)^{N}
    \label{eqn:availability}
\end{equation}
where $P$ is the stability of the resources used to deploy the object. LTAG replicates the object data and uses the DHT to manage them. However, it keeps only one object replica, called \textit{primary}, active at a time. To enforce consistency, the primary object handles all state modifications and then commits the results across all replicas. If the primary replica fails, Oparaca chooses one of the remaining replicas as the new primary. 

\subsection{Oparaca Architecture}
\label{sec:oaas-architecture}

\begin{figure}
    \centering
    \includegraphics[width=0.8\linewidth]{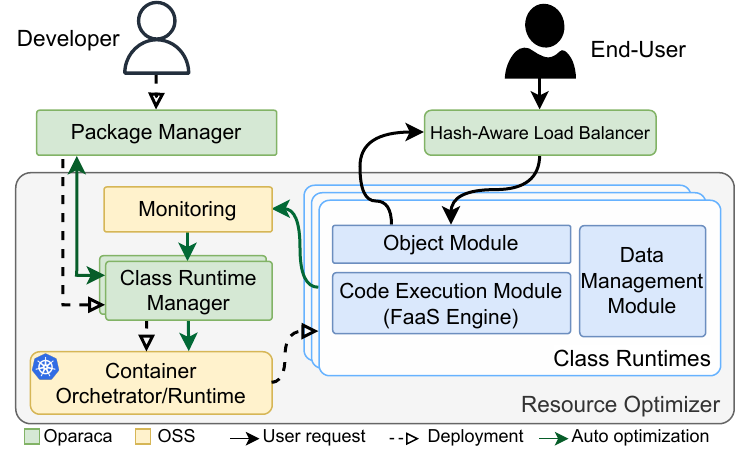}
    \caption{A bird-eye view of Oparaca's NFR-driven architecture}
    \label{fig:overall-architecture}
\end{figure}

Oparaca's architecture, shown in Figure \ref{fig:overall-architecture}, includes the following key components: \textit{(1) Package Manager}: responsible for managing classes registered in Operaca and their corresponding deployment packages. This component also acts as a gateway and offers APIs to develop and deploy OaaS-based applications. \textit{(2) Class Runtime}:    turns the class descriptions and corresponding packages into the actual object deployments on the cloud. \textit{(3) Class Runtime Manager}: create dynamic class runtime from existing templates (e.g., LTAG). It is also responsible for class runtime deployment and management. \textit{(4) Monitoring System}: gathers the performance metrics from class runtime. \textit{(5) Hash-aware Load Balancer} and \textit{Container Runtime}: responsible for scheduling and managing function execution. Once a function invocation is issued, the hash-aware load balancer routes the request to the corresponding class runtime by using consistent hashing that, in turn, forwards the request to the corresponding container for execution. 

Given the interface and architecture, the application lifetime on the cloud now consists of two phases: 

\textbf{(a) Registration:} The developer registers their class to Oparaca. Upon registration, the \textit{package manager} unpacks the deployment, extracting the class logic (e.g., functions), state (e.g., data schema), and non-functional requirements (e.g., NFR specifications and constraints). The extracted information is then forwarded to the \textit{class runtime manager} to find an appropriate class runtime template to generate a dedicated \textit{class runtime} to handle the object realization for the class.

\textbf{(b) Execution:} Once a \textit{class runtime} is created, it is responsible for managing the execution and state of all objects generated from the associated class. Every interaction with the application users is handled through the class runtime, independent from other Oparaca components. To ensure reliability, the \textit{class runtime manager} periodically collects monitoring metrics from class runtime. Based on the information, Oparaca can adjust the \textit{Container Orchestrator/Runtime} to improve efficiency and take administrative actions (e.g., to recover from failure, etc.) if needed.

Note that the above procedures are performed solely by Oparaca platform. Application developers do not have to intervene or refine their configuration for both functional and non-functional requirements. This greatly simplifies application deployment.

\section{Performance Evaluation}
\label{sec:evaluation}

% In this section, we seek to learn the performance of Oparaca in the following aspects: scalability (Section \ref{sec:evaluation-scalability}), efficiency to achieve QoS (Section \ref{sec:evaluation-efficiency}), and ease of use (Section \ref{sec:evaluation-ease-of-use}). 
In this section, we seek to learn the performance of Oparaca in the following aspects: 
non-functional requirement enforcement (Section~\ref{sec:evaluation-non-functional}), 
implementation efficiency (Section \ref{sec:evaluation-implementation}), 
% efficiency to achieve QoS (Section \ref{sec:evaluation-efficiency}); 
deployment productivity (Section \ref{sec:evlt:deployment}), and development productivity (Section \ref{sec:evaluation-ease-of-use}).

\subsection{Experimental Setup}
\label{sec:evaluation-setup}

\begin{figure} [t]
    \centering
    \includegraphics[width=0.75\columnwidth]{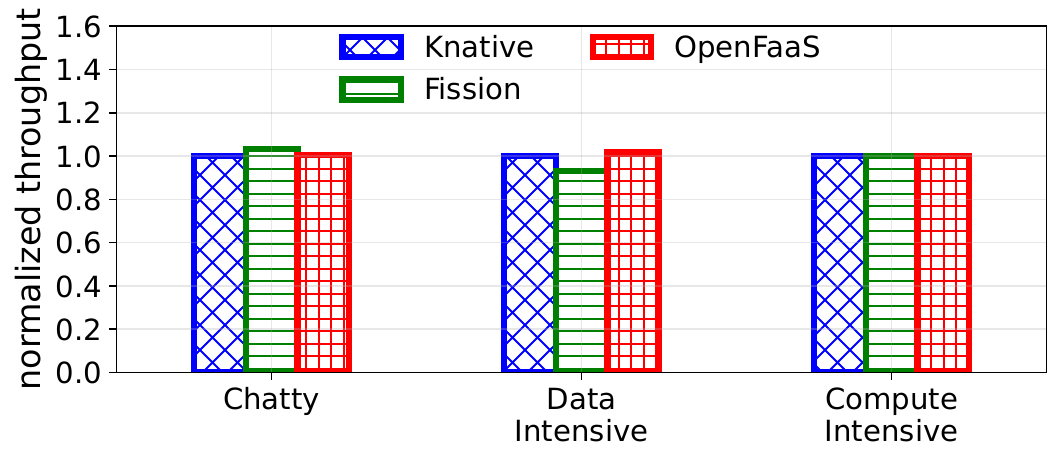}
    \caption{Oparaca does not significantly differ in throughput performance across the FaaS engines.}
    \label{fig:faas-engines}
\end{figure}

\begin{figure*}[ht]
    \centering
    % \subfloat[JSON randomization]
    \subfloat[Chatty]{\includegraphics[width=0.49\textwidth]{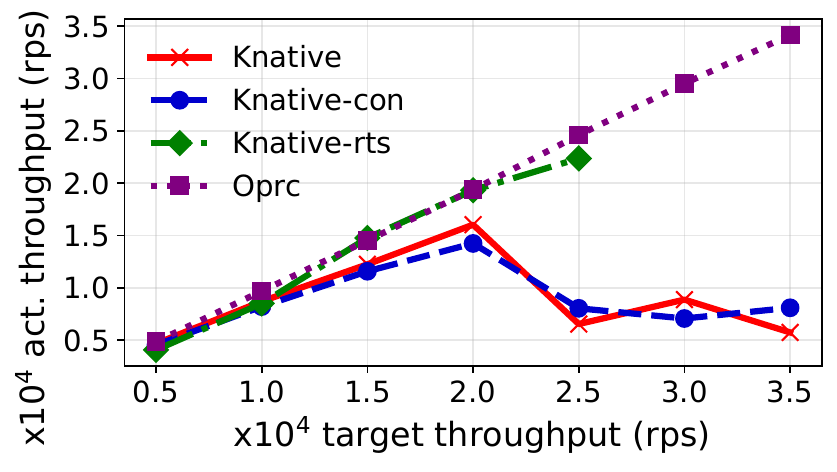}\label{fig:evlt:tp:json}} 
    \hfill
    % \subfloat[Image resizing]
    \subfloat[Data Intensive]{\includegraphics[width=0.49\textwidth]{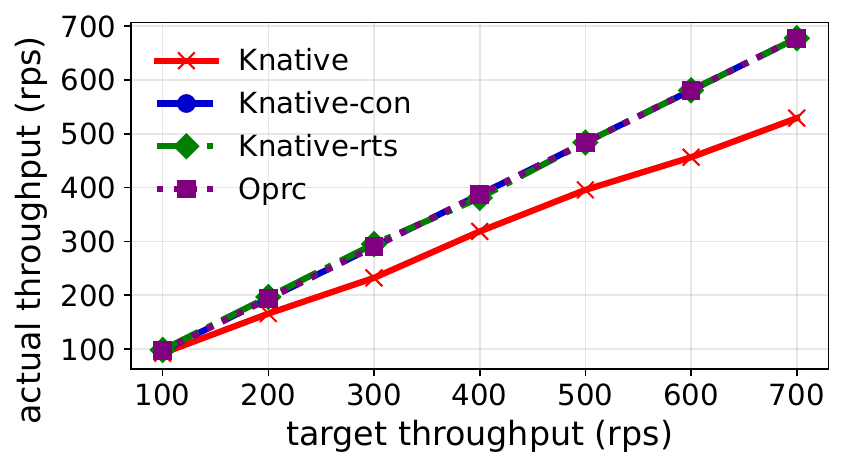}\label{fig:evlt:tp:image}}
    \hfill 
    % \subfloat[Video transcoding]
    \subfloat[Compute Intensive]{\includegraphics[width=0.49\textwidth]{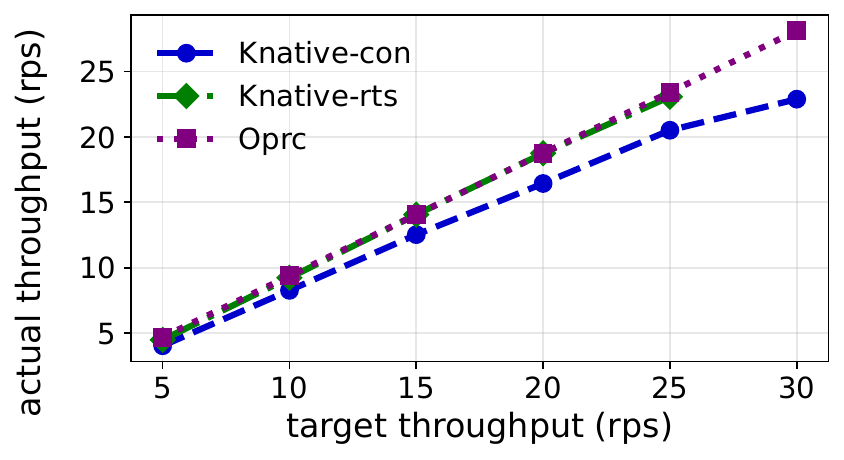}\label{fig:evlt:tp:video}}

    \caption{Achievable throughput varying target throughput. Oparaca ensures the actual throughput matches the target one across settings, while the other approaches fail to do so at high throughput targets.
    }
    \label{fig:evlt:tp-enforcement}
    % \vspace{-2mm}
\end{figure*}

\begin{figure}
    \centering
    \includegraphics[width=0.7\columnwidth]{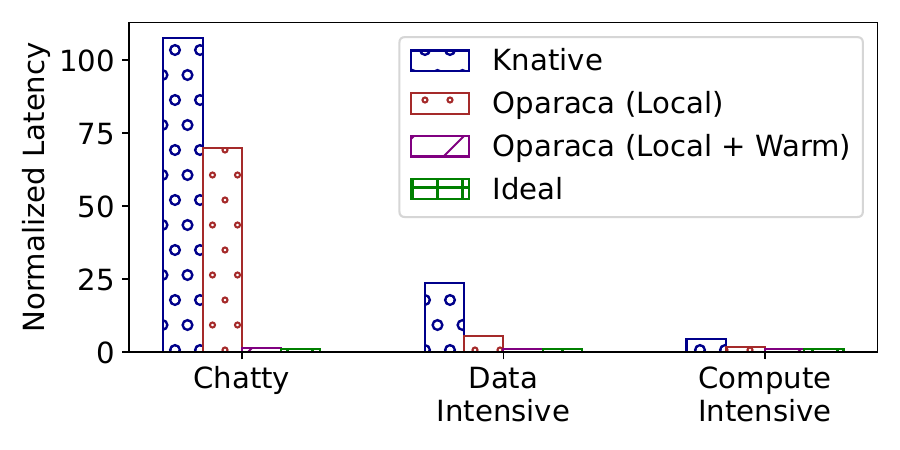}
    \caption{Oparaca can exploit data locality to provide various latency guarantees.}
    \label{fig:latency-enforcement}
\end{figure}

\begin{figure} [ht]
    \centering
    \includegraphics[width=0.6\linewidth]{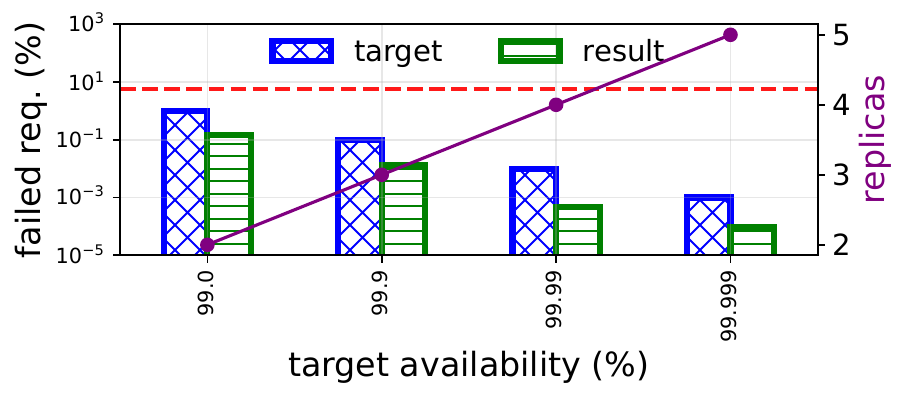}
    \caption{
    Successful invocation rate at different availability targets with availability enforcement. Resource stability ($P$) is 94.36\% (red line).
    }
    \label{fig:evlt:availability}
\end{figure}

We prepare the experimental environment on 4 machines on Chameleon Cloud \cite{chameleon_cloud}, each with 2 sockets of Intel(R) Xeon(R) Gold 6240R CPU processors that collectively have 192 cores, 768 GB memory, and SSD SATA storage. We use 3 machines to install the Kubernetes cluster (RKE2~\cite{rke2}) for deploying applications. The last machine generates load using Gatling~\cite{gatling}. Regarding data management, we use Minio~\cite{minio} (S3-compatible storage) for unstructured data and  ArangoDB~\cite{arangodb} (document database) for structured data.

\paragraph{\textbf{Workloads}}
To make sure our evaluation is comprehensive, we consider the following three classes of applications that exhibit different behaviors: 
\begin{itemize}[leftmargin=*, noitemsep, topsep=0pt]
    \item \textit{Chatty}: characterized by frequent small communications that impose significant overhead on network transmission \cite{chatty-app}. As a representative workload for the application class, we utilize \textit{JSON randomization} \cite{lloyd2018serverless}, which involves a sequence of ten invocation requests, each randomly updates a JSON key-value pair to the document database. 
    \item \textit{Data Intensive}: characterized by substantial data access operations \cite{hoseinyfarahabady2021data}. We use an \textit{image resizing} workload \cite{solaiman2020wlec, balla2021estimating}, which resizes images stored in object storage through FaaS invocations, to represent this class of applications.
    \item \textit{Compute Intensive}: demand extensive computational resources throughout their lifecycle (e.g., ML \cite{chahal2020migrating} and HPC \cite{nguyen2020motivating} applications). To represent this class, we use \textit{video transcoding} \cite{moina2023event, wu2020descriptive}, which involves changing the resolution of a video file stored in object storage.
\end{itemize}

% \vspace{1mm}
% \noindent\textbf{FaaS Engine Selection.} We 

% In order to choose the FaaS engine for Oparaca's class runtime, we conducted experiments to measure the throughput of Oparaca on multiple FaaS engines: . Since OpenFaaS has a resource limit on free licenses, we made sure all three FaaS engines operated under the same limitation to ensure fairness. The results are shown in Figure~\ref{fig:faas-engines}. Our findings indicate that different FaaS engines do not significantly impact Oparaca's performance. We have decided to go with Knative as the FaaS engine for Oparaca due to its extensive configuration options, well-defined documentation, and large community support.

\paragraph{\textbf{Approaches}}
To ensure generality, we integrated Oparaca with various FaaS engines---Knative~\cite{knative}, Fission~\cite{fission}, and OpenFaaS~\cite{openfaas}, all backed by Kubernetes---to host object functions. Figure \ref{fig:faas-engines} shows the maximum throughput achieved by workloads mentioned above when deployed over Oparaca using these different FaaS backends under identical resource configurations (each deployment can scale up to five Kubernetes pods, each with 4 CPUs). The throughputs, normalized to Knative, are nearly equivalent across all FaaS engines for all three workloads. This confirms that Oparaca can be configured to work with various FaaS engines with negligible performance differences, making it flexible for deployment across different cloud environments.
Thus, due to space limits, we report only the experimental results for Oparaca's Knative variant. Also, for fair comparison, we use Knative with various deployment configurations as experiment baselines:
\begin{itemize}[leftmargin=*, noitemsep, topsep=0pt]
    \item \textit{Knative:} Default Knative configuration that declares only per-container resource requirements (i.e., CPU and memory) and leaves the rest to the auto-scaling system.
    \item \textit{Knative-con:} Default Knative configurations plus applying per-container concurrency limit to avoid overloading.
    \item \textit{Knative-rts:} adopt Real-time Serverless resource management \cite{nguyen2019real} to enforce throughput guarantee.
    \item \textit{Oprc} is Oparaca, which allows the applications to enforce their throughput, latency, and availability in their class definitions. Since Oparaca needs to learn the workload metrics before properly optimizing the class runtime, we perform one more extra round of load generating in each experiment. The first round acts as the warm-up for Oparaca to properly gather the metrics. 
\end{itemize} 
Beyond ensuring a fair comparison, we choose Knative as a FaaS baseline because it offers a rich set of configuration options to capture diverse deployment scenarios often unsupported by other engines. Additionally, varying Knative settings demonstrate how current FaaS implementations address non-functional requirements---by adjusting low-level configurations (e.g., per-container concurrency) in a best-effort manner. Configuring Knative allows us to explore a broad range of FaaS deployment configurations, whether these adjustments are made by developers (if the FaaS engine exposes the configurations) or by cloud providers (if it does not---for example, Microsoft Azure doesn't allow developers to configure per-container concurrency). Thus, although our evaluation results are specific to Knative, the insights and implications are generalizable to other FaaS engines.

\begin{figure*}[ht]
    \centering
    % \subfloat[JSON randomization]
    \subfloat[Chatty]{\includegraphics[width=0.33\textwidth]{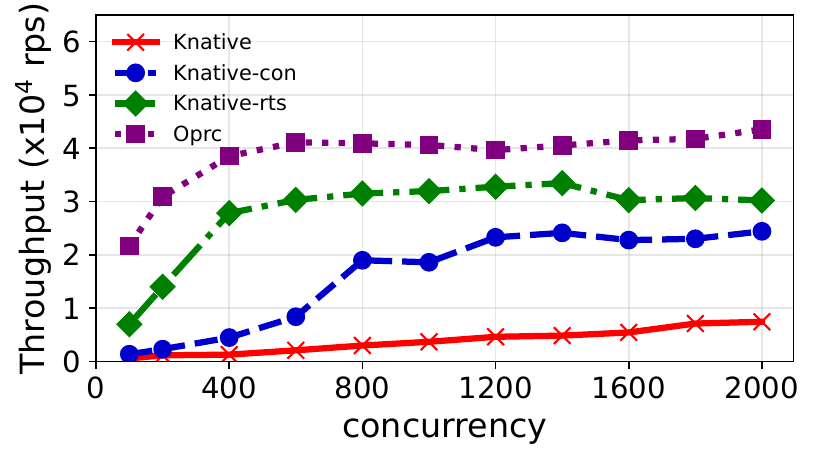}\label{fig:evlt:scalability:json}} 
    \hfill
    % \subfloat[Image resizing]
    \subfloat[Data Intensive]{\includegraphics[width=0.33\textwidth]{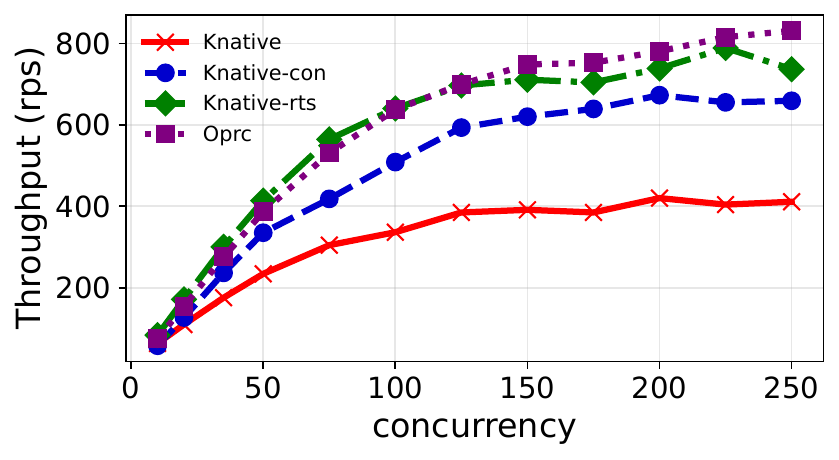}\label{fig:evlt:scalability:image}}
    \hfill 
    % \subfloat[Video transcoding]
    \subfloat[Compute Intensive]{\includegraphics[width=0.33\textwidth]{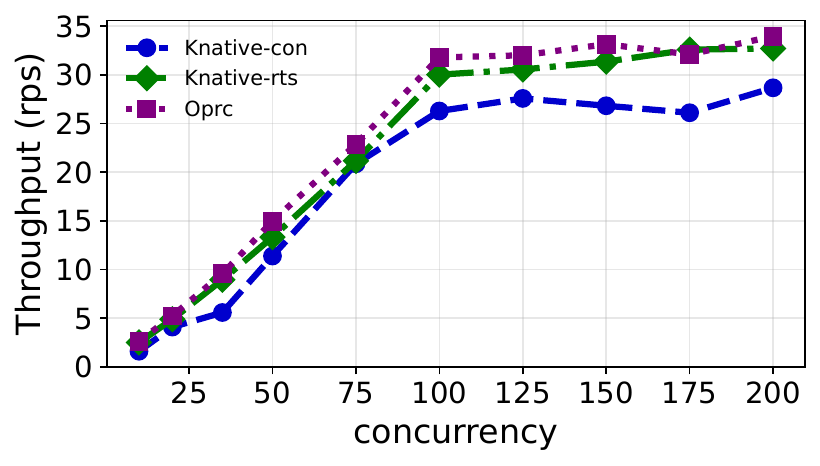}\label{fig:evlt:scalability:video}}

    \caption{
    Achievable throughput under various request concurrency.
    % The throughput of each system given different levels of request concurrency on the three applications.
    Concurrency is defined as the number of clients that concurrently generate requests for the system.
    % The vertical axis presents the system's throughput in response per second (rps).
    }
    \label{fig:evlt:scalability}
    
\end{figure*}

\begin{figure*}[ht]
    \centering
    % \subfloat[JSON randomization]
    \subfloat[Chatty]{\includegraphics[width=0.33\textwidth]{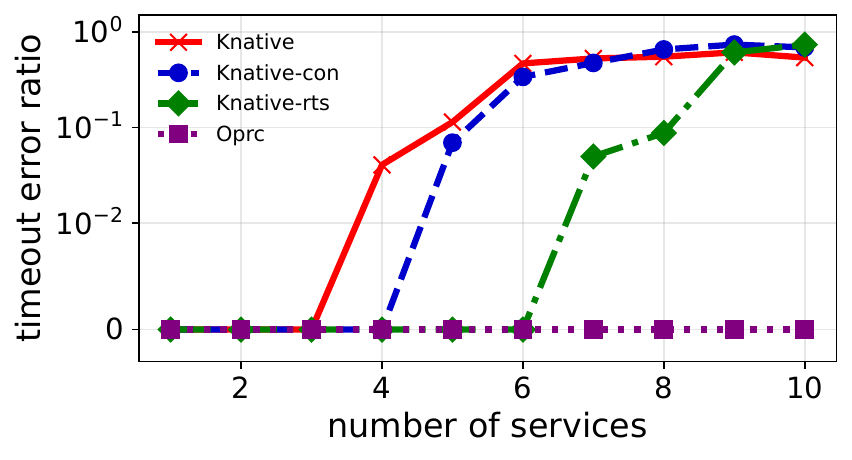}\label{fig:evlt:efficiency:json}} 
    \hfill
    % \subfloat[Image resizing]
    \subfloat[Data Intensive]{\includegraphics[width=0.33\textwidth]{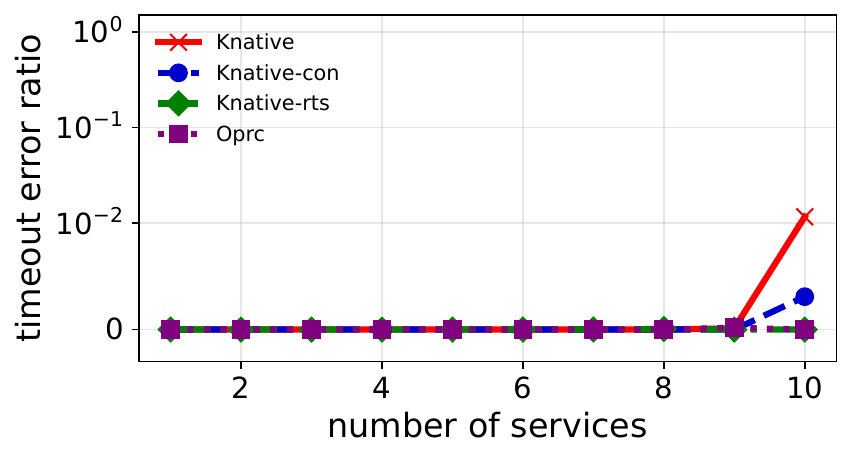}\label{fig:evlt:efficiency:image}}
    \hfill 
    % \subfloat[Video transcoding]
    \subfloat[Compute Intensive]{\includegraphics[width=0.33\textwidth]{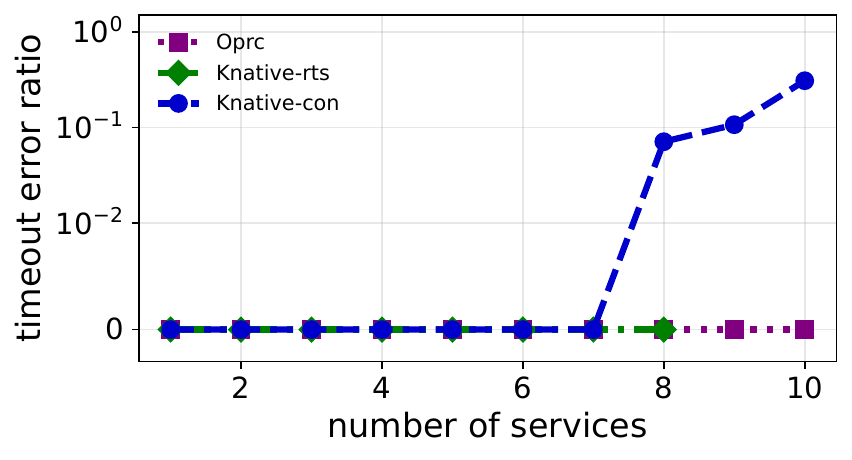}\label{fig:evlt:efficiency:video}}

    \caption{Error response ratio of different solutions upon deploying them with the different number of services.
    % for the three applications.
    % The vertical axis presents the ratio of the number of timeout errors to the number of total requests.
    }
    \label{fig:evlt:efficiency}

\end{figure*}

% In Oparaca, these applications can use the data access mechanism provided by the Oparaca platform.
In the following experiments, Oparaca deploys and manages workloads using class runtime derived from the LTAG template. Thus, data access is automated via the invoker.
% In the Knative variants, these applications have to manually access the data by connect to the storage or database directly
In the Knative variants, however, these applications have to implement direct data access to storage or database manually.
% We note that JSON randomization retrieves the test data from the document database, performs its operation, and stores the result back in the document database. The second and third applications perform a similar process but use object storage instead.

\subsection{Non-functional Requirement Enforcement}
\label{sec:evaluation-non-functional}

We validate the NFR enforcement capability of Oparaca by deploying applications mentioned in Section \ref{sec:evaluation-setup} using the LTAG class runtime template as described in Section \ref{sec:architecture-realization-ltag}.

% \color{blue}
% \begin{itemize}
%     \item GOAL: show the effectiveness of throughput enforcement
%     \item CONCLUSION: Oparaca can meet all the target throughputs and mitigate the problem of unpredictable performance of auto-scaling.
% \end{itemize}
% \color{black}

\paragraph{\textbf{Throughput}}
To validate Oparaca's throughput enforcement, we deployed the three applications with various target throughputs. Then, we configured the load generator to send the request at the same rate as the target throughput and measured the actual throughput on each system. The results are reported in Figure~\ref{fig:evlt:tp-enforcement}.

Overall, Oparaca can guarantee the throughput for all three applications. \textit{Knative-rts} only meets low throughput targets and fails at higher ones due to over-provisioning.
% and the underlying hardware is running out of resources to meet the requirement.
The other two \textit{Knative} variances fail to meet the targets since they only rely on auto-scaling without the awareness of the target throughput.
In the chatty workload, with the high request arrival rate, the internal queue cannot hold requests long enough to wait for the new pod to be spawned. Meanwhile, in the compute-intensive application, it takes longer for each request to be processed, making it easier to time out. Only the data-intensive application that \textit{Knative-con} can meet the target throughput.

The results also demonstrate the complexity of FaaS configuration. Even when utilizing the same backend services (i.e., Knative), varying FaaS deployment configurations result in significantly different performance outcomes. Thus, manual adjustment of FaaS deployment, while daunting, is often required to achieve the desired throughput. In contrast, Oparaca simplifies and automates this process with its high-level interface.  
% Since Knative does not support throughput enforcement, the experiment shows evidence that, at most, one of the Knative variants can meet all target throughputs. This indicates that developers have to manually adjust their deployment, depending on the application, to achieve the target throughput. However, Oparaca offers a high-level interface that simplifies this process and automates it for developers.  

\paragraph{\textbf{Latency}}
We deployed all three applications over Operaca under the \textit{locality} and \textit{throughput} guarantee. We let the applications run under bursty loads by configuring the load generator to remain idle most of the time but occasionally create sudden bursts that send requests at a rate equal to the application throughput guarantee for a very short duration.
We compare Oparaca against two baselines: (i) \textit{Knative} with the data storage deployed at a separate data center from the FaaS deployment, representing a typical scenario of FaaS deployment \cite{aws-security-resilient}, and (ii) \textit{Ideal} where functions and data storage are deployed together on a dedicated machine with excessive resources, representing an ideal execution environment where the invocation execution latency depends solely on the application itself. 

Figure \ref{fig:latency-enforcement} shows the average execution time of the three applications across different deployments. The latency is normalized to the case of the ideal deployment. Knative is the worst among approaches, with the latency can be as high as 60$\times$ the ideal. The reason is two-fold. First, Knative needs external storage to keep the application data, but the actual data location is hidden under the storage abstraction, causing significant data transmission latency. Oparaca does not have this limitation as it encapsulates the data and invocations under a unified object abstraction, enabling locality enforcement, i.e., \textit{Oparaca (Local)}, that allows invocations to execute at the same machine with their data, significantly reducing the latency by 1.5$\times$ (Chatty) to 4$\times$ (Data Intensive). Second, Knative scales resources allocated to FaaS functions based on concurrency. That makes invocations suffer from cold-start under bursty loads. Applications can workaround this issue with Oparaca via throughput guarantee, enforcing the cloud to execute invocations without cold-start up to a certain rate, i.e., \textit{Oparaca (Local + Warm)}. This configuration further reduces the latency by 1.7$\times$ (Compute Intensive) to 46.5$\times$ (Chatty)! Enforcing these two non-functional requirements together allows applications deployed over Oparaca to minimize their invocation overhead (as low as 7\% of execution time), achieving invocation execution latency that is very close to the ideal execution.
% (at most 10\% longer).

\paragraph{\textbf{Availability}}
Next, we validate Oparaca's availability enforcement. We have created a failure emulator that injects failures by deleting the platform container according to a predefined Mean Time Between Failures (MTBF). Whenever a failure is injected, Kubernetes automatically recovers the container. The emulator then waits for MTBF, which is also supplemented by a random value from a normal distribution, before introducing the next failure. The emulator carries out these operations on each container individually. 
To select the MTBF, we use the reference MTBF of the Intel server boards~\cite{intel-board} that have around 50K hours on average. To speed up the experiment process, we scaled this number down by a million, setting the MTBF to 180s, which makes each container only operate for 94.36\% of the time. We then use 94.36\% as the resource stability ($P$) to configure Oparaca. We deploy the application according to the different target availability, generate the load to test the actual application availability with a rate of 200 requests per second for 1.5 hours, and measure the ratio of the requests being processed unsuccessfully. 

The results of this experiment are reported in Figure \ref{fig:evlt:availability}. 
When availability enforcement is on, Oparara deploys classes and objects with replications, significantly reducing the failure rate to meet the availability targets. The actual failed request ratio is slightly lower than each predefined target because Oparaca adds just enough replicas to meet the target, minimizing availability enforcement overhead. Notably, increasing the availability from 99\% to an exceptional rate of 99.999\% (1000$\times$ better) incurs only 2.5$\times$ extra resource cost. This is a 50$\times$ improvement versus the current industry standard that necessitates an SLA on availability of 99.95\% \cite{aws-lambda-sla} with only 1.67$\times$ cost increment.

% Given the current resource stability of 94\%, to achieve an exceptional availability rate of 99.999\% (or 50$\times$ better), Oparara would need to increase the minimum instance from 3 to 5 instances, representing a 1.67x increase in terms of cost.

% Overall, Oparaca can provide higher availability than the predefined target through replication. %If developers have to do this method manually, they have to manually estimate the number of replications. In contrast, Oparaca can save the developer from this burden. 
% Each increment of the availability objective implies one more replication of the data and runnable component (i.e., invoker). If the request fails to reach the target invoker, the hash-aware load balancer will reroute the request to the available invoker.

% \vspace{2mm}
% \noindent
% \colorbox{blue!10}{
% \parbox{0.96\linewidth}{
% \underline{\textbf{Takeaway}:} \emph{
% %Oparaca can effectively enforce the availability through the replication. %It provides a high-level interface to save the developer from the daunting task of fine-tuning.
% Oparaca replicates the class runtime to the minimum extent required to achieve the target availability.
% }}}

\vspace{2mm}
\noindent
\colorbox{blue!10}{
\parbox{0.96\linewidth}{
\underline{\textbf{Takeaway}:} \emph{
Unlike traditional FaaS deployments, Oparaca can automatically reconfigure to enforce various non-functional requirements for different classes of applications, eliminating the need for manual refinement.   
}}}
\vspace{2mm}

\subsection{Efficiency of Oparaca}
\label{sec:evaluation-implementation}

In this subsection, we examine Oparaca efficiency, running various experiments on a fixed quantity of resources to see how well the implementation handles various workloads under different operation scenarios.

\paragraph{\textbf{Function Invocation Efficiency.}}
To evaluate Oparaca invocation efficiency, we compare its maximum throughput with Knative variants; all are under limited resources. The throughput measurement takes multiple runs with an increasing number of clients (i.e., concurrency). We measure the mean throughput achieved in each run and report them in Figure \ref{fig:evlt:scalability}.
% To study Oparaca's scalability, we measure its maximum throughput for each application type and compare it against the other baselines. In this part, we define the target throughput using a number that is surely higher than what the system can practically achieve. In our setup, we set the target throughputs of JSON randomization, image resizing, and video transcoding to 50,000, 1,000, and 50, respectively. 

% As shown in Figure~\ref{fig:evlt:scalability}.
In general, the throughput becomes steady after increasing the concurrency to a certain level. Oparaca provides a higher throughput compared to other baselines, especially for the chatty workload (Figure~\ref{fig:evlt:scalability:json}) because Oparaca relies on the internal in-memory distributed hash table (DHT) to store the object data; thereby, it speeds up the data access and reduces the database operation. For the chatty workload, \textit{Knative-con} and \textit{Knative} yield significantly lower throughput compared to \textit{Knative-rts}. This is because this workload performs little computation compared to its network I/O operation, which makes the Knative auto-scaler inaccurately adapt the acquired resources to the workload.

% For the image resizing application (Figure~\ref{fig:evlt:scalability:image}), Oparaca offers a lower throughput compared to \textit{Knative-rts}. This is because image resizing is more I/O-intensive compared to the other studied applications. JSON randomization is also I/O intensive; however, it mostly requires the network I/O and is more likely to have a bottleneck in the database. In this figure, we observe that \textit{Knative} performs poorly because the auto-scaler cannot accurately adjust acquired resources to the increasing workload without per-container concurrency declaration. In contrast, by only declaring per-container concurrency, \textit{Knative-con} can perform with a similar performance as \textit{Knative-rts}.

For the data-intensive workload (Figure~\ref{fig:evlt:scalability:image}), \textit{Knative} performs poorly because the auto-scaler cannot accurately adjust acquired resources to the increasing workload without per-container concurrency declaration. In contrast, by only declaring per-container concurrency, \textit{Knative-con} can perform with a little less performance than \textit{Knative-rts}.

For the compute-intensive workload (Figure~\ref{fig:evlt:scalability:video}), because it is computationally intensive and the invocation rate is also less than the other workloads, all of the solutions can provide similar performance. Only \textit{Knative} cannot be used for this workload because without controlling the concurrency, each function container has to handle more concurrent invocations than it can. As a result, they fail to handle requests continually. Oparaca can perform slightly better than the others because it eliminates the need to fetch and deserialize the record (i.e., metadata) from the database on each function container.

% In summary, Oparaca demonstrates competitive throughput performance across different applications, particularly excelling in JSON randomization due to its efficient internal mechanism. However, it lags behind Knative variants in image resizing, likely due to I/O bottlenecks, while video transcoding sees comparable performance among all systems due to its computation-intensive nature. Overall, Oparaca's scalability showcases promising potential, especially in scenarios with only structured data.

% \vspace{2mm}
% \noindent
% \colorbox{blue!10}{
% \parbox{0.96\linewidth}{
% \underline{\textbf{Takeaway}:} \emph{Oparaca demonstrates competitive throughput performance across different applications, particularly excelling in I/O-intensive applications with only structured data due to its ability to utilize its internal DHT. %\textcolor{red}{However, it lags behind Knative variants in storage-intensive applications, mainly due to storage I/O bottlenecks. For CPU-intensive applications, Oparaca offers comparable performance to its FaaS counterparts.}
% }}}

\paragraph{\textbf{Throughput Enforcement Efficiency.}}
Our primary objective in this experiment is to examine the resource efficiency of Oparaca against other baselines and ensure its throughput is not attained with the cost of lavishly allocating resources. The other objective is to investigate Oparaca's behavior in the face of services with different throughput expectations. To that end, we configure multiple services of the same type, each with its own target throughput. To achieve this, we started by testing on a single service and gradually increasing the number of services to ten. We set the target throughput of each replicated service to be 1/10th of the maximum throughput we found in the previous experiment. We chose these numbers so that the target throughput is not too low and scaling remains relevant. The experiment is performed by generating invocation requests to each service, with the request rate capped to the target throughput, and then measuring the ratio of the number of timeout errors to the total number of requests. 

As shown in Figures~\ref{fig:evlt:efficiency}, overall, Oparaca outperforms other baselines for almost all workloads. For the chatty workload (Figure~\ref{fig:evlt:efficiency:json}), Oparaca can handle all of the requests with zero error rate because of its ability to readjust its allocated resources and its internal DHT structure. \textit{Knative-rts} also performs well at the beginning; however, after 6 services, the external document database starts to slow down, leading to a sharp increase in the error rate. The poorer performance of \textit{Knative} and \textit{Knative-con} is mainly because their independent scaling of services and lack of awareness of performance objectives lead to resource contention among co-existing services.

For the data-intensive workload (Figure~\ref{fig:evlt:efficiency:image}), all baselines are capable of handling requests up to 9 services. Nonetheless, for 10 services, only \textit{Knative-rts} and Oparaca remain error-free. \textit{Knative-con} and \textit{Knative} still suffer from the resource contention. Similarly, for compute-intensive workload (Figure~\ref{fig:evlt:efficiency:video}), \textit{Knative-rts} and \textit{Knative-con} only have enough resources to meet the target throughput up to 8 and 7 services without any error, respectively. Oparaca, however, can handle all of the requests for up to 10 services.

\vspace{2mm}
\noindent
\colorbox{blue!10}{
\parbox{0.96\linewidth}{
\underline{\textbf{Takeaway}:} \emph{
Being cognizant of performance objectives is crucial for Oparaca to deliver competitive efficiency for both the user and the system across different applications while also offering a high-level abstraction to the user.
}}}
\vspace{2mm}

\subsection{Deployment Productivity Using Oparaca}
\label{sec:evlt:deployment}

\begin{figure} [ht]
    \centering
    % \vspace{-4mm}
    \includegraphics[width=0.7\linewidth]{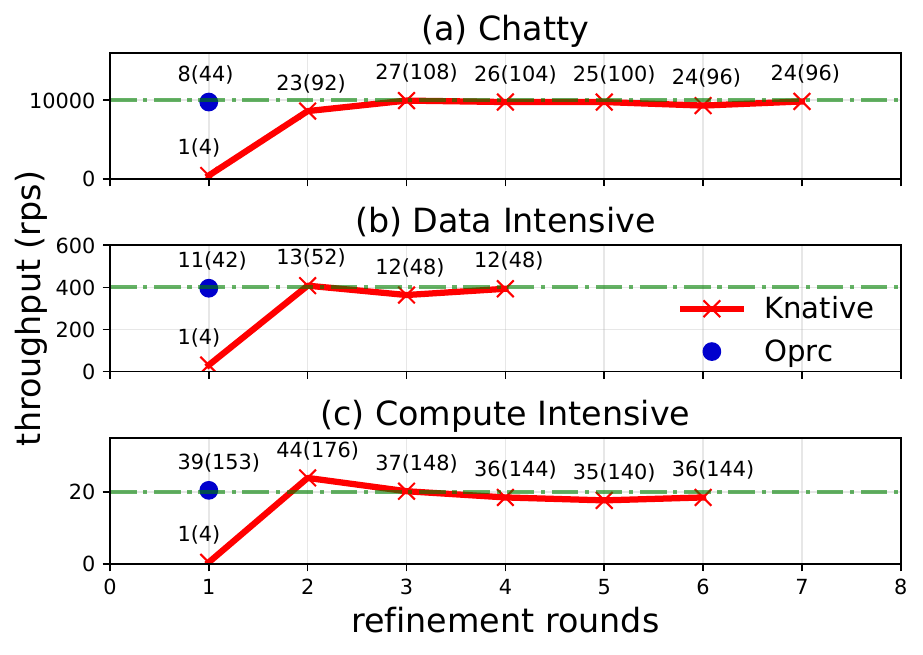}
    \caption{Rounds of refinement for Knative to enforce the target throughput (green lines) versus Oparaca. Data points are annotated by \texttt{\#pods(\#cores)}, including invoker pods.}
    \label{fig:evlt:deployment}
\end{figure}

\begin{figure} [h]
    \centering
    % \vspace{-2mm}
    \includegraphics[width=0.75\linewidth]{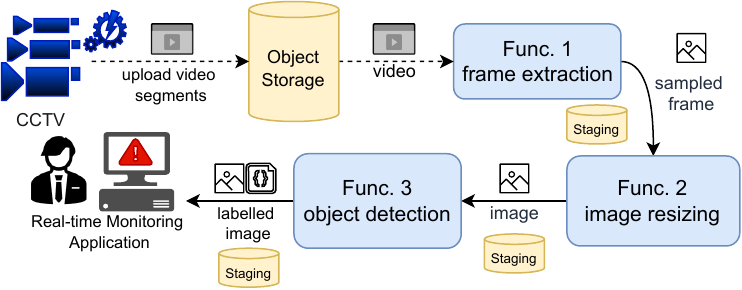}
    \caption{The case study of developing video and image processing for a real-time monitoring system}
    \label{fig:evlt:case-study}
    % \vspace{-2mm}
\end{figure}

To show the productivity improvement of Serverless application deployment, we present the experiment on the refinement steps using Knative on three application deployments with the requirement to enforce the throughput of 10k, 400, and 20 requests per second for chatty, data-intensive, and compute-intensive, respectively. The manual refinement strategy consists of three phases. First, we want to find the number of pods that roughly provide throughput that is equal to our objective. We deploy the application with a single pod and then perform load testing to find the throughput. Then, we scale it up using the formula below and repeat this process until the throughput matches the objective. 
\[ pods_{next} = \dfrac{throughput_{target}}{throughput_{current}}\times{pods_{current}} \]
The second phase reduces the pods until they cannot satisfy the target. The last phase increases the container-level concurrency but reduces the number of pods to improve utilization. 

As shown in Figure~\ref{fig:evlt:deployment}, the manual refinement method needs at least 4 rounds to find the optimal number of pods to meet the target throughput, while we only need to give the Oparaca the number, and it will automatically adjust the deployment when we feed the load. Furthermore, Oparaca improves application performance while reducing the required resource allocation to meet the target throughput. For IO-intensive workloads focused on structured data like the chatty workload, Oparaca reduces resource usage from 100 cores to 44 cores. This is because OaaS unlocks cross-domain optimization---in this case, data locality---to speed up invocation execution time, quickly freeing up FaaS pods for higher concurrency and significantly reducing resource requirements compared to Knative. Even for the compute-intensive application, where locality is not an issue, Oparaca automatic refinement still achieves the throughput target at a comparable cost (153 cores) versus Knative (144 cores,  only 6\% higher), which requires much more effort in manual tuning (6 rounds of refinement versus one).

\vspace{2mm}
\noindent
\colorbox{blue!10}{
\parbox{0.96\linewidth}{
\underline{\textbf{Takeaway}:} \emph{
% Oparaca not only automates the deployment configuration refinement process, but it also improves the application performance and reduces the resource requirement to meet the same objective.
Oparaca's OaaS abstraction improves deployment productivity and performance enforcement effectiveness.
}}}
\vspace{2mm}

\subsection{Development Productivity Using Oparaca}
\label{sec:evaluation-ease-of-use}

In this part, we provide two cloud application developments representing common cloud applications at different scales, non-functional requirements, and complexities. We will deploy these applications using the OaaS paradigm and recommended FaaS deployment practices to demonstrate how OaaS can make the development of cloud-native serverless applications more productive.

% \textbf{The first use case is a video processing application for real-time monitoring.}

\paragraph{\textbf{\underline{Case Study \# 1. Real-time Monitoring System.}}}
Figure \ref{fig:evlt:case-study} shows a CCTV system uploading video segments to object storage, waiting to be processed by a workflow of function that includes \texttt{extractFrame()} that splits a video segment into multiple frames; \texttt{resizeImg()} whose job is to resize the image frame to be usable by the next function in the pipeline; and \texttt{detectObject()} is in charge of performing the object detection on an image and generating label in the \texttt{JSON} format. These functions must persist their output data so that the following function in the workflow can consume it. Because the entire workflow is latency sensitive, the execution rate of the whole workflow (i.e., throughput) has to be guaranteed. Developers can calculate the throughput by the number of cameras and the object detection frequency. 

\noindent\textbf{FaaS implementation.}
The developer must repeat the following steps for each function deployment: (i) Configuring cloud-based object storage, database and maintaining the credential access token for the functions to use. (ii) Implementing the functions' business logic. (iii) Data management within the functions that itself involves three steps: (a) allocating the storage addresses to fetch or upload data; (b) authenticating access to the object storage via the access token; and (c) implementing the fetch and upload operations on the allocated addresses.
Upon implementing these functions, the developer must connect them as a workflow via a function orchestrator service (e.g., AWS Step Functions~\cite{aws-sf}). Finally, upon arrival of a new video segment, the event triggers the workflow to put the result into the database, waiting to be processed by the monitoring system. To ensure the target throughput, developers have to go through multiple rounds of testing and refinement to get the final configuration for each function.

\noindent\textbf{OaaS implementation.} The developer defines three classes: 
\begin{itemize}[leftmargin=*, noitemsep, topsep=0.5pt]

\item \texttt{\textbf{Video}} class with \texttt{extractFrame()} function that produces \texttt{LabeledImage} as the output, and \texttt{wfDetectObject(freq)} workflow function that has a detection frequency as the input. This class also has \texttt{video} file as an unstructured state.
    
\item \texttt{\textbf{Image}} class contains \texttt{resize} function and \texttt{image} file as an unstructured state (see Listing~\ref{lst:image-yaml}). 
    
\item \texttt{\textbf{LabeledImage}} class inherits from the \texttt{Image} class and has its own \texttt{objectDetection()} function and \texttt{labels} data (state) in JSON format (see Listing~\ref{lst:image-yaml}).

\end{itemize}
Upon uploading a new video to the Oparaca platform by the CCTV system, it creates a ``video'' object and invokes \texttt{video}. \texttt{wfDetectObject(freq)} that outputs a \texttt{LabeledImage} object that is consumed by the real-time monitoring application. We note that, in developing the class functions, the developer does not need to be involved in the data locating and authentication steps. To ensure the application performance, developers only need to declare the target throughput within the class definition (see example in Listing~\ref{lst:image-yaml}); then, the Oparaca can transparently create the suitable class runtimes and their configuration.

\begin{figure}[t]
    \centering
    \subfloat[FaaS-based \cite{aws-searchable-document-repo}]{
        \includegraphics[width=0.6\linewidth]{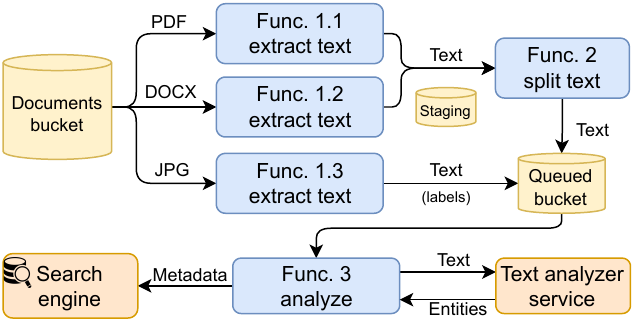}
        \label{fig:evlt:case-study2-faas}
    }
    
    \subfloat[OaaS-based]{
        \includegraphics[width=0.65\linewidth]{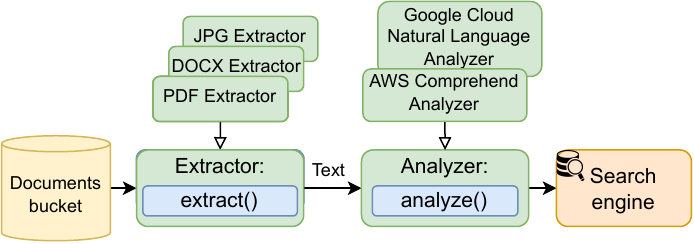}
        \label{fig:evlt:case-study2-oaas}
    }
    \caption{The searchable enterprise document repository implemented based on FaaS and OaaS paradigm.}
    \label{fig:evlt:case-study2}
\end{figure}

\paragraph{\textbf{\underline{Case Study \# 2. Searchable Document Repository.}}}
Retrieving and processing at scale the vast repositories of valuable documents, images, and media from enterprise customers is a common practice in the cloud \cite{nguyen2020motivating, wang2018efficient}. In this case study, we first present how the application is deployed with traditional FaaS on the cloud, the limitations of this approach, and how to resolve them with OaaS/Oparaca. 

\vspace{1mm}
\noindent\textbf{FaaS implementation.}
Figure~\ref{fig:evlt:case-study2} shows the serverless workflow to analyze the document in various formats and update the metadata to the search engine recommended by AWS \cite{aws-searchable-document-repo}. Upon the document uploads to the document bucket (object storage), the storage triggers the event to invoke \texttt{extractText()} based on the type of the document. If the document is in \texttt{PDF} or \texttt{DOCX} format, the function extracts the text and sends the text to be split by the next function \texttt{splitText()}. The result will be put into the \texttt{Queued bucket}. Alternatively, if the document is in \texttt{JPG} format, the \texttt{extractText()} function analyzes the image to get labels and puts them in the \texttt{Queued bucket}. In the next step, the \texttt{analyze()} function loads text from the \texttt{Queued bucket} to analyze it via the external text analyzer service (e.g., AWS Comprehend) and then saves the metadata result to the search engine.

The FaaS implementation has two main drawbacks. First, developers must explicitly manage application state and data using separate storage services, which increases complexity and makes it difficult to configure non-functional requirements as in the previous case study. Second, functionalities may require numerous and heterogeneous FaaS deployments—for example, needing separate extraction functions for each document type, where some (like \texttt{PDF} and \texttt{DOCX}) require staging and others (like \texttt{JPG}) do not. These drawbacks complicate development, deployment, and management as the application evolves to handle various document types and integrates more functionalities and options (e.g., using multiple text analyzer services instead of one).

\noindent\textbf{OaaS implementation.} To demonstrate the feasibility of OaaS in production, we transform the given FaaS-based solution into OaaS with minimal effort to resolve the previously mentioned drawbacks. The transformation involves three steps.

\begin{itemize}[leftmargin=*, noitemsep, topsep=0.5pt]
    \item \textbf{Workflow Construction.} We encapsulate related FaaS functions, states, and key data into objects representing two key functionalities: \texttt{\textbf{Extractor}} to extract text from the document repository and \texttt{\textbf{Analyzer}} to analyze the extracted text. The two classes form the critical path of the application processing pipeline, as shown in Figure \ref{fig:evlt:case-study2-oaas}.
    \item \textbf{Object Encapsulation.} We apply inheritance and polymorphism to promote software reuse by wrapping corresponding FaaS functions and states into classes derived from the two base classes. This approach hides the need for storage services behind the object abstraction and outsources their implementation to the cloud. It also simplifies development, as developers only need to construct the processing pipeline once in the base class definitions and then focus on implementing functionalities for specific cases with their derived classes, avoiding repetitive pipeline construction and implementation whenever a new document type or analyzer service is added.
    \item \textbf{Integration of Non-Functional Requirements.} Developers integrate appropriate non-functional requirements into the corresponding objects to meet application needs for performance, availability, and cost. With Oparaca, non-functionality requirement enforcement, as shown in previous experiments, is achieved without any additional refinement effort from the developers.
\end{itemize}

\vspace{2mm}
\noindent
\colorbox{blue!10}{
\parbox{0.96\linewidth}{
\underline{\textbf{Takeaway}:} \emph{Oparaca accelerates development by abstracting low-level infrastructure concerns and automating runtime configurations through a high-level interface.
}}}
\vspace{2mm}

\section{Summary}

This chapter presented declarative Non-functional Requirement (NFR) management as a critical capability of the Object-as-a-Service paradigm. By encapsulating application logic, data, and performance specifications into unified class definitions, OaaS enables developers to express desired outcomes—throughput guarantees, latency targets, availability requirements—through intuitive declarations rather than low-level configurations. The Oparaca platform demonstrates this approach through the LTAG class runtime template, which automatically translates NFR specifications into optimized deployments. Evaluation across diverse workloads confirms that Oparaca enforces declared NFRs with comparable or superior resource efficiency versus state-of-the-art approaches while dramatically reducing deployment complexity and eliminating manual refinement cycles. Case studies further illustrate how OaaS abstractions simplify real-world development by hiding infrastructure complexity behind object-oriented interfaces, allowing developers to focus on business logic.

While this chapter focused on centralized cloud deployments, modern applications increasingly demand distributed execution across edge and cloud environments. IoT applications, mobile computing, and latency-sensitive services require deployment flexibility beyond single data centers. However, the edge introduces new challenges: resource heterogeneity, intermittent connectivity, and geographic distribution complicate NFR enforcement strategies designed for centralized infrastructure. Chapter~\ref{chapter5} addresses these challenges by extending OaaS across the edge-cloud continuum, introducing EdgeWeaver—an architecture that adapts declarative NFR management to distributed, geo-dispersed deployments.
%%%%%%%%%%%%%%%%%%%%%%%%%%%%%%%%%%%%%%
\chapter{OaaS-IoT: Extending Object as a Service to the Edge--Cloud Continuum\protect\footnotemark}
\label{chapter5}
%%%%%%%%%%%%%%%%%%%%%%%%%%%%%%%%%%%%%%

\footnotetext{This chapter is based on: P.~Lertpongrujikorn, H.~D.~Nguyen, and M.~Amini Salehi, ``EdgeWeaver: Seamless Edge--Cloud Integration for Stateful Serverless Applications,'' in \textit{Proceedings of the 40th IEEE International Parallel and Distributed Processing Symposium (IPDPS '26)}, 2026 (under review). The version included here is the author's dissertation version and may differ from a final published version, if accepted.}

\section{Overview}
\label{ch5-sec:overview}

Chapters~\ref{chapter3} and~\ref{chapter4} established the OaaS paradigm and SLA-driven deployment within centralized cloud environments. However, the empirical findings in Chapter~\ref{chapter1} revealed that 43\% of practitioners face responsiveness challenges, particularly in distributed and edge computing scenarios where low-latency, localized decision-making is critical for IoT, healthcare, finance, and robotics applications. Modern applications increasingly demand deployment across the edge-cloud continuum—spanning resource-constrained edge devices, regional edge datacenters, and centralized cloud infrastructure—yet current serverless platforms lack abstractions that gracefully span these heterogeneous, intermittently connected tiers.

This chapter extends OaaS to the edge-cloud continuum through OaaS-IoT, a paradigm that maintains the unified object abstraction while introducing SLA-driven placement, connectivity-aware invocation, and graceful degradation mechanisms tailored for geo-distributed, resource-heterogeneous environments. We present \name, a platform that realizes OaaS-IoT by automatically deploying objects across edge and cloud tiers based on declarative SLA specifications, managing state consistency despite intermittent connectivity, and adapting execution strategies to network conditions. The chapter is organized as follows: Section~\ref{ch5:sec:approach} presents the OaaS-IoT abstraction and its extensions for the edge-cloud continuum; Section~\ref{sec:realization} describes the \name architecture, including comprehensive object abstraction, SLA-driven deployment across tiers, and connectivity-aware enforcement mechanisms; and Section~\ref{ch5:sec:evaluation} evaluates \name through case studies and human studies demonstrating 44.5\% reduction in lines of code, $10\times$ fewer configuration requirements, and 31\% faster development completion time compared to traditional FaaS-based approaches while maintaining robust SLA enforcement across the continuum.

\section{OaaS-IoT: Abstraction for the Continuum}
\label{ch5:sec:approach}

To overcome the limitations of existing approaches in addressing the complexity and connectivity challenges of IoT deployment across the Edge–Cloud continuum, we present OaaS-IoT, an extension of the OaaS paradigm that brings unified object abstraction to heterogeneous edge–cloud environments. We realize OaaS-IoT through the \name platform, which provides a \textit{unified abstraction} that \textit{decouples} the application from the underlying infrastructure, reducing the effort required to develop, adapt, and maintain IoT services in heterogeneous environments.

\subsection{Comprehensive Object Abstraction}

Building on the OaaS abstraction introduced in Chapter~\ref{chapter3}, OaaS-IoT applies the same object-oriented model across the edge–cloud continuum. Applications are composed of \textit{distributed objects} whose \emph{attributes} capture IoT state and whose \emph{methods} (serverless functions) encapsulate logic; objects communicate via method invocations. QoS constraints are declared as SLAs (Table~\ref{ch5:tab:non-functional-requirements}) and enforced by the platform. Unlike the cloud-only setting, objects may span intermittently connected tiers and heterogeneous devices, so the abstraction emphasizes location awareness (locality), resilience to disconnection, and smooth integration with device endpoints.

Developers reuse familiar OOP constructs (abstract classes, inheritance, polymorphism) to build modular IoT services while staying decoupled from infrastructure details. The same class definitions can be deployed at the edge or in the cloud without per-environment configuration, preserving portability while \name (the OaaS-IoT implementation) manages placement and communication.

\subsection{Declarative SLA-Driven Deployment}
Extending the declarative NFR management of Chapter~\ref{chapter4}, OaaS-IoT abstracts deployment complexity through SLA specifications attached to classes, attributes, and methods. SLAs drive automated placement and configuration across edge and cloud, so applications meet their QoS targets without manual tuning. With a unified view of logic and state, \name co-locates functions with their data to minimize latency and transfer overhead, and adapts to changing conditions and heterogeneous resources to sustain SLA-compliant execution across the continuum.

\subsection{Application Development and Deployment}

\begin{figure*} [t]
    \centering
    \includegraphics[width=\textwidth]{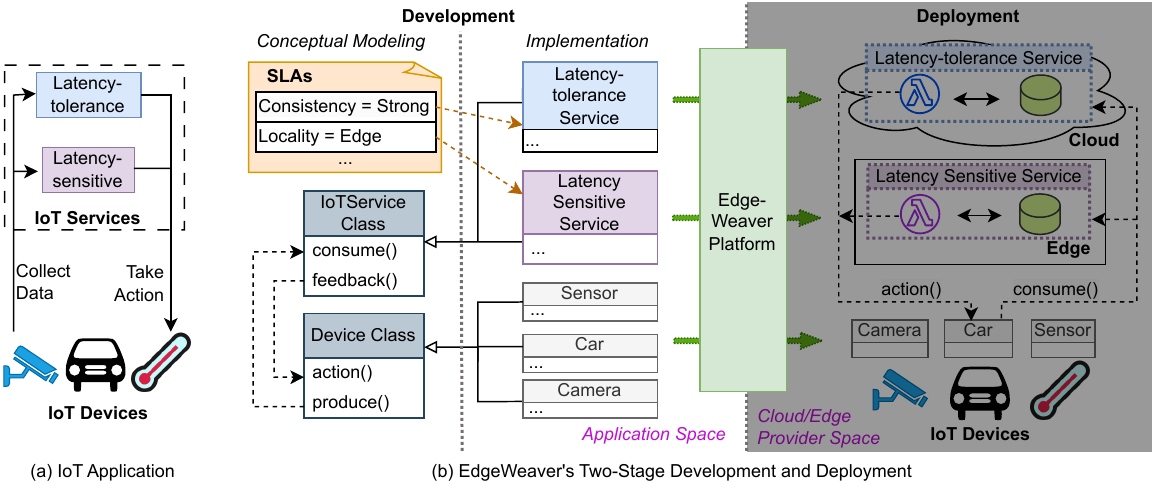}
    \caption{Overview of \name (\name implements the OaaS-IoT paradigm) to resolve Edge-Cloud challenges for IoT applications}
    \label{fig:oaas-iot}
\end{figure*}

As illustrated in Figure \ref{fig:oaas-iot}, OaaS-IoT establishes a novel two-stage abstraction for developing and deploying IoT applications across Edge-Cloud continuum:

\underline{\textit{Conceptual Modeling}}: Developers begin by modeling essential components and workflows using OaaS-IoT's unified abstraction. For example, heterogeneous IoT devices are modeled by a \texttt{Device} class, defining their basic operations (e.g., produce data and take action) with other IoT services, which are also modeled as separate classes. This high-level blueprint abstracts away the complexities of the underlying infrastructure, offering a clear and unified design framework.
    
\underline{\textit{Implementation}}: Developers extend these conceptual classes to capture the specifics of actual IoT devices and services, associating them with SLAs. The enriched class definitions are then submitted to the \name platform (the OaaS-IoT implementation). The platform extracts the embedded logic, state, and communication patterns to instantiate concrete Edge-Cloud components (e.g., FaaS functions, databases, and event pipelines) for deployment.

\textit{Deployment is fully automated} within the provider ecosystem, with the declared SLAs driving each component placement, scheduling, and management. For example, as shown in Figure \ref{fig:oaas-iot}b, the latency-sensitive service is implemented with a ``\texttt{locality=Edge}'' SLA. During deployment, \name places its corresponding FaaS functions and state at the edge nodes to reduce data access and device communication latency. Conversely, a latency-tolerant service that requires strong consistency (e.g., linearization) is deployed on the cloud, where robust infrastructure can meet its SLA demands. With this SLA-driven deployment, \name ensures diverse application QoS requirements are met across Edge-Cloud continuum without any tuning effort from the developers, thereby, mitigating their deployment effort.   

In sum, OaaS-IoT overcomes the limitations of existing approaches by providing a unified framework that simplifies IoT application development and deployment by abstracting both functional and non-functional aspects, ensuring seamless operation across intermittent Edge–Cloud environments.

\begin{figure}[t]
    \centering    
    \includegraphics[width=0.85\linewidth]{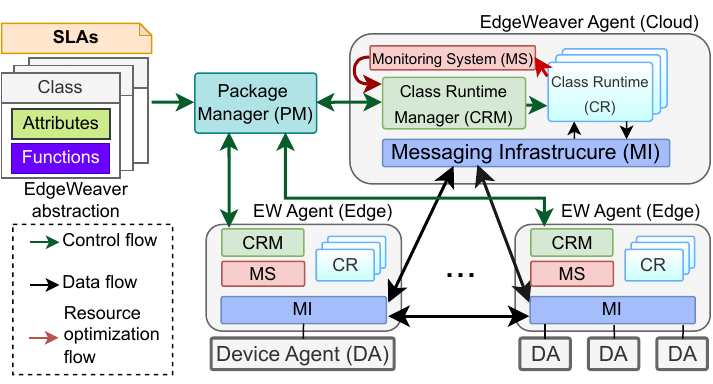}
    \caption{\name Architecture. The cloud-based Package Manager (PM) coordinates deployment across edge and cloud tiers, each hosting EdgeWeaver Agents with Class Runtimes (CR) for object execution, Class Runtime Managers (CRM) for lifecycle management, Monitoring Systems (MS) for SLA metrics, and Messaging Infrastructure (MI) for communication. Device Agents (DA) integrate IoT devices. Control flows (green), data flows (black), and resource optimization flows (red) enable SLA-compliant execution across the continuum.}
    \label{fig:oaas_arch}
\end{figure}

\section{\name: Realizing OaaS-IoT}
\label{sec:realization}

\subsection{Design Goals and Architecture}

We design the \name platform, following the architecture shown in Figure~\ref{fig:oaas_arch}, to realize the OaaS-IoT paradigm and meet three key requirements:

\vspace{2mm}
\noindent\textbf{a. Comprehensiveness.} To reduce development complexity and provide a unified view of IoT applications, we leverage object-oriented programming (OOP) concepts to define abstractions that capture both functional and non-functional aspects of application logic without exposing developers to underlying infrastructure details. We implement the \textit{Object Abstraction}, which provides APIs and tools for modeling applications as classes, along with SLAs to specify requirements for performance, availability, and consistency. Additionally, we introduce lightweight \textit{Device Agents} (DA) that run on IoT devices, exposing platform-specific APIs to integrate these devices into the \name environment. Together, these components create a unified abstraction layer that supports full lifecycle of application design and development.
% that simplifies application design and supports full lifecycle development.

\vspace{2mm}
\noindent\textbf{b. Adaptability.} To handle the dynamics of execution environments (e.g, network failures), \name enables automated, SLA-driven adaptation. For that purpose, each tier across Edge–Cloud hosts an \textit{\agent}, which includes a \textit{Monitoring System} (MS in Figure~\ref{fig:oaas_arch}) that collects SLA-related metrics and a \textit{Class Runtime Manager (CRM)} that adjusts deployments in real-time. This forms a localized control loop that continuously adapts to workload fluctuations, resource availability, and network conditions without manual tuning.

\vspace{2mm}
\noindent\textbf{c. Applicability.} 
To ensure broad adoption and scalable operation across heterogeneous infrastructures, \name uses a modular architecture. Objects are deployed via \textit{Class Runtimes (CR)}, configured by the Class Runtime Manager (CRM) according to both SLA specifications and capabilities of the hosting datacenter. Each \agent includes a \textit{Messaging Infrastructure (MI)} that abstracts inter-object communication into a topic-based protocol-agnostic model. At the global level, \textit{Package Manager} coordinates deployment and synchronization across tiers, enabling platform-wide consistency and scalable, hybrid Edge-Cloud deployments.

% To ensure broad adoption and scalable operation across heterogeneous infrastructures, \name supports a wide range of applications and system configurations. Objects are deployed via \textit{Class Runtimes (CR)}, configured by the Class Runtime Manager (CRM) according to both SLA specifications and capabilities of the hosting datacenters. Each \agent includes a \textit{Messaging Infrastructure (MI)} that abstracts inter-object communication into a topic-based protocol-agnostic model. At the global level, a centralized \textit{Package Manager} coordinates deployment and synchronization across all tiers, ensuring platform-wide consistency and enabling scalable, hybrid Edge-Cloud deployments.

By fulfilling these design requirements, the \name architecture delivers its original vision: simplifying application development; enabling flexible and automated deployment; and supporting robust, SLA-compliant execution in heterogeneous and intermittently connected environments. In the remainder of this section, we describe how the architectural components interact to support end-to-end IoT application conceptualization, development, and deployment---demonstrating how \name meets its design goals in practice.

\begin{table*}[ht]
    \centering
    \small
    \begin{tabular}{|p{2.2cm}|p{4.5cm}|p{1.2cm}|p{7.5cm}|}
        \hline
        \textbf{SLA} & \textbf{Value Type} & \textbf{Unit} & \textbf{Definition}\\ \hline \hline
        % \multicolumn{4}{|l|}{\textit{QoS Requirements}}\\ \hline
        Throughput & Integer & RPS & Minimum number of invocations guaranteed to be executed per second\\ \hline
        %\hline
        % \textbf{Latency} & \textbf{Enumerate} & \textbf{N/A} & \textbf{How close should the actual invocation latency be to the ideal latency}\\ \hline\hline
        Locality & Preferred datacenters & N/A & Preferred data centers that will be used for deployment\\ \hline
        Availability & Real & \% & The percentage of time an object/function must be available for service\\ \hline
        Consistency 
        & - Read your Write (RYW) & N/A & Ensure a client's next read includes its most recent write\\
        & - Bounded Staleness ($\Delta$) & sec & Read can lag behind the latest write, but only within $\Delta$ seconds\\
        & - Strong & N/A & Read always reflects the latest write\\
        \hline
    \end{tabular}
    \caption{SLAs supported by OaaS-IoT}
    \label{ch5:tab:non-functional-requirements}
\end{table*}

\begin{table*}[ht]
    \centering
    \small
    \begin{tabular}{|p{2.2cm}|p{4.5cm}|p{8.3cm}|}
        \hline
        \textbf{Categories} & \textbf{API} & \textbf{Explanation} \\ \hline \hline
        \multirow{3}{2.2cm}{Object APIs}
            & \texttt{CLASS.create()} & Create a new object of class \texttt{CLASS} and return its ID \\ \cline{2-3}
            & \texttt{CLASS.get(ID)} & Retrieve an object of class \texttt{CLASS} by ID \\ \cline{2-3}
            & \texttt{CLASS.delete(ID)} & Delete an object of class \texttt{CLASS} by ID \\ \cline{2-3} \hline
        \multirow{2}{2.2cm}{Attribute APIs}
            & \texttt{commit(obj, attr)} & Write local changes of \texttt{attr} in \texttt{obj} to storage \\ \cline{2-3}
            & \texttt{refresh(obj, attr)} & Read the latest value of \texttt{attr} in \texttt{obj} from storage \\ \cline{2-3} \hline
        \multirow{2}{2.2cm}{Function APIs}
            & \texttt{trigger(func, src, e)} 
                & Trigger \texttt{func} when event \texttt{e} occurs on \texttt{src}. Events: \texttt{OnComplete} or \texttt{OnFailure} if \texttt{src} is a function; \texttt{OnCreate}, \texttt{OnUpdate}, or \texttt{OnDelete} if \texttt{src} is an attribute \\  \cline{2-3}
            & \texttt{suppress(func, src, e)} & Disable trigger on \texttt{func} from \texttt{src} on event \texttt{e} \\ \hline
    \end{tabular}
    \caption{\name's API}
    \label{tab:oaas-iot-api}
\end{table*}

\begin{figure} [t]
    \centering
    \includegraphics[width=0.8\linewidth]{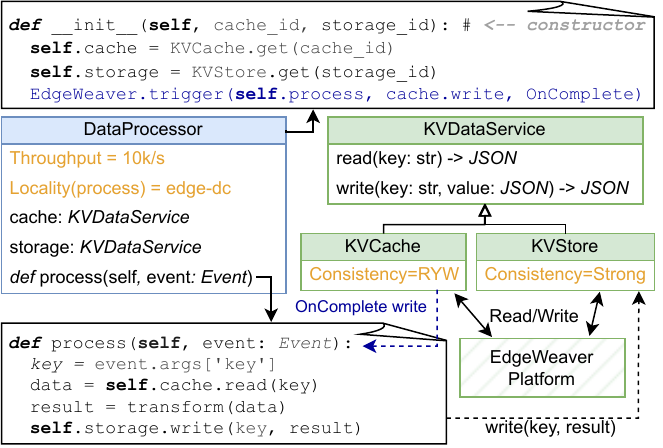}
    \caption{Modeling and implementing a simple IoT processing service with \name}
    \label{fig:oaas-abstraction-example}
\end{figure}

\subsection{Object Abstraction Realization}
\label{sec:realization-abstraction}

At its core, an OaaS-IoT application is structured around classes that define the blueprint for independently executable objects. Each class encapsulates attributes (representing state or data) and methods (implemented as serverless functions), following object-oriented programming principles. Upon deployment, \name automatically instantiates and manages these objects using distributed \textit{Class Runtimes}, as described earlier. The abstraction also supports inheritance and polymorphism, allowing developers to create extensible, reusable components, promoting modular design and reducing code duplication.
\name allows developers to associate SLAs with classes, methods, or attributes to specify QoS requirements, including locality, throughput, availability, and consistency (see Table \ref{ch5:tab:non-functional-requirements}). It also provides a high-level API (Table~\ref{tab:oaas-iot-api}) for inter-object communication, event-driven triggers, and system interactions---all without handling low-level details like networking protocols or deployment scripts.

Figure \ref{fig:oaas-abstraction-example} illustrates how the abstraction supports comprehensive, infrastructure-independent application development. This example has an IoT data processing service, implemented by the \texttt{DataProcessor} class that consumes data from a short-term cache, processes it, and writes it to a long-term store. Since both expose a key-value interface, developers define a base class (\texttt{KVDataService}) and extend it into \texttt{KVCache} and \texttt{KVStore} through inheritance. 
\name manages all object instantiations and data handling internally. Developers simply attach SLA annotations, for example, \texttt{Consistency=Strong} on \texttt{KVStore} to ensure strict linearizability without manually configuring consensus protocols (e.g., Raft \cite{ongaro2014search}). The \texttt{DataProcessor} interacts with these services via class attributes, which are automatically injected by the platform at object creation. Developers can also register event-driven triggers (e.g., executing a function upon adding data to the cache). Furthermore, performance requirements can be specified either globally (e.g., \texttt{throughput=10k/s}) or per method (e.g., \texttt{Locality(process)=edge-dc}). All are managed entirely by \name; without any custom orchestration, messaging setup, or deployment scripting.

\begin{figure}
    \centering
    \includegraphics[width=0.7\linewidth]{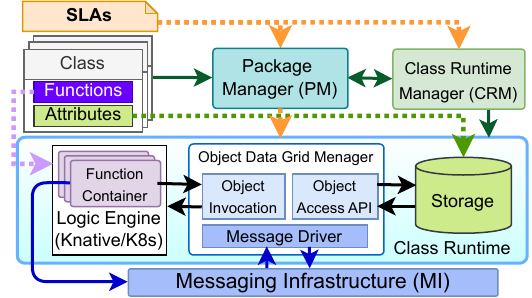}
    \caption{Class Runtime Internal Architecture. Each Class Runtime (CR) contains three core components: the Logic Engine executes class methods in isolated function containers (via Knative/K8s); the Storage System manages object attributes with appropriate backends based on data types and consistency requirements; and the Object Data Grid Manager (ODGM) orchestrates all operations through its Object Invocation module (routes method calls), Object Access API (enforces consistency protocols), and Message Driver (handles cross-datacenter communication).}
    \label{fig:oaas_cr}
\end{figure}

\begin{figure}
    \centering
    \includegraphics[width=0.85\linewidth]{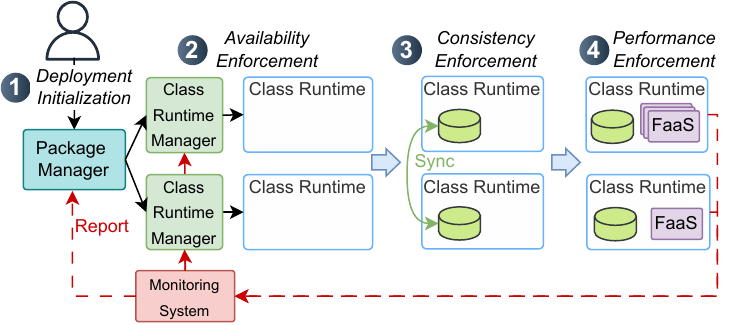}
    \caption{SLA Enforcement Workflow. The Package Manager initializes deployment (1) by distributing Class Runtime Managers to selected datacenters. Availability enforcement (2) determines replication factor and placement using failure probability metrics. Consistency enforcement (3) provisions storage with Raft (strong) or anti-entropy (bounded staleness) protocols. Performance enforcement (4) deploys pre-warmed containers to meet throughput requirements. The Monitoring System tracks metrics and triggers dynamic adaptation when violations occur.}
    \label{fig:oaas-object-slas}
\end{figure}

\subsection{Class Deployment}
\label{sec:realization-object}

Upon development completion, developers submit their applications to \name as a collection of class definitions annotated with SLAs.
Figure~\ref{fig:oaas_cr} illustrates how \name transforms this submission into concrete deployments.
First, the deployment package, including class definitions and SLAs, is submitted to the Package Manager. Based on the list of data centers the application is authorized to access, the Package Manager forwards the relevant class definitions to the corresponding \agents operating in those target data centers. Each destination \agent invokes its Class Runtime Manager to process the submission and instantiate a dedicated Class Runtime for each class. These class Runtimes manage the lifecycle of all object instances associated with the class and enforce their SLA during their lifetime.

Each Class Runtime is composed of three key components: (i) \textit{Logic Engine} that executes class methods; (ii) \textit{Storage System} that instantiates appropriate storage backends for managing object attributes based on their data types and consistency requirements; and (iii) \textit{Object Data Grid Manager (ODGM)} that orchestrates invocations and data access using modules for invocation routing, data consistency, and cross-datacenter communication (via Zenoh~\cite{liang2023performance}).

% \vspace{-3mm}
\section{SLA Enforcement}
\label{sec:realization-sla}

As in Chapter~\ref{chapter4}, SLAs define target semantics; here we adapt enforcement to cross-datacenter deployments and intermittently connected environments. Once a class is submitted and deployed, developers can instantiate objects whose lifecycles must comply with the SLAs in their class definitions. Figure~\ref{fig:oaas-object-slas} summarizes how \name enforces SLAs during deployment and execution, detailed next.
% When a class is deployed via Package Manager, the system uses the associated SLAs to choose suitable deployment and runtime strategies.

\subsection{Availability Enforcement}

At class deployment time, the Package Manager selects appropriate tiers (a.k.a. datacenters) to host Class Runtimes, guided by the availability SLA.
It estimates the failure probability of each data center using metrics (e.g., uptime, network reliability) collected from the Monitoring System. Based on these estimates and the desired availability target, it calculates the required replication factor using 0the Meroufel and Belalem method~\cite{meroufel2013managing}. The Package Manager then uses the replication factor to select the necessary number of data centers that satisfy developer-specified constraints (e.g., locality). If no constraints are given, it defaults to a round-robin strategy for load balancing. Class Runtimes are then deployed across these selected sites to host object replicas and ensure SLA-compliant availability.

\subsection{Consistency Enforcement}
\label{sec:realization-sla:consistency}

When Class Runtimes are deployed, they provision storage and coordinate with each other to enforce the consistency SLA. For \textbf{Strong Consistency}, the Raft consensus protocol~\cite{ongaro2014search} is integrated into the ODGM’s \textit{Object Access API} to ensure that all replicas agree on the latest object states before processing reads or writes. \textbf{Bounded-Staleness Consistency} allows stale reads within a time-bound. Class Runtime's ODGM employs anti-entropy techniques with Merkle-Search Trees\cite{auvolat2019merkle} to detect and repair inconsistencies within the defined window, along with CRDTs \cite{simic2020crdts} to manage out-of-order changes. Read/write access is blocked if network partitions exceed the allowed staleness window. \textbf{Read-Your-Write (RYW) Consistency} allows reads to see a recent write from the same source, even under network partition. Class Runtime enforces this by routing reads and writes through the object access API to the same local storage.
% \vspace{-2mm}

\subsection{Performance Enforcement}
Each class’s methods are deployed by the Logic Engine into isolated containers. If a throughput SLA is specified, containers are pre-warmed with sufficient compute resources---calculated using techniques from real-time Serverless~\cite{nguyen2019real}---to meet the required invocation rate. If a locality SLA is present, the preferred data center must reserve enough resources to meet both throughput and co-location requirements. The Class Runtime ensures that containers are deployed on the same machine as the object’s storage to reduce access latency and pre-warms containers to avoid cold starts. When multiple replicas exist and locality is not a constraint, resource allocation is balanced across them, proportional to their resource availability.
% \vspace{-2mm}
\subsection{SLA-compliance Execution}
After deployment, users interact with objects via the API provided in Table~\ref{tab:oaas-iot-api}. API calls are routed through the \textit{Messaging Infrastructure}, which transparently directs each request to the appropriate ODGM instance based on the object ID embedded in the calls.
Upon receiving a request, the ODGM’s \textit{Object Invocation} module triggers the corresponding function on the local Logic engine. If the function call targets a remote object, the ODGM leverages the \textit{Message Driver} to relay the request to the corresponding location. This mechanism enables seamless, location-transparent invocation across the Edge–Cloud continuum.
When a function needs to access object attributes, the attribute ID is passed through the \textit{Messaging Infrastructure} to the \textit{Object Access API} of the relevant ODGM. Before any data operation is executed, the ODGM enforces consistency guarantees by running the necessary replication and consistency protocols (see \S\ref{sec:realization-sla:consistency}), to enforce SLA-compliant execution.
% \vspace{-1mm}
\subsection{SLA Monitoring and Lifecycle Management}
Monitoring System continuously collects SLA-related metrics (from Class Runtimes) that are reported to the Package Manager and the Class Runtime Manager. Upon SLA violation or runtime failure, \name automatically initiates corrective actions, such as reallocating resources or instantiating new runtimes to maintain SLA compliance  for all objects during their lifecycle.

\section{Performance Evaluation}
\label{ch5:sec:evaluation}
\subsection{Methodology}

\textbf{Goals.}
We evaluate \name across realistic settings to assess whether it fulfills its design objectives (\S\ref{sec:realization-object}) and thus, effectively addresses the challenges of IoT application development and deployment across the Edge–Cloud continuum. %(\S\ref{sec:background}).
 Specifically, we aim to answer the following key questions:
(i) \textit{\underline{Comprehensiveness and Productivity}:} Does the unified object abstraction and declarative SLA interface provide a high-level view of IoT applications to support diverse QoS requirements and simplify development, ultimately improving developer productivity? (\S\ref{sec:evaluation-results-case-study})
(ii) \textit{\underline{Efficiency for Practice Uses}:} Can \name implement its abstractions and enforcement mechanisms efficiently and at scale, matching or even exceeding the performance of state-of-the-art systems, thus developers enjoy higher productivity without incurring significant trade-offs? (\S\ref{sec:evaluation-results-applicability})
(iii) \textit{\underline{Adaptability for Reliable Execution}:} Can \name dynamically respond to workload and infrastructure changes to preserve QoS with minimal developer effort? (\S\ref{sec:evaluation-results-adaptability})

\subsection{Experimental Setup}
We conduct experiments on Chameleon Cloud \cite{chameleon_cloud}, using two clusters to represent the cloud and edge tiers. The \textit{cloud cluster} consists of machines equipped with dual-socket Intel(R) Xeon(R) Platinum 8380 CPUs (240 cores total) and 768 GB of memory. The \textit{edge cluster} uses machines with dual-socket Intel(R) Xeon(R) Gold 6240R CPUs (96 cores total) and 256 GB of memory. To reflect realistic deployment scenarios, we deploy the two clusters in geographically dispersed data centers: TACC (Texas) for the cloud and UC (Illinois) for the edge. The clusters communicate over a standard Internet connection with an average round-trip latency of 33 ms.
Cloud cluster runs a full-fledged Kubernetes distribution using rke2~\cite{rke2} while the edge cluster emulates resource-constrained edge environments with K3d \cite{k3d}, a lightweight Kubernetes distribution, in Docker containers. We configure the cloud Kubernetes with unlimited scaling, while the edge k3d consists of 8 K3d clusters, each of which has access to 8 vCPUs and 16 GB of memory, plus one machine acts as an IoT gateway, generating synthetic data and invocation requests targeting services deployed at both the edge and cloud levels.
All machines in the setup communicate via Zenoh, a low-latency, publish/subscribe and query-based protocol that enables efficient coordination across the Edge–Cloud continuum. To simulate real-world network disruptions, we use Chaos Mesh \cite{chaosmesh2025} to inject intermittent connectivity faults. We install \name \footnote{The source code is available at \url{https://github.com/hpcclab/OaaS-IoT}.} alongside other baselines across both cloud and edge clusters to help deploy IoT applications. \name uses OpenRaft~\cite{openraft} to implement the Raft protocol while its ODGM maintains application state inside its embedded in-memory storage.

\subsection{Comprehensiveness and Productivity}
\label{sec:evaluation-results-case-study}
We show how \name improves the application development and deployment productivity via case studies and human evaluations.

\begin{figure}[ht]
    \centering
    \includegraphics[width=0.75\linewidth]{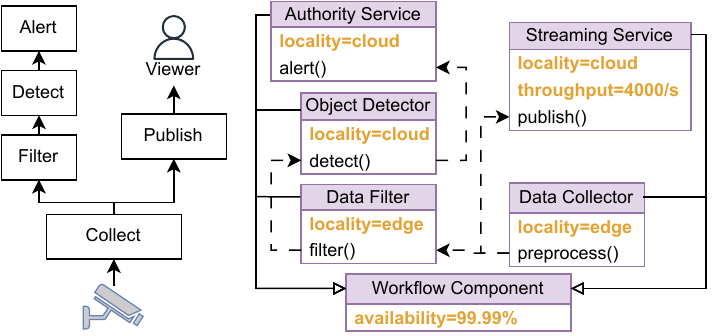}
    \caption{Real-time video analysis case study and the corresponding \name Implementation}
    \label{fig:evaluation-case-study-streaming}
\end{figure}

\paragraph{\textbf{Real-time Video Analysis (VA).}}
We begin with a representative use case in stream processing—a core building block of many modern IoT applications \cite{sasaki2021survey}. Figure~\ref{fig:evaluation-case-study-streaming} illustrates a real-time video analysis workflow \cite{nguyen2019real} and its implementation using \name. The application \textit{collects} live video from CCTV cameras, streams it to a traffic control center for monitoring purposes, and simultaneously runs an object detection pipeline to detect suspicious activity (e.g., overspeeding). The pipeline includes a \textit{filter} to extract relevant frames for an analyzer to \textit{detect} suspicious actions and possibly issue \textit{alert}.

On the right-hand side of Figure~\ref{fig:evaluation-case-study-streaming} is how this workflow maps naturally into the object abstraction. Each processing component is implemented as an object, with functions encapsulating logic and attributes maintaining internal state. Component interactions are expressed through function chaining, easily configured using \name’s event-trigger API (Table~\ref{tab:oaas-iot-api}). \name further simplifies QoS configuration through high-level constructs. Developers can define an abstract class (e.g., \texttt{Workflow Component}) with a shared availability SLA, and extend it for individual modules, enabling SLA inheritance without duplication. Components with intensive computation demands, such as the object detector, can specify \texttt{locality=cloud} to prefer cloud placement. For video streaming, stable throughput is critical and can be declared with an SLA: \texttt{throughput=4,000} invocations/sec with each invocation processes one second of video (i.e., 30 frames) for a camera. This enables smooth playback for up to 4000 concurrent connected devices without manual tuning.

% Crucially, all configurations (functionality and QoS) are expressed solely through the object abstraction, aligned with the application's logical structure. Developers are not required to understand or manipulate system-level details. To quantitatively evaluate the improvement, we implement the workflow using \name and a pure FaaS baseline (Knative) in Python.  
% While we acknowledge that Lines of Code (LoC) is not a comprehensive measure of productivity, it serves as a practical and objective indicator of the verbosity and integration overhead imposed on a developer. In this context, a significant reduction in the code required to achieve the same functionality points to a more efficient and less error-prone development process.
% \name implementations required just 256 lines of code (LoC), a 26.2\% reduction compared to 347 LoC with Knative. More impressively, if we count only the extra integration code introduced by each approach by excluding the common application logic implementation (107 LoC), \name reduces development effort by up to 37.9\%.
% This illustrates the power of \name’s comprehensive abstraction: it unifies logic, state, and deployment concerns, making IoT application development intuitive, concise, and ultimately, productive.

\begin{figure*}[t]
    \centering
    \subfloat[FaaS-based solutions using AWS and Azure services \cite{inventory-managerment-azure}]{
        \includegraphics[width=0.6\linewidth]{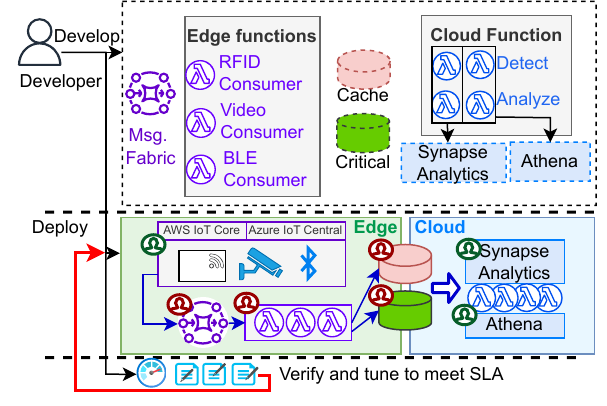}
        \label{fig:evaluation-case-study-faas}
    }
    \hfill    
    \subfloat[Using \name]{
        \includegraphics[width=0.65\linewidth]{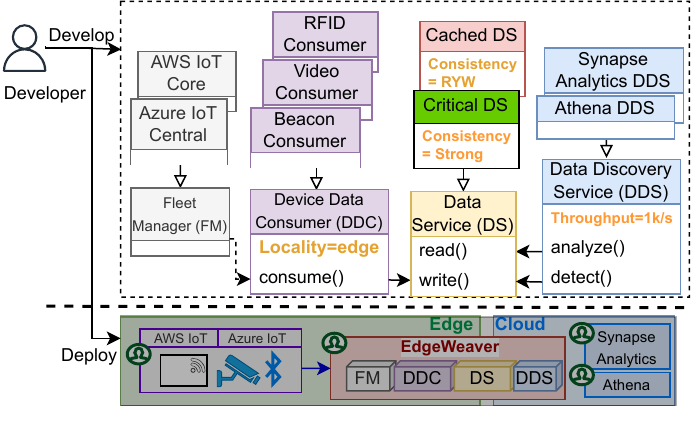}
        \label{fig:evaluation-case-study-oaas-iot}
    }
    \caption{Developing and deploying a real-time inventory management system with FaaS and \name. head and should icons depict system components the developer has to interact during the application life cycle (green: high-level interaction, red: direct configuration). \name needs fewer component interactions and doesn't require the \textbf{verify and tune} loop to meet desired SLAs.
    % highlights the manual integration and \textbf{verify and tune} loop required by the FaaS approach, which is eliminated by \ names's automated, declarative abstractions.
    }
    \label{fig:evaluation-case-study}
\end{figure*}

\paragraph{\textbf{Real-time Inventory Management (IM).}}
Figure~\ref{fig:evaluation-case-study} shows how developers could use traditional FaaS and \name to develop and deploy a Real-time Inventory Management (IM) system, a common workflow pattern of modern IoT applications.
The FaaS-based architecture, recommended by Azure IoT~\cite{inventory-managerment-azure} and AWS IoT~\cite{dalba2023reference}, ingests heterogeneous data streams (e.g., RFID tags, beacons, video) from devices registered through Azure IoT Central. Each device type requires a dedicated FaaS function tailored to its protocol (e.g., \texttt{RFIDConsumer} over MQTT, \texttt{VideoConsumer} over TCP). Processed data are stored in a fast \textbf{cache} and later queried by analytics functions (e.g., Analyze) via services like \textbf{Azure Synapse Analytics} for analysis, and results are persisted in a \textbf{critical} database. 

This FaaS-based approach is \textit{complex} and \textit{fragmented}. Developers must manage numerous protocol-specific functions, analytics modules, and data stores, while integrating multiple services (e.g., AWS IoT Core, Azure IoT Central) for cross-platform support. It also lacks native QoS enforcement, forcing a manual verify-and-tune cycle of adjusting function placement, resource allocation, and network settings to meet SLAs.

In contrast, \name \textit{unifies} and \textit{automates} this entire process (Figure~\ref{fig:evaluation-case-study-oaas-iot}). Developers work with high-level object abstractions instead of low-level components. Specialized handlers (e.g., \texttt{RFIDConsumer}) are derived from a reusable Device Data Consumers (DDC) class; cloud-specific integrations are encapsulated within a polymorphic Fleet Manager (FM) class; and data management is streamlined via unified Data Service (DS) and Data Discovery Service (DDS) interfaces. SLAs are attached declaratively (e.g., \texttt{Locality=edge} for latency-sensitive tasks, \texttt{Consistency=strong} for critical data), eliminating the need for manual tuning.

Guided by these declarative policies, the \name runtime automatically handles provisioning, placement, and networking, automatically deploying components like FM and DDC at the edge to meet locality requirements. This full-stack automation removes the manual, error-prone configuration cycle inherent in traditional FaaS systems, enabling faster, more reliable, and maintainable IoT deployments.

\paragraph{\textbf{Productivity Improvement.}}
To quantitatively assess EdgeWeaver’s productivity gains over FaaS approach, we implemented two prototypes of the inventory management application: one using Knative (FaaS-based) and one with \name. Development effort was measured using three metrics: Lines of Code (LoC), Lines of Configuration Code (LoCC), and the number of developer-facing interfaces.

The results show a dramatic reduction in development overhead with \name. The Knative implementation required 666 LoC, while \name achieved equivalent functionality in only 363 LoC (44.5\% reduction). The improvement in configuration effort was even more pronounced: \name needed just 39 LoCC, nearly $10\times$ fewer than Knative (417 LoCC), which involves configuring multiple external services such as RabbitMQ, databases, and triggers. This highlights \namens's strength in abstracting complex infrastructure management.

\begin{figure}[ht]
    \centering
    \includegraphics[width=0.85\linewidth]{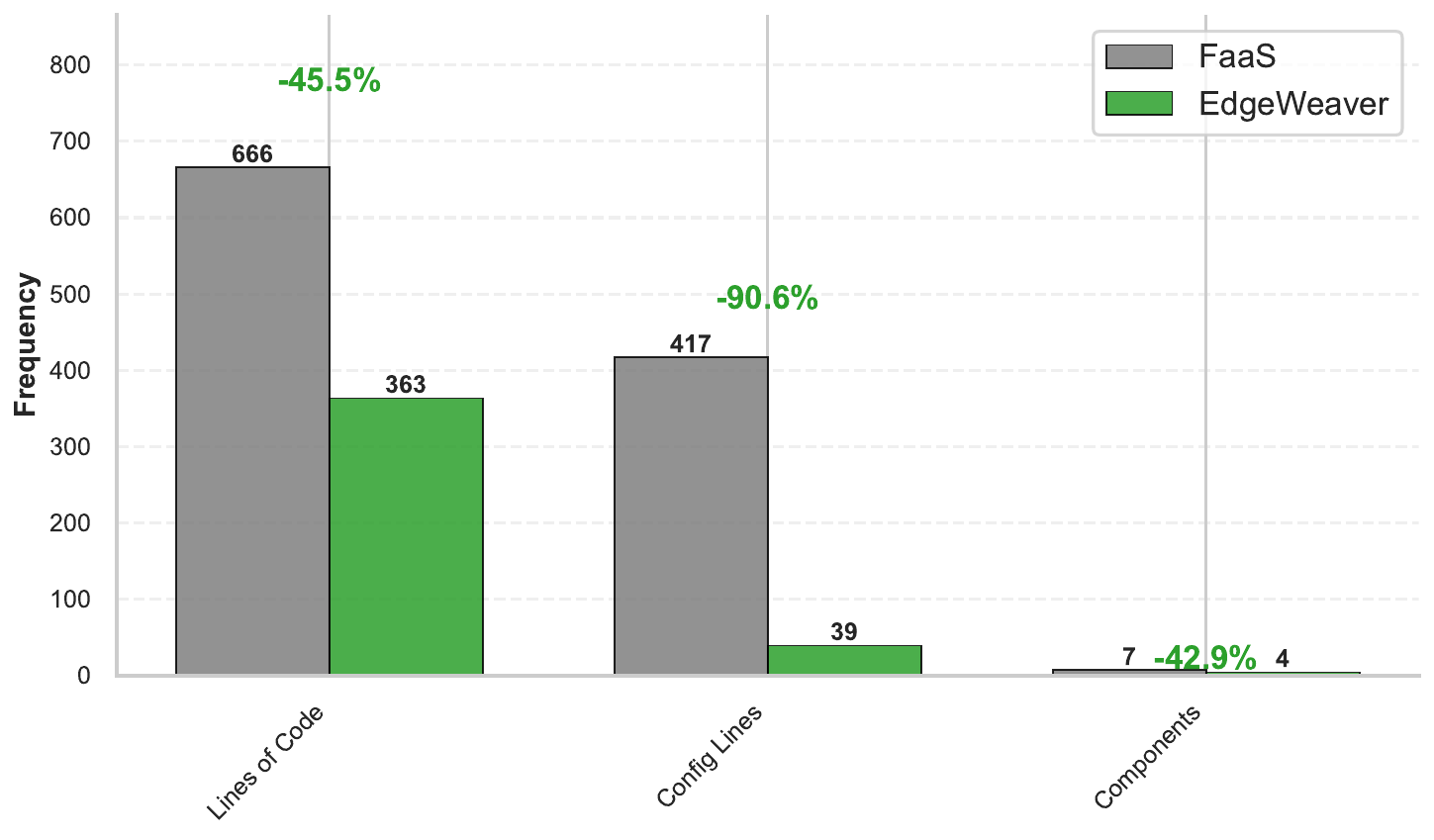}
    \caption{EdgeWeaver productivity improvements in the inventory management case study}
    \label{fig:productivity-gains}
\end{figure}

Furthermore, as shown in Figures~\ref{fig:evaluation-case-study} and \ref{fig:productivity-gains}, the FaaS-based design forces developers to manage at least seven distinct components, whereas \name consolidates these into only four. Together, these results demonstrate that \namens's unified, declarative interface and automated orchestration significantly reduce development complexity—making it far more productive, maintainable, and developer-friendly than traditional FaaS-based solutions.

\begin{table}[h]
  \centering
  \small
  \setlength{\tabcolsep}{5pt}
  \begin{minipage}[t]{0.56\linewidth}
    \centering
    \hspace*{-8mm}
    \begin{tabular}{|l|c|c|}
      \hline
      \textbf{Cloud fam.} & \textbf{EW} & \textbf{FaaS} \\ 
      \hline
      \hline
      Unfamiliar & 84.6\% & 81.5\% \\
      Basic      & 90.0\% & 80.0\% \\
      Competent  & 92.0\% & 88.0\% \\
      \hline
    \end{tabular}
  \end{minipage}
  \hspace{-0.06\linewidth}
  \begin{minipage}[t]{0.43\linewidth}
    \centering
    \begin{tabular}{|l|c|c|}
      \hline
      \textbf{Metrics} & \textbf{EW} & \textbf{FaaS} \\
      \hline
      \hline
      Time (min.)      & 22.43 & 32.40 \\
      Score (\%)  & 52.85 & 53.55\\
      \hline
    \end{tabular}
  \end{minipage}
  \caption{Human Study results (average): Quiz (left) and Programming (right). (EW=\name)}
  \label{tab:quiz_results}
\end{table}

\paragraph{\textbf{Developer Experience.}}
To evaluate how \namens's comprehensiveness and productivity translate into better developer experience, we conducted a human study with 39 college students. Participants received short (15 min.) tutorials on both the FaaS and \name abstractions, followed by a quiz to assess conceptual understanding and a programming assignment to measure practical performance.
As shown in Table~\ref{tab:quiz_results} (left), participants scored consistently higher on \name-related quiz questions than on FaaS ones, regardless of prior cloud experience. This indicates that \name is more intuitive and easier to learn. For the programming task (Table~\ref{tab:quiz_results}, right), out of the group who can complete the task in both platforms in time, students completed the assignment 31\% faster using \name while achieving nearly identical code quality (\name: 53.6\% vs FaaS: 52.9\%). These results demonstrate that \namens’s unified abstractions and automation not only simplify development but also deliver a measurably better, faster, and more accessible.

% The results (Table~\ref{tab:quiz_results}) showed strong comprehension overall. Participants scored  higher on \name than FaaS, regardless of prior cloud experience, suggesting \name was slightly more intuitive.
% All participants performed a programming task; 12 completed at least one paradigm, and 7 completed both. Completion time was the main productivity metric. Among those who finished both, \name took only 68\% of the time required for FaaS, with comparable scores (\name 53.6\% vs. FaaS 52.9\%). To reduce order bias, we separately averaged groups by starting paradigm; this balanced analysis showed \name at 74\% of FaaS time, indicating a remarkable productivity improvement. 

% All participants then proceeded to the programming task; however, due to time constraints, only 12 were able to complete it for at least one paradigm. Of those students, 7 completed the design challenge using both paradigms. We measured the time required to complete the design for each paradigm as a primary indicator of productivity. For the 7 participants who completed both designs, the average completion time for the \name paradigm was 68\% of the time required for the FaaS paradigm. To account for potential learning effects from the task order, we normalized the data by separately averaging the results from the group that started with FaaS and the group that began with \name. This balanced analysis revealed that the \name design consumed 74\% of the completion time of the FaaS design, indicating a significant productivity improvement.

% \color{black}

\vspace{2mm}
\noindent
\colorbox{blue!10}{
\parbox{0.96\linewidth}{
\underline{\textbf{Takeaway}:} \emph{
\name provides a unified, comprehensive abstraction that streamlines development and deployment of IoT applications across the Edge–Cloud continuum.
}}}
\vspace{1mm}

\begin{figure}[t]
    \centering
    \subfloat[Availability = 4 nines]{\includegraphics[width=0.46\linewidth]{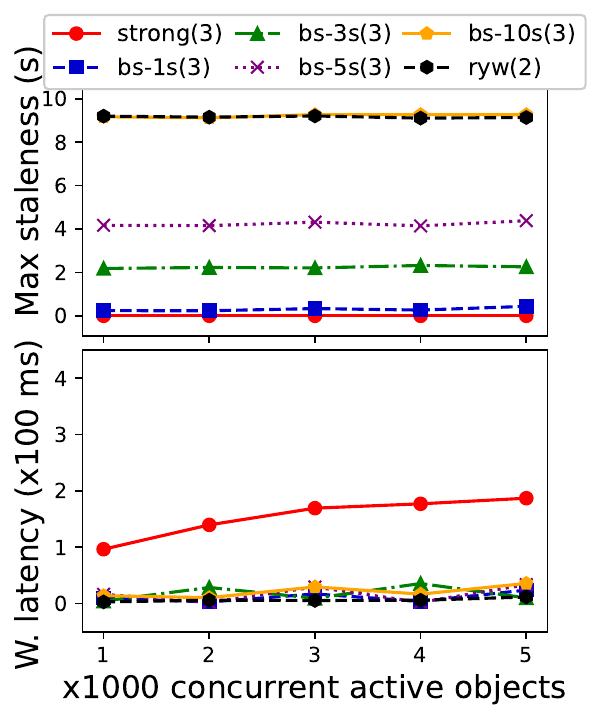}\label{fig:evlt:consistency-4}}
    % \hfill
    \subfloat[Availability = 9 nines]{\includegraphics[width=0.44\linewidth]{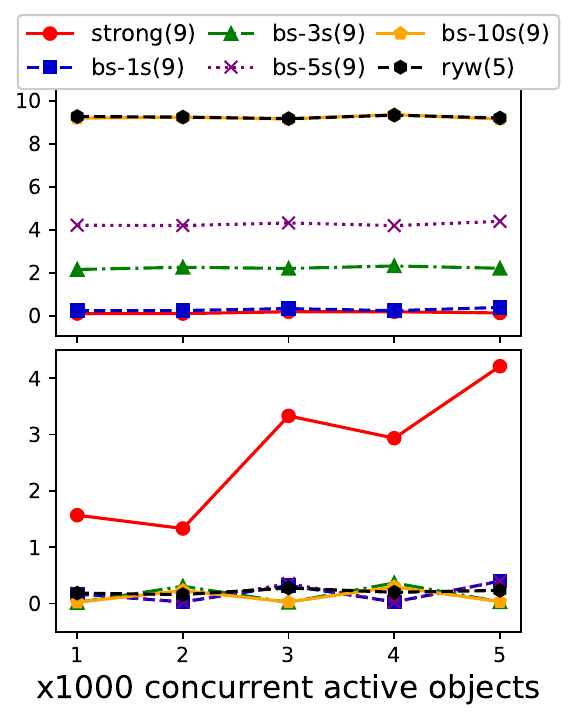}}\label{fig:evlt:consistency-9}
    \caption{Maximum read staleness (top) and average write latency (bottom) under different consistency–availability configurations. Numbers in parentheses show the replica count required to meet both guarantees.}
    \label{fig:evlt:consistency}
\end{figure}

\subsection{Applicability and Efficiency}
\label{sec:evaluation-results-applicability}
We evaluate applicability of \name by demonstrating its enforcement of diverse SLA combinations through case studies. We assess its efficiency in handling applications with different computational demands, showing its suitability for diverse IoT scenarios.

\paragraph{\textbf{SLA Enforcement.}}
We evaluate the ability to enforce various consistency levels: Read-Your-Write (\textit{ryw}), \textit{strong}, and bounded staleness (\textit{bs}) under varying staleness bounds and high availability targets. According to Fig.~\ref{fig:evlt:consistency}, we deploy multiple concurrent \texttt{DataService} objects (from the Inventory case study), each issuing reads and writes. We set availability to 99.99\%, comparable to leading FaaS SLAs (e.g., AWS Lambda’s 99.95\% \cite{aws-lambda-sla}) and Tier-4 datacenters (99.995\% \cite{hpe-cloud-tiers}).

Across all configurations, \name consistently enforces the specified consistency guarantees: Under bounded staleness, observed staleness remains well below set thresholds. Even when stateless is relaxed in RYW, the stateless is consistently below 10 seconds. Notably, under strong consistency, zero staleness is detected, validating \textit{reliable} consistency enforcement of \name.
These guarantees hold at scale: With 1,000 concurrent objects, strong consistency maintains an average write latency of 100 ms, which only increases by $1.87\times$ when scaling to 5,000 objects. RYW and bounded staleness achieve significantly lower latency ($<$ 20 ms), over $9\times$ faster than strong consistency, highlighting promising performance–consistency trade-offs that developers can leverage for various QoS needs.

To test robustness, we increase the SLA target to nine nines (i.e., 99.999999999\%)---five orders of magnitude higher than the current standard. \name continues to satisfy all consistency requirements: bounded staleness remains under 10s, and strong consistency still achieves zero staleness. The write latency for strong consistency increases by at most $2\times$, while weaker models show a negligible impact. Finally, \name achieves these guarantees cost-effectively. At four nines, enforcing availability requires just 2–3 replicas. Even at nine nines, the system needs no more than nine replicas, a $3\times$ cost increase for $10,000\times$ higher reliability.

% \color{blue}

\paragraph{\textbf{Implementation Efficiency.}}
We evaluate the implementation overhead introduced by \namens’s object abstraction and SLA enforcement by comparing it against equivalent FaaS-based implementations. Since \name builds on standard FaaS engines and employs Pub/Sub protocols for communication, we benchmark it using combinations of Knative~\cite{knative} and Fission~\cite{fission} with MQTT~\cite{rabbitmq} and Zenoh~\cite{corsaro2023zenoh}.

\begin{figure}[t]
    \centering
    \includegraphics[width=0.85\linewidth]{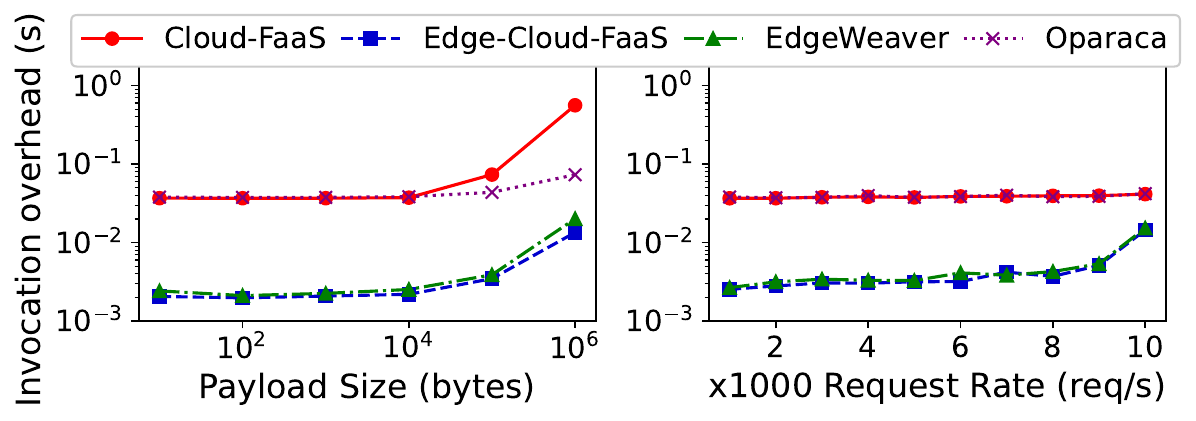}
    \caption{Latency of function invocation with increasing request rate and payload size in various Edge-Cloud deployment options.}
    \label{fig:evlt:efficiency-edge}
\end{figure}

We use a lightweight \texttt{echo} function that returns immediately upon invocation. This allows us to measure the combined overhead of network latency and platform runtime, independent of application logic.
First, we evaluate the ability to minimize invocation overhead in the Edge-Cloud setting.
Figure~\ref{fig:evlt:efficiency-edge} shows the invocation overhead across four Edge-Cloud deployment options varying the request rate and payload. Note that this experiment uses a different set of baselines: Cloud-FaaS (Knative deployed in cloud only), Edge-Cloud-FaaS (Knative with manual edge placement), \name (EdgeWeaver with \textit{Locality=edge}), and Oparaca (cloud-only OaaS deployment). For \name, we deploy the function with \textit{Locality=edge}, forcing it to dispatch invocations at the closest edge to the IoT gateway. For Edge-Cloud-FaaS, we mimic this enforcement by manually rerouting invocation requests to the edge only.

Both \name and Edge-Cloud-FaaS outperform cloud-only baselines by at least 6.9$\times$, confirming the benefits of edge deployment. In certain configurations, Edge-Cloud-FaaS can outperform \name by up to 6\% due to its direct invocation path versus \name's SLA-driven routing. However, this minor difference is outweighed by \name automation. Unlike Edge-Cloud-FaaS, which requires manual tuning for optimal performance, \name automatically manages SLA-based placement to deliver comparable performance. This confirms that \name delivers a significant productivity boost without sacrificing performance or requiring additional developer effort.

\begin{figure}[t]
    \centering
    \includegraphics[width=1\linewidth]{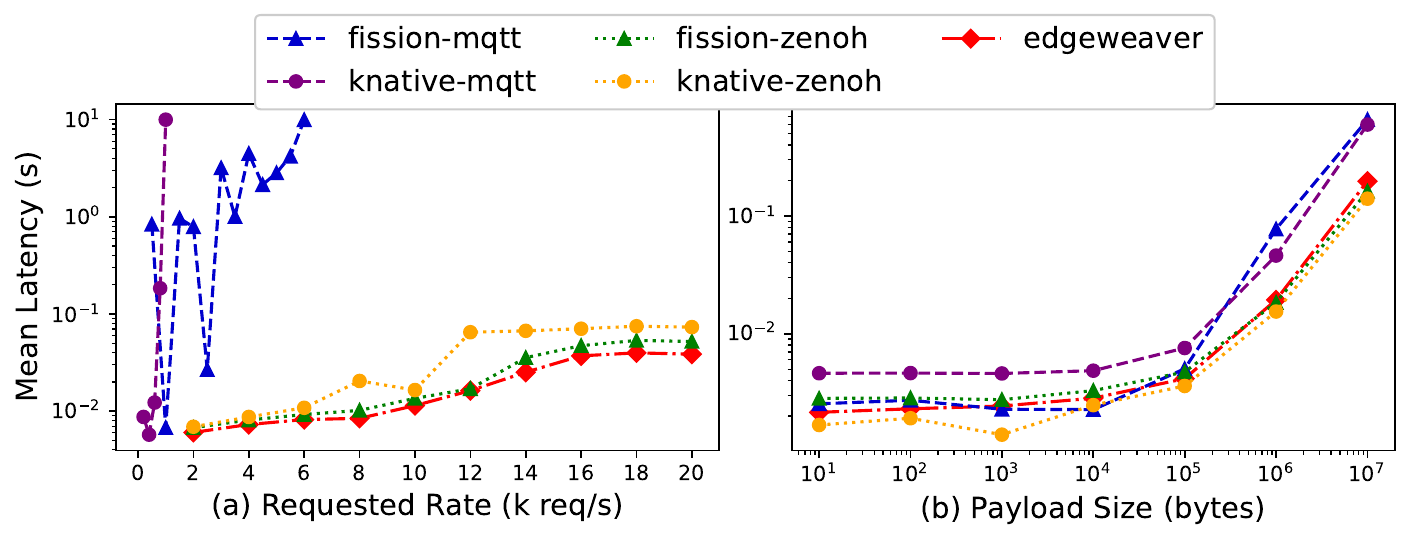}
    \caption{Latency of function invocation with increasing request rate and payload size in various baseline systems.}
    \label{ch5:fig:evlt:efficiency}
\end{figure}

More broadly, we compare the latency across different FaaS platforms and transport protocols.
Figure~\ref{ch5:fig:evlt:efficiency} shows the results across various baseline systems. Although \name introduces additional mechanisms for object abstraction and SLA enforcement, this overhead is negligible. Across different request rates and payload sizes, \name consistently delivers performance on par with or better than the baselines, sometimes even outperforming them (e.g., Knative-MQTT). These results confirm that \namens's adaptable runtime realizes its object abstraction and declarative SLA enforcement without compromising its performance, providing strong evidence of implementation efficiency in practice.

\begin{figure}[t]
    \centering
    \includegraphics[width=0.85\linewidth]{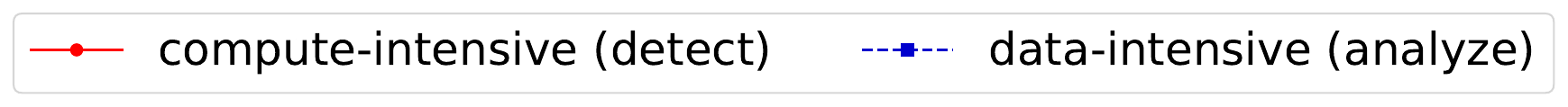}
    \subfloat[Edge-Cloud]{\includegraphics[width=0.28\linewidth]{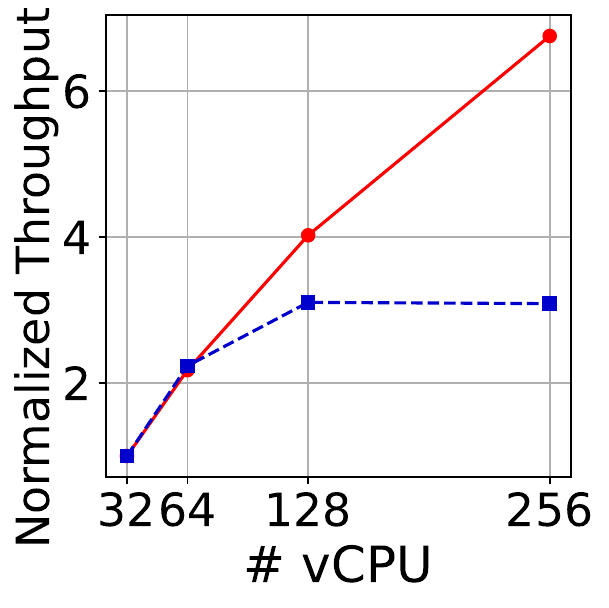}\label{fig:evlt:scale-ec}}
    \hfill
    \subfloat[Cloud]{\includegraphics[width=0.42\linewidth]{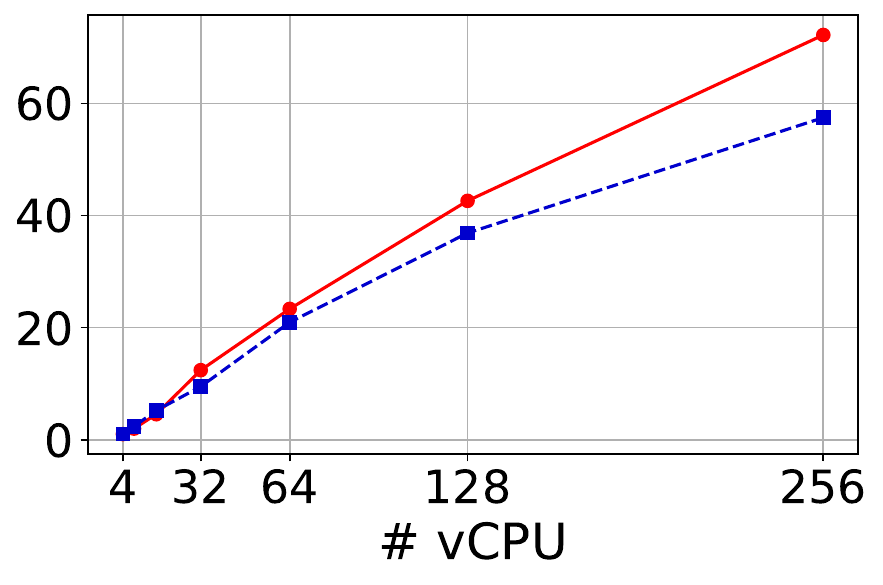}\label{fig:evlt:scale-cloud}} 
    \hfill
    \subfloat[Edge]{\includegraphics[width=0.28\linewidth]{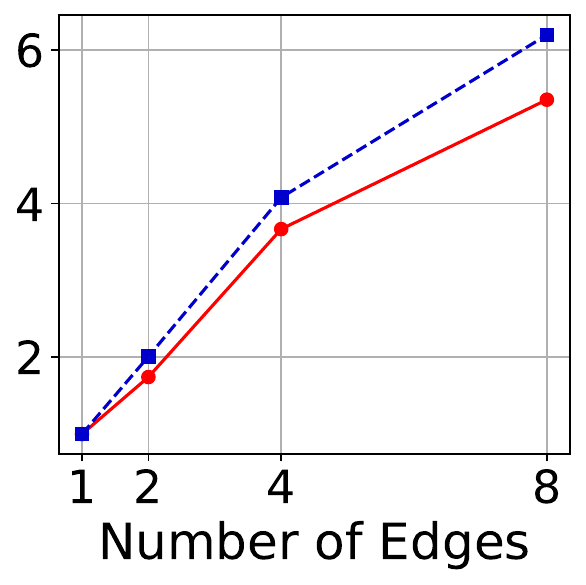}\label{fig:evlt:scale-edge}}
    \caption{Scalability analysis of \name across different deployment configurations. Throughput is normalized to the baseline allocation and measured for two workloads: data-intensive JSON processing (\textit{analyze}) and compute-intensive object detection (\textit{detect}).}
    \label{ch5:fig:evlt:scalability}
\end{figure}

\paragraph{\textbf{Scalability.}}
We evaluate the scalability of \name by examining whether its implementation efficiency, observed in earlier experiments, holds under workload and resource scaling. To reflect realistic IoT usage, we consider two representative workloads: JSON document processing as a \textit{data-intensive} task and image object detection using the YOLO model as a \textit{compute-intensive} task. Both are implemented as the \textit{analyze} and \textit{detect} functions of the \textit{Data Discovery Service} (DDS) in the inventory management case study (Figure \ref{fig:evaluation-case-study}). For each workload, we deploy multiple object instances across edge and cloud nodes, allowing \name to automatically determine their placement without explicit Locality constraints. Each instance is driven by a dedicated load generator that repeatedly invokes its corresponding function.
% function: \textit{analyze} (processing JSON documents) for DDS and \textit{detect} (video object detection) for Object Detector.

Figure~\ref{fig:evlt:scale-ec} shows the throughput as we scale the total number of vCPUs across edge and cloud--from 8 vCPUs at the edge and 24 in the cloud, doubling the capacity incrementally. Throughput is normalized to the lowest allocation. Both workloads show strong scalability: throughput increases nearly linearly up to 128 vCPUs for \textit{analyze} and 256 for \textit{detect}. Beyond those points, performance plateaus due to network saturation between the TACC and UC data centers.
To isolate the network impact, we rerun the experiments independently within each site. Figure~\ref{fig:evlt:scale-cloud} presents the cloud-only results. The \textit{detect} workload, being compute-intensive, scales nearly linearly—achieving a $70\times$ throughput gain from 4 to 256 vCPUs, peaking at 144 invocations/sec. \textit{analyze}, which is more data- and I/O-intensive, scales more moderately, reaching approximately 200,000 invocations/sec at 256 vCPUs.
We observe similar patterns for edge-only (Fig.~\ref{fig:evlt:scale-edge}):  throughput scales proportionally with the number of edges (8 vCPU per edge), confirming \name sustains high throughput and scalability across diverse workloads and deployments.

\vspace{1mm}
\noindent
\colorbox{blue!10}{
\parbox{0.96\linewidth}{
\underline{\textbf{Takeaway}:} \emph{
\name demonstrates strong applicability by providing efficient implementation to reliably enforce diverse SLAs and scale efficiently across Edge–Cloud.
}}}
\vspace{1mm}

\begin{figure} [t]
    \centering
    \includegraphics[width=0.8\linewidth]{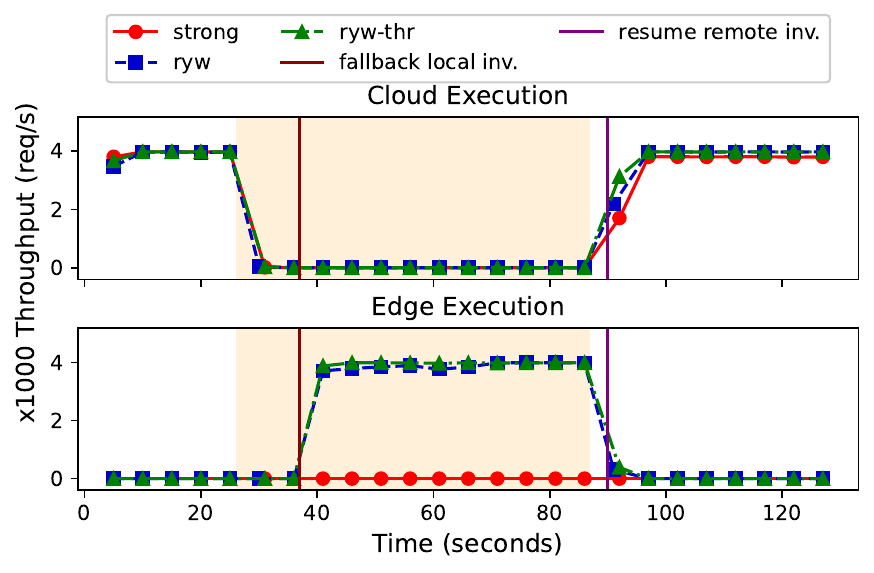}
    \caption{Impact of network partitioning for classes with: RYW, RYW with throughput, and Strong Consistency.}
    \label{fig:evlt:net-partition}
\end{figure}

\subsection{Adaptability}
\label{sec:evaluation-results-adaptability}

We evaluate the adaptability of \name by testing its ability to maintain QoS under dynamic network conditions. We deploy three functions with different SLA configurations: (i) \texttt{consume} (from the Device Data Consumer) with \textit{Read-Your-Write} consistency (\texttt{ryw}), (ii) \texttt{write} (from the Data Service) with \textit{RYW + throughput} guarantee (\texttt{ryw-thr}, 4,000 RPS), and (iii) \texttt{detect} (from the Data Discovery Service) with \textit{strong} consistency. We assign high-availability SLAs, prompting \name to place replicas across edge-cloud. We emulate \textit{network partitioning} period between cloud and edge using Chaos Mesh~\cite{chaosmesh2025} (yellow area in Fig.~\ref{fig:evlt:net-partition}), where injected faults disrupt connectivity and trigger \namens's runtime adaptation.

Initially, all functions continuously issue 4,000 RPS write requests to their associated data services. During the partition, \name detects the disruption and redirects invocations to the edge whenever possible.
For the \textit{ryw}, it permits continued execution by relaxing consistency, but since the underlying Logic engine is not inherently prepared for this scenario, its throughput is unstable. In contrast, the \textit{ryw-thr} function benefits from the throughput SLA, maintaining stable performance. For strong consistency, \name enforces quorum strictly; as consensus cannot be achieved across the partition, throughput drops to zero, preserving correctness.
Upon network restoration, \textit{ryw-thr} quickly recovers full throughput, while \textit{ryw} experiences a brief delay due to reactive scaling. The results show \name fine-grained adaptability, enabling  dynamic balancing of QoS desires in response to changes.

\vspace{2mm}
\noindent
\colorbox{blue!10}{
\parbox{0.96\linewidth}{
\underline{\textbf{Takeaway}:} \emph{
SLA-driven deployment enables \name to adapt automatically to dynamic environments to consistently meet application needs.
}}}
\vspace{2mm}

\section{Summary}

In this chapter, we presented OaaS-IoT, an extension of the Object as a Service paradigm to the Edge-Cloud continuum, and its realization through the \name platform. Inspired by OOP principles, OaaS-IoT offers object abstraction that encapsulates application state, functions (logics), and SLAs, thereby providing a holistic view across the continuum. \name implements this paradigm by transparently handling user-defined consistency and availability trade-offs in the presence of network failure. Importantly, the benefits of OaaS-IoT as realized by \name do not come with any significant overhead to the system.

With the OaaS paradigm implemented for various contexts and use cases, in the next chapter, we explore the challenges and gaps to productize OaaS as a new cloud-native application development paradigm.
%%%%%%%%%%%%%%%%%%%%%%%%%%%%%%%%%%%%%%
\chapter{Exploration and Analysis of the OaaS Productization}
\label{chapter6}
%%%%%%%%%%%%%%%%%%%%%%%%%%%%%%%%%%%%%%

\section{Overview}
\label{sec:ch6-overview}

The proliferation of cloud, edge, and IoT computing has created a significant commercial opportunity born from a critical market pain point: profound infrastructural complexity. While Chapters~\ref{chapter3}--\ref{chapter5} presented the technical foundations of Object-as-a-Service (OaaS), understanding the real-world market need and commercial viability requires systematic customer discovery. This complexity imposes a substantial tax on developer productivity, inflates operational costs, and acts as a barrier to innovation---representing a multi-billion-dollar market gap that OaaS is positioned to address.

This chapter presents commercialization validation from the NSF I-Corps National program (Summer 2025 Cohort 3), where we conducted 101 interviews across 86 organizations. Our customer discovery quantitatively confirmed a significant gap between modern infrastructure capabilities and teams' practical ability to utilize them effectively, validating the commercial opportunity OaaS addresses.

Section~\ref{sec:ch6-methodology} describes our customer discovery approach and participant demographics. Section~\ref{sec:ch6-findings} presents quantitative validation of market pain points and expectations, with deployment complexity (38.6\%), onboarding difficulty (35.6\%), and system complexity (24.8\%) as the top challenges. Section~\ref{sec:ch6-value-prop} examines how OaaS technical capabilities directly address these validated pain points. Section~\ref{sec:ch6-implications} identifies production readiness requirements and validates three target market segments, with technology SMEs and startups as the primary early-stage focus given their lower compliance barriers, while enterprise accounts require addressing production gaps first. Section~\ref{sec:ch6-business-model} presents the Business Model Canvas synthesizing customer insights into a staged commercialization strategy prioritizing SMEs and research institutions before enterprise expansion. Section~\ref{sec:ch6-summary} summarizes the chapter's contributions.

\section{Methodology}
\label{sec:ch6-methodology}

\subsection{Customer Discovery Approach}

We adopted the evidence-based customer discovery framework prescribed by the NSF I-Corps program \cite{blank2020four}, which emphasizes getting out of the building to test hypotheses about customer problems, needs, and willingness to pay. Our approach consisted of semi-structured interviews designed to explore current workflows, pain points, attempted solutions, and desired outcomes rather than pitching our technology.

\subsection{Interview Protocol}

Each interview followed a consistent protocol:
\begin{itemize}
    \item \textbf{Duration}: 30+ minutes per interview
    \item \textbf{Format}: Semi-structured conversations via video conference or phone or in-person
    \item \textbf{Focus Areas}: Conversations explored current cloud/edge-cloud development workflows, deployment practices, primary challenges (technical and organizational), performance and cost concerns, prior attempted solutions, expectations for improvement, and potential commercialization opportunities including preferred deployment models and pricing sensitivities.
    \item \textbf{Data capture}: For each interview, we systematically recorded participant role, company industry and size, and coded thematic data on specific pain points and desired outcomes. Any quantitative metrics mentioned by participants regarding time savings or performance improvements were also explicitly captured.
\end{itemize}

\subsection{Participant Demographics}

Over the course of the I-Corps program, we conducted 101 interviews that 86 unique organizations (some organizations had multiple interviewees).

\begin{figure}[t]
    \centering
    % Source: report.md Section 3c
    % Shows: Developer/Engineer (29), Executive Leadership (24), DevOps/Infra (19), 
    %        Research/Academia (17), Data/ML (6), Entrepreneur (3), Security (3), Consulting (2)
    \includegraphics[width=0.85\textwidth]{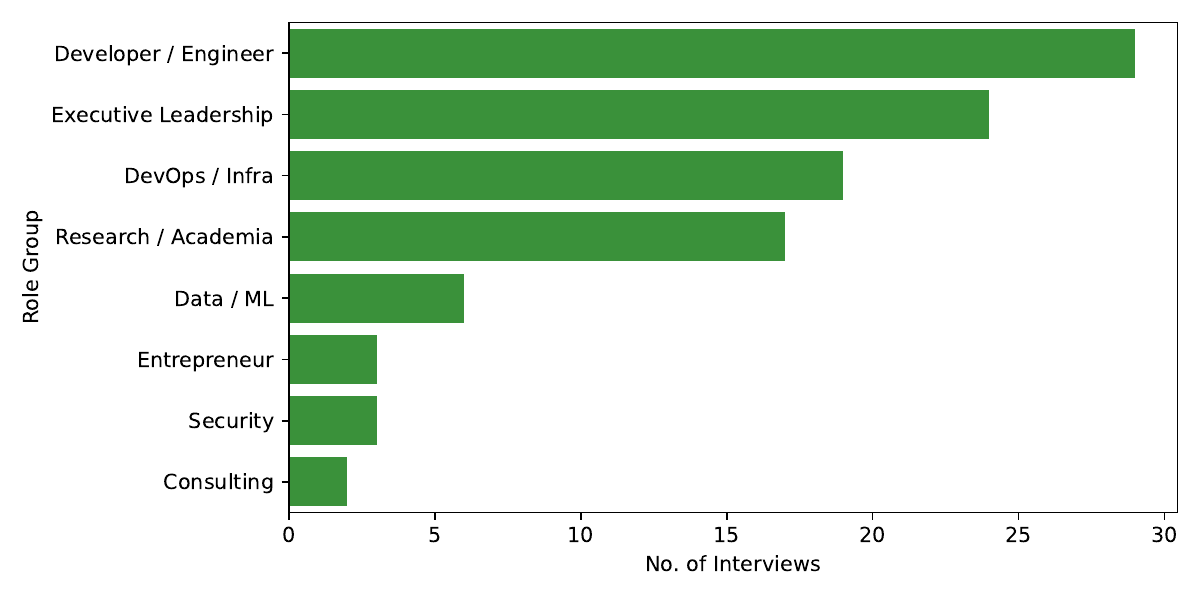}
    \caption{Distribution of interview participants by role group (N=101 interviews). Developer/Engineer and DevOps/Infra roles comprised nearly half of all interviews, reflecting our focus on technical practitioners directly involved in cloud-native development and deployment.}
    \label{fig:ch6-role-distribution}
\end{figure}

\textbf{Role Distribution}: As shown in Figure~\ref{fig:ch6-role-distribution}, participants spanned diverse technical and leadership roles. Table~\ref{tab:ch6-role-distribution} summarizes the distribution of interview participants by role category.

\begin{table}[ht]
    \centering
    \small
    \begin{tabular}{|p{4cm}|c|p{7.5cm}|}
        \hline
        \textbf{Role Category} & \textbf{\#Interviews} & \textbf{Specific Roles} \\ \hline \hline
        Developer / Engineer & 29 & Software Engineers, Full-stack Developers, Product Engineers, QA Engineers \\ \hline
        Executive Leadership & 24 & Managers, Founders, CEOs, CTOs, VPs of Engineering, Product Managers \\ \hline
        DevOps / Infrastructure & 19 & DevOps Engineers, Infrastructure Engineers, SREs, IT Administrators \\ \hline
        Research / Academia & 17 & Professors, Researchers, Research Scientists, PhD Students \\ \hline
        Data / ML & 6 & Data Scientists, ML Engineers, Research Scientists \\ \hline
        Security & 3 & Security Engineers, DevSecOps Engineers \\ \hline
        Entrepreneur & 3 & Independent entrepreneurs and startup founders \\ \hline
        Consulting & 2 & Technical consultants \\ \hline
    \end{tabular}
    \caption{Distribution of interview participants by role category (N=101 interviews)}
    \label{tab:ch6-role-distribution}
\end{table}

\begin{figure}[t]
    \centering
    % Source: report.md Section 4a
    % Shows: Education & Academia (20), Software & Technology (14), Financial Services (8),
    %        Consulting & Services (8), Cloud & Infrastructure (7), Public Safety (5), etc.
    \includegraphics[width=0.85\textwidth]{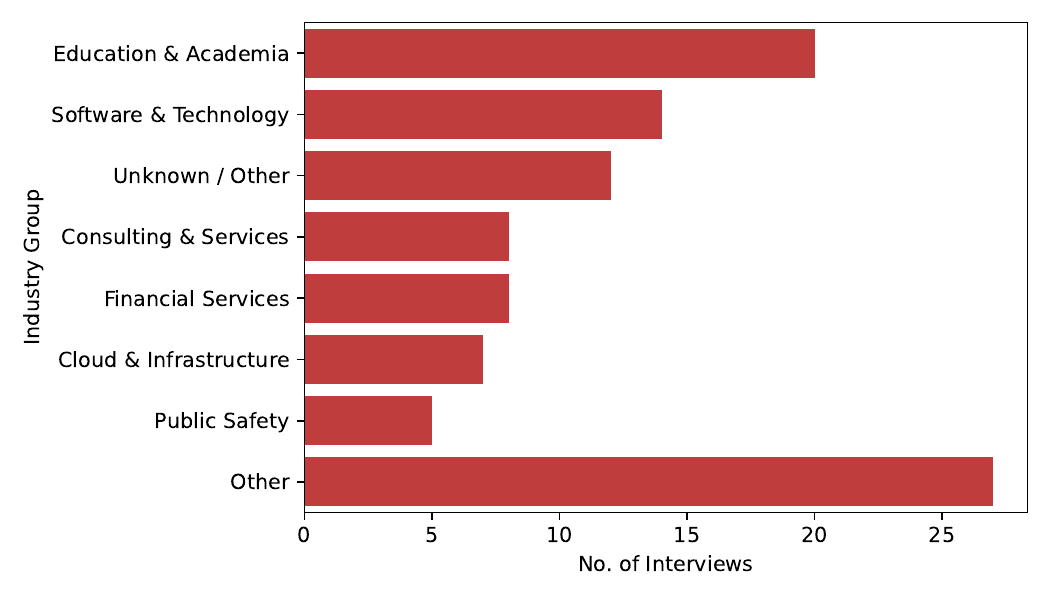}
    \caption{Distribution of interviews by industry group (N=101 interviews across 86 companies). Education, Software \& Technology, and Financial Services represented the largest segments, providing diverse perspectives on cloud-native development challenges.}
    \label{fig:ch6-industry-distribution}
\end{figure}

\textbf{Industry Coverage}: Figure~\ref{fig:ch6-industry-distribution} shows the breadth of industries represented. Table~\ref{tab:ch6-industry-distribution} summarizes the distribution of interviews across industry sectors.

\begin{table}[ht]
    \centering
    \small
    \begin{tabular}{|p{4.5cm}|c|c|p{5.5cm}|}
        \hline
        \textbf{Industry Sector} & \textbf{\#Interviews} & \textbf{\#Companies} & \textbf{Description} \\ \hline \hline
        Education \& Academia & 20 & 13 & Universities and research institutions \\ \hline
        Software \& Technology & 14 & 14 & Software development firms and technology companies \\ \hline
        Financial Services & 8 & 7 & Banking, insurance, and fintech \\ \hline
        Consulting \& Services & 8 & 7 & IT consulting and managed services \\ \hline
        Cloud \& Infrastructure & 7 & 4 & Cloud providers and infrastructure companies \\ \hline
        Public Safety & 5 & 4 & Public safety technology and emergency services \\ \hline
        Healthcare \& Life Sciences & 4 & 4 & Digital health and pharmaceutical companies \\ \hline
        Research \& HPC & 4 & 4 & High-performance computing and research facilities \\ \hline
        Other Industries & 12 & 12 & Logistics \& Transportation, Marketing \& Advertising, Cybersecurity, Retail, Telecommunications, and more \\ \hline
    \end{tabular}
    \caption{Distribution of interviews by industry sector (N=101 interviews across 86 companies)}
    \label{tab:ch6-industry-distribution}
\end{table}

\begin{figure}[t]
    \centering
    % Source: report.md Section 4c
    % Shows: 10,001+ (38), 11-50 (21), 51-200 (17), 1-10 (12), 501-1000 (8), 1001-5000 (3)
    \includegraphics[width=0.9\textwidth]{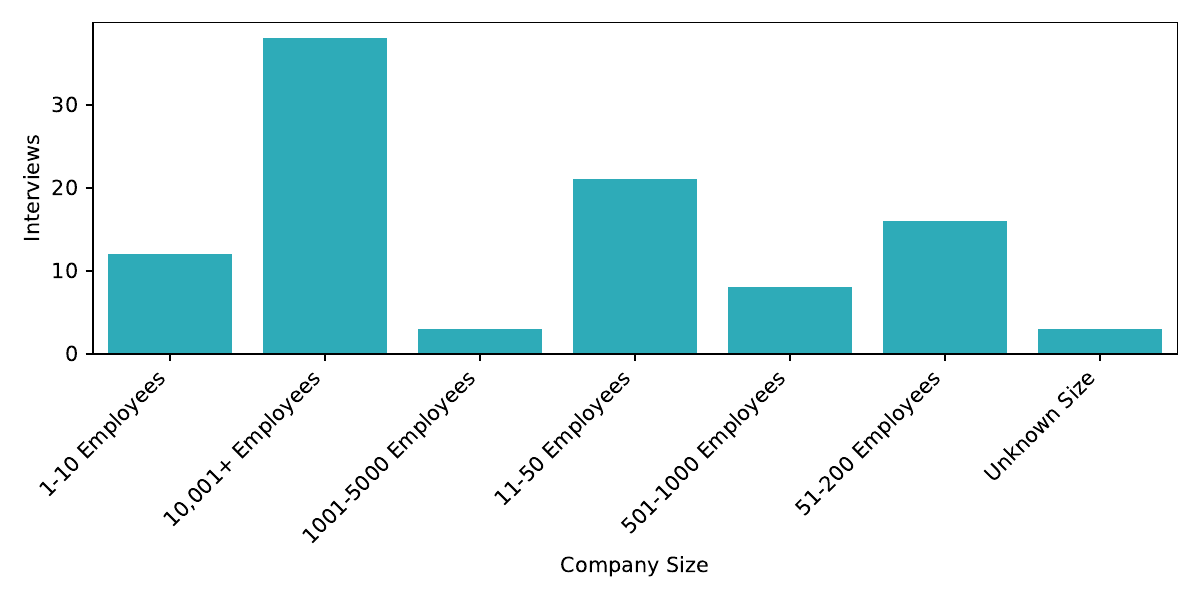}
    \caption{Distribution of interviews by company size. The sample included significant representation from both large enterprises (10,001+ employees) and small-to-medium companies (1--200 employees), providing insights across organizational scales.}
    \label{fig:ch6-company-size}
\end{figure}

\textbf{Company Size Distribution}: Figure~\ref{fig:ch6-company-size} shows the sample spanned organizations of all sizes, from startups with fewer than 10 employees to large enterprises with over 10,000 employees. This distribution ensured that our findings captured challenges across different organizational contexts, from resource-constrained startups to large enterprises with established platform teams.

\subsection{Data Analysis}

Following each interview, we conducted systematic thematic coding to identify recurring pain points and expectations for improvement. We categorized pain points into eleven primary themes: deployment complexity, onboarding difficulty, system complexity, serverless-specific concerns, security concerns, integration challenges, observability limitations, cost management, scaling issues, documentation gaps, and responsiveness issues. We separately categorized expectations for improvement into thirteen dimensions: productivity, automation, onboarding, maintainability, availability, security, performance, cost optimization, programmability, integration, compliance, observability, and scalability. For interviews where participants made quantitative claims about potential productivity improvements or time savings, we extracted and recorded these metrics for aggregate analysis. This dual coding approach—capturing both current pain points and desired future states—enabled us to validate problem-solution fit between market needs and OaaS capabilities.

\section{Market Validation: Pain Points and Customer Expectations}
\label{sec:ch6-findings}

Our customer discovery revealed a consistent set of critical, data-supported challenges that quantitatively confirm the market opportunity OaaS addresses. The analysis demonstrates a significant gap between infrastructure capabilities and teams' practical ability to utilize them effectively---a challenge creating substantial operational drag on development organizations.

\subsection{Pervasive Complexity and Operational Overhead}

\begin{figure}[t]
    \centering
    % Source: report.md Section 2
    % Shows frequencies: deployment (81.2%), onboarding (62.4%), cost (53.5%), 
    %                    security (34.7%), documentation (34.7%), complexity (33.7%),
    %                    integration (22.8%), observability (12.9%), scaling (10.9%), serverless_cons (7.9%)
    \includegraphics[width=\textwidth]{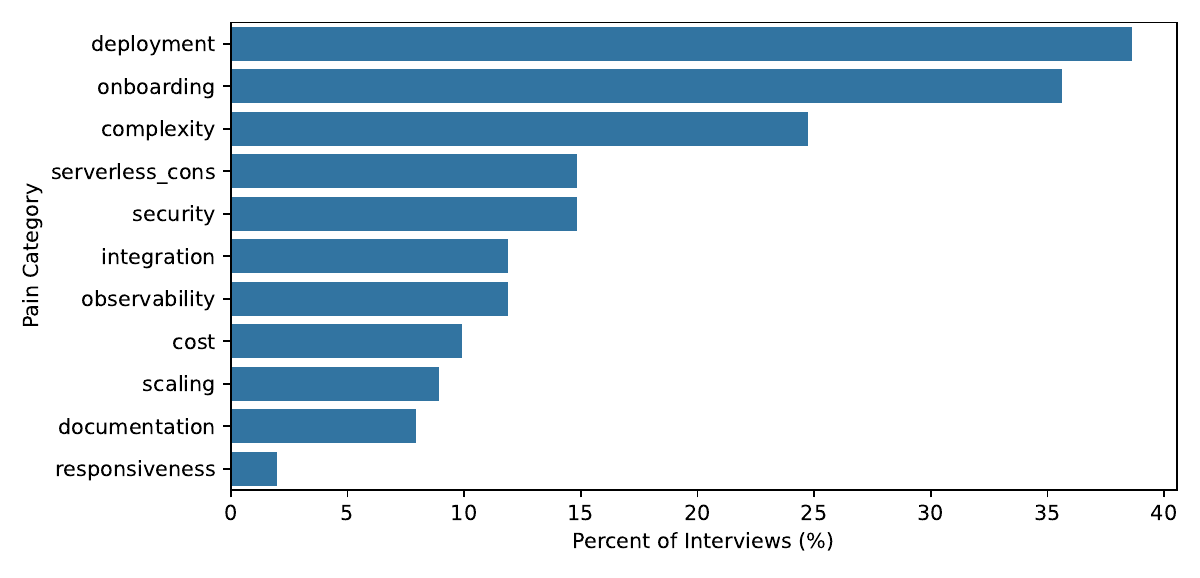}
    \caption{Frequency of pain points mentioned across 101 interviews. Deployment complexity and onboarding difficulty dominated, while cost management and security also featured prominently.}
    \label{fig:ch6-pain-points}
\end{figure}

Figure~\ref{fig:ch6-pain-points} summarizes the frequency of pain points across all 101 interviews. The top challenges were:

\begin{enumerate}
    \item \textbf{Deployment complexity (38.6\%)}: The most prevalent pain point involved fragmented tooling, configuration sprawl, and the burden of orchestrating multiple services (compute, storage, networking, monitoring) to deploy applications. Practitioners described spending significant time on ``glue code'' and infrastructure-as-code templates, with deployments often requiring coordination across multiple teams and tools.
    
    \item \textbf{Onboarding difficulty (35.6\%)}: Steep learning curves for new team members and long time-to-productivity were frequently cited. The proliferation of cloud services, each with distinct APIs, configuration patterns, and best practices, created a significant barrier to entry.
    
    \item \textbf{System complexity (24.8\%)}: The cognitive load of managing microservices architectures with numerous dependencies, distributed state, and failure modes overwhelmed many teams.
    
    \item \textbf{Serverless-specific concerns (14.9\%)}: Cold start latency, execution time limits, and vendor lock-in specific to Function-as-a-Service platforms.
    
    \item \textbf{Security concerns (14.9\%)}: Security configuration complexity, credential management, and compliance requirements added significant overhead, especially in regulated industries like finance and healthcare.
    
    \item \textbf{Integration challenges (11.9\%)}: Connecting heterogeneous services (databases, message queues, APIs, functions) required significant custom integration code and debugging effort.
    
    \item \textbf{Observability limitations (11.9\%)}: Inadequate visibility into application performance, resource utilization, and failure modes hindered troubleshooting and optimization.
    
    \item \textbf{Cost management (9.9\%)}: Unpredictable costs and difficulty correlating resource usage with business value emerged as concerns, particularly for organizations running serverless or autoscaling workloads. The opacity of cost-performance trade-offs and lack of intuitive budget controls made it challenging to optimize spending.
    
    \item \textbf{Scaling issues (8.9\%)}: Difficulty managing autoscaling policies and handling load spikes, particularly with stateful workloads.
    
    \item \textbf{Documentation gaps (7.9\%)}: Inadequate or outdated documentation for cloud services and internal platforms hindered developer productivity and increased reliance on tribal knowledge.
\end{enumerate}

Importantly, these pain points often co-occurred. Our analysis revealed that complexity and onboarding challenges frequently appeared together, suggesting an underlying issue of \textit{cognitive overload}---the sheer amount of knowledge required to effectively use modern cloud platforms.

\subsection{Desired Outcomes and Expectations for Improvement}

\begin{figure}[t]
    \centering
    \includegraphics[width=\textwidth]{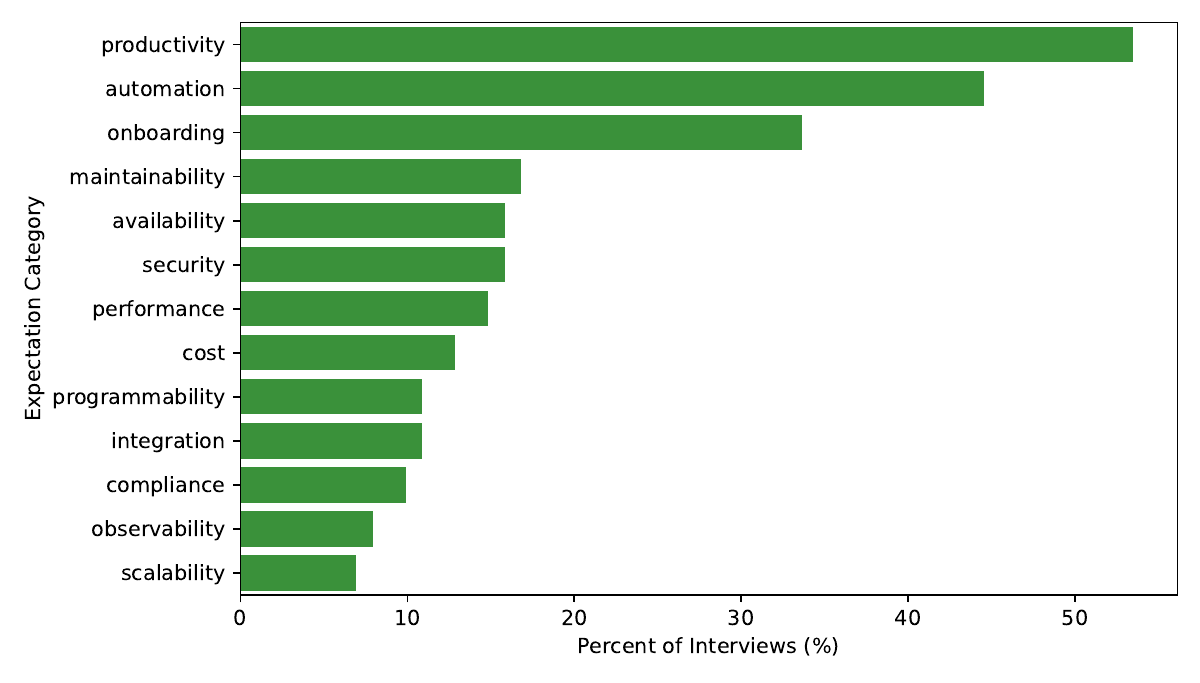}
    \caption{Frequency of expectations for improvement mentioned across 101 interviews. Productivity improvements, automation, and onboarding dominated, revealing that practitioners prioritize developer experience and operational efficiency over raw performance optimization.}
    \label{fig:ch6-expectations}
\end{figure}

Beyond identifying pain points, we asked participants about their desired outcomes and expectations for improvement. We systematically categorized these expectations across thirteen dimensions, revealing clear priorities (Figure~\ref{fig:ch6-expectations}):

\begin{enumerate}
    \item \textbf{Productivity improvements (53.5\%)}: The most prevalent expectation focused on overall time-to-value improvements in development, deployment, and operations. Participants wanted faster time-to-market, reduced development cycles, and quantifiable efficiency gains across the entire software delivery lifecycle.
    
    \item \textbf{Automation (44.6\%)}: Strong demand for increased automation in CI/CD pipelines, infrastructure provisioning, auto-scaling, and rollout/rollback processes. Participants wanted declarative, infrastructure-as-code approaches that eliminate manual steps and approval bottlenecks.
    
    \item \textbf{Onboarding improvements (33.7\%)}: Expectations for significantly reduced ramp time for new engineers, from months to weeks. Participants emphasized the need for intuitive interfaces, comprehensive examples, and streamlined environment setup.
    
    \item \textbf{Maintainability (16.8\%)}: Desire for reduced operational toil, safer updates and rollbacks, simplified patching, and fewer flaky tests or pipelines. Participants wanted systems that are easier to operate and maintain over time.
    
    \item \textbf{Availability and reliability (15.8\%)}: Expectations for improved uptime, faster recovery from failures (lower MTTR), and more resilient systems with fewer outages.
    
    \item \textbf{Security (15.8\%)}: Expectations for stronger security controls, better access management, data locality enforcement, and encryption—particularly acute in regulated industries.
    
    \item \textbf{Performance (14.9\%)}: Improved runtime characteristics including lower latency, higher throughput, reduced cold starts, and better tail latency (p95/p99).
    
    \item \textbf{Cost optimization (12.9\%)}: Expectations for lower cloud and operational costs, better ROI, reduced idle capacity, and more predictable spending.
    
    \item \textbf{Programmability (10.9\%)}: Better developer ergonomics through simpler APIs and SDKs, reduced glue code, unified frameworks, and richer documentation with examples.
    
    \item \textbf{Integration (10.9\%)}: Easier integration with legacy systems, third-party services, and internal platforms, with fewer API mismatches and simpler service composition.
    
    \item \textbf{Compliance (9.9\%)}: Smoother alignment with regulatory requirements (PCI, HIPAA, ITAR, GDPR), faster governance approvals, and better audit trail capabilities.
    
    \item \textbf{Observability (7.9\%)}: Improved end-to-end visibility through better tracing, logging, metrics, and dashboards for debugging and root cause analysis.
    
    \item \textbf{Scalability (6.9\%)}: Better elasticity, horizontal/vertical scaling, multi-region support, and improved capacity management during traffic spikes.
\end{enumerate}

Participants were notably less interested in raw performance gains than in \textit{productivity}, \textit{automation}, and \textit{simplified onboarding}. This pattern reveals that practitioners prioritize developer experience and operational efficiency over technical performance optimization.

\vspace{2mm}
\noindent
\colorbox{blue!10}{
\parbox{0.96\linewidth}{
\underline{\textbf{Takeaway}:} \emph{
Customer discovery reveals deployment complexity, onboarding difficulty, and system complexity as dominant pain points, while practitioners prioritize productivity and automation expectations over performance—demonstrating a clear demand for developer experience improvements.
}}}
\vspace{2mm}

\section{OaaS Technical Capabilities Addressing Market Needs}
\label{sec:ch6-value-prop}

The customer discovery findings demonstrate compelling alignment between validated market pain points, practitioner expectations, and the technical contributions of OaaS presented in Chapters~\ref{chapter3}--\ref{chapter5}. This section examines how OaaS's core architectural innovations---unified object abstraction, declarative non-functional requirements, and edge-cloud orchestration---directly address the challenges identified in Section~\ref{sec:ch6-findings}.

\subsection{Unified Object Abstraction for Deployment and Onboarding}

The top pain points---deployment complexity (38.6\%), onboarding difficulty (35.6\%), and system complexity (24.8\%)---stem from fragmented tooling, service sprawl, and the cognitive load of coordinating multiple infrastructure abstractions. Practitioners reported spending significant time on glue code, configuration management, and cross-team coordination just to deploy applications.

The object abstraction presented in Chapter~\ref{chapter3} directly addresses these challenges by consolidating compute (methods as serverless functions), state (attributes backed by databases or object storage), and workflow (dataflow/macro functions) into a unified deployment unit. Developers define application components using familiar object-oriented programming concepts in declarative specifications, while the platform automatically handles provisioning, configuration, and orchestration across FaaS runtimes, storage backends, and messaging infrastructure.

This architectural consolidation eliminates the fragmentation that drives deployment complexity, reduces the learning curve for new team members by providing a single consistent abstraction, and decreases system complexity by hiding heterogeneous infrastructure details behind a unified programming model. The declarative specification approach further reduces cognitive load by allowing developers to focus on application logic rather than infrastructure mechanics.

\subsection{Declarative Non-Functional Requirements for QoS Guarantees}

Practitioners identified performance (14.9\%), availability and reliability (15.8\%), and cost optimization (12.9\%) as critical expectations, though cost management appeared less frequently as a direct pain point. The underlying concern across these dimensions was unpredictable behavior and opacity in resource-to-outcome mapping, particularly for serverless and autoscaling workloads where teams struggle to correlate spending with performance.

The non-functional requirement (NFR) interface presented in Chapter~\ref{chapter4} addresses these challenges through declarative QoS specifications. Developers specify desired outcomes---throughput targets, availability levels, and consistency requirements---while the platform's Class Runtime Manager continuously monitors actual performance against declared SLAs and dynamically adjusts resource allocations to meet targets.

The \textit{accept-or-reject} admission control model provides upfront validation: if an SLA cannot be met within specified constraints, the system rejects deployment with actionable feedback and alternative configurations rather than allowing deployment and discovering issues in production. This mechanism directly addresses availability concerns (preventing outages from under-provisioning), performance expectations (guaranteeing latency/throughput targets), and cost predictability (budget-aware admission control). The declarative approach eliminates manual performance tuning and resource provisioning, reducing operational toil identified in maintainability expectations (16.8\%).

\vspace{2mm}
\noindent
\colorbox{blue!10}{
\parbox{0.96\linewidth}{
\underline{\textbf{Takeaway}:} \emph{
OaaS core technical innovations---unified object abstraction and declarative NFR interface---directly map to validated market pain points, addressing deployment complexity through architectural consolidation and performance unpredictability through automated QoS management.
}}}
\vspace{2mm}

\section{Production Readiness and Commercialization Pathway}
\label{sec:ch6-implications}

While Section~\ref{sec:ch6-value-prop} established technical validation, transforming OaaS from research prototype to production-ready commercial platform requires addressing critical gaps identified during customer discovery. This section outlines production requirements, target market positioning, and the pathway from research contribution to market-ready product.

\subsection{Production Readiness Requirements}

Customer interviews revealed three priority areas for adoption. \textbf{Security and compliance} emerged as particularly acute in regulated industries like finance and healthcare, requiring defense-in-depth strategies including robust identity and access management (IAM), multi-tenant isolation, encryption at rest and in transit, and audit logging. Several financial services participants explicitly stated that security certification (SOC 2, PCI-DSS) would be prerequisites for evaluation.

\textbf{Developer experience} demands go beyond the unified abstraction to include idiomatic SDKs for popular languages (Python, JavaScript, Java, Go), comprehensive documentation with real-world examples, quickstart templates, and a library of reusable object patterns. Onboarding friction (35.6\% of pain points) can only be fully addressed through polished tooling and learning resources that enable developers to be productive within hours rather than days.

\textbf{Observability and integration} requirements reflect operational realities. Observability limitations (11.9\%) and integration challenges (11.9\%) necessitate distributed tracing across object invocations, structured logging with correlation IDs, metrics dashboards for resource utilization and SLA compliance, and anomaly detection for performance degradation. Integration expectations (10.9\%) demand seamless connectivity with existing CI/CD pipelines, monitoring tools (Datadog, Prometheus, Grafana), and legacy systems through standard protocols and adapters.

\subsection{Validated Target Market Segments}

Interview data validates three primary market segments for commercialization. Technology sector SMEs and startups (33 interviews across 20+ organizations) demonstrated acute sensitivity to deployment complexity and onboarding friction, representing the most suitable early-stage adoption target given their tolerance for emerging platforms and lower security/compliance barriers. Research institutions (17 interviews, 13 organizations) prioritized rapid prototyping and multi-cloud flexibility over enterprise operational features, offering a complementary early adopter segment. Enterprise accounts in regulated industries (12 interviews in finance and healthcare) emphasized security, compliance, and observability alongside operational efficiency—requirements that necessitate addressing the production readiness gaps identified in Section~\ref{sec:ch6-implications} before large-scale enterprise pursuit. This staged approach focuses initial commercialization efforts on technology SMEs and research institutions while building the enterprise-grade capabilities required for regulated industry adoption. Section~\ref{sec:ch6-business-model} presents the complete Business Model Canvas that operationalizes these validated segments into a coherent go-to-market strategy with detailed customer personas, acquisition channels, and phased execution plans.

\vspace{2mm}
\noindent
\colorbox{blue!10}{
\parbox{0.96\linewidth}{
\underline{\textbf{Takeaway}:} \emph{
Transforming OaaS from research prototype to commercial platform requires addressing security, developer experience, and observability gaps for enterprise adoption, while validated target segments---technology SMEs, research institutions, and regulated enterprises---provide evidence-based foundation for the commercialization strategy detailed in Section~\ref{sec:ch6-business-model}.
}}}
\vspace{2mm}

\section{Business Model and Go-to-Market Strategy}
\label{sec:ch6-business-model}

Building on the validated pain points and value propositions from customer discovery, this section presents the business model framework that translates OaaS technical capabilities into a sustainable commercial venture. The Business Model Canvas (BMC)~\cite{osterwalder2010business} synthesizes insights from our 101 interviews into a coherent strategy for market entry, revenue generation, and customer acquisition.

\subsection{Business Model Canvas}
\label{sec:ch6-bmc}

\begin{figure}[ht]
    \centering
    \includegraphics[width=\textwidth]{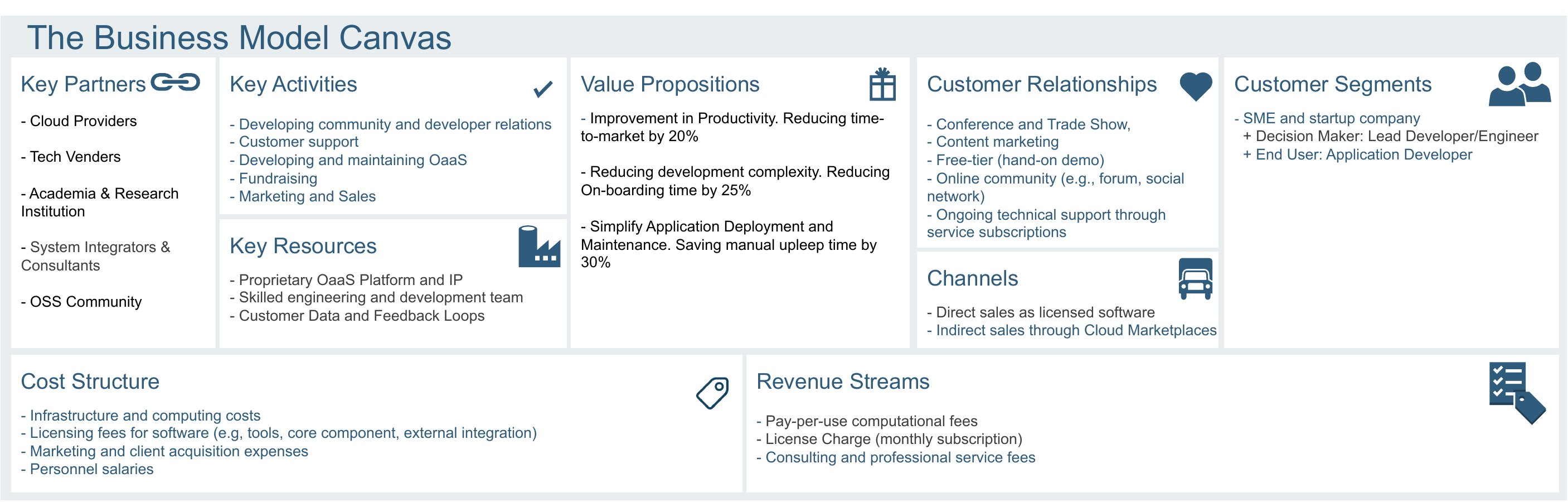}
    \caption{OaaS Business Model Canvas developed through NSF I-Corps customer discovery.}
    \label{fig:ch6-business-model-canvas}
\end{figure}

Figure~\ref{fig:ch6-business-model-canvas} presents the OaaS Business Model Canvas, developed iteratively throughout the I-Corps program through hypothesis testing and customer feedback. The canvas articulates nine interdependent components that define how OaaS creates, delivers, and captures value.

\begin{figure}[ht]
    \centering
    \includegraphics[width=0.8\textwidth]{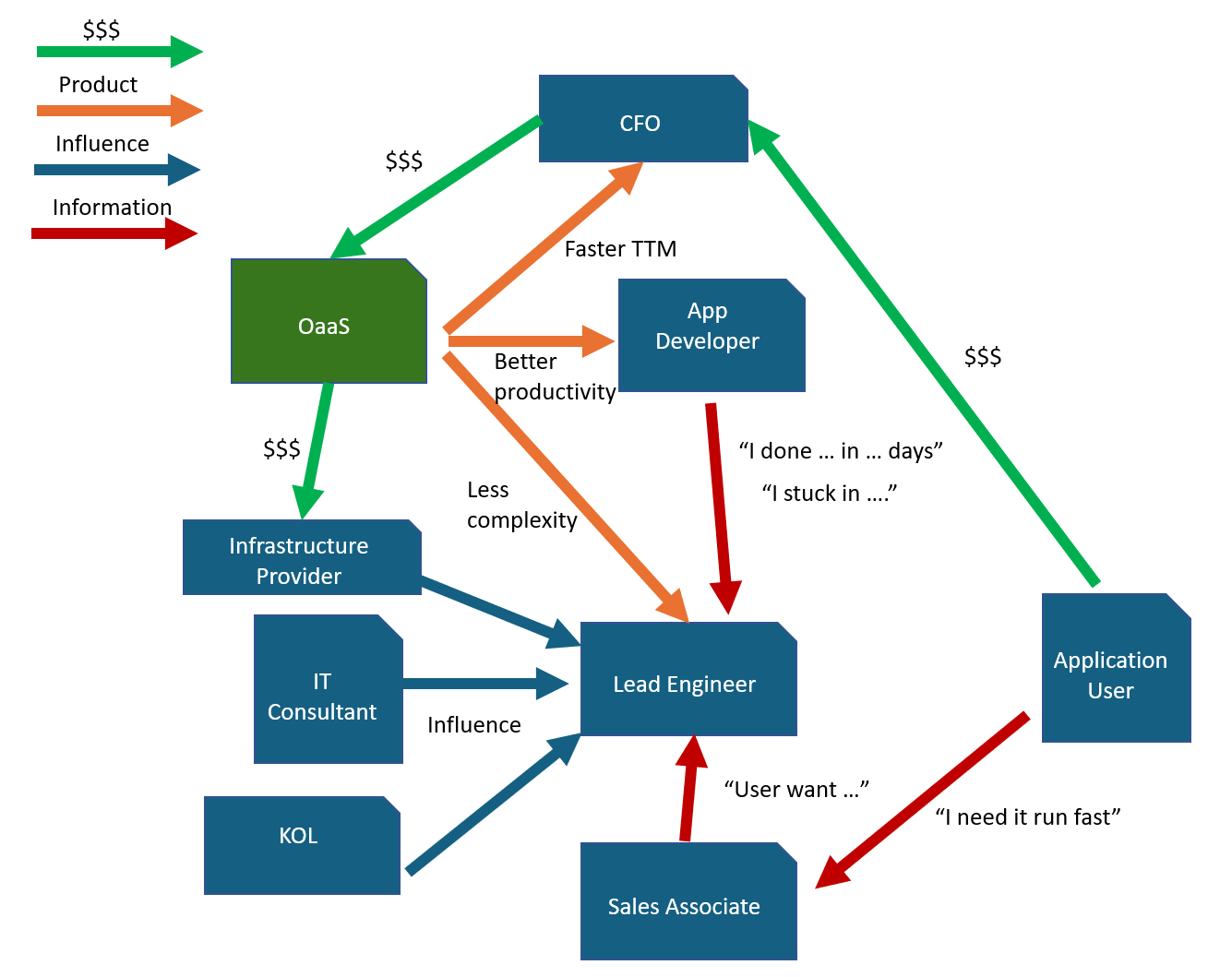}
    \caption{OaaS customer ecosystem showing stakeholder relationships and value flows. Green arrows indicate financial flows, orange arrows represent value delivery (faster time-to-market, improved productivity, reduced complexity), blue arrows show influence patterns, and red arrows indicate requirements and feedback flows from end users.}
    \label{fig:ch6-customer-ecosystem}
\end{figure}

Figure~\ref{fig:ch6-customer-ecosystem} visualizes the OaaS customer ecosystem, revealing the multi-sided market dynamics and stakeholder relationships that shape platform adoption and value delivery. The diagram illustrates how OaaS creates value for multiple interconnected personas while navigating influence patterns from infrastructure providers, consultants, and key opinion leaders who shape purchasing decisions.

\paragraph{\textbf{Customer Segments.}} Our customer discovery identified SME and startup companies as the primary target segment, with a complex buying ecosystem involving multiple interconnected stakeholders (Figure~\ref{fig:ch6-customer-ecosystem}). The ecosystem reveals three internal buyer personas and four external influencer categories, each shaping platform adoption through distinct evaluation criteria and decision authority.

\textit{Financial Decision Makers} (CFOs or budget holders) control purchasing authority and require quantified ROI evidence—minimum 20\% productivity improvements, and reduced time-to-market. \textit{Technical Decision Makers} (Lead Engineers, Engineering Managers) evaluate technical fit, validate feasibility, and champion adoption based on deployment complexity reduction and toolchain integration. Their approval is prerequisite to financial sign-off. \textit{End Users} (Application Developers) prioritize developer experience, intuitive APIs, and reduced cognitive load. While developers rarely control purchasing decisions, negative experiences can derail adoption even after executive approval.

External influencers include \textit{Infrastructure Providers} (AWS, Azure, Google Cloud) creating partnership requirements, \textit{IT Consultants and System Integrators} recommending platforms during modernization projects, \textit{Key Opinion Leaders} shaping developer mindshare through thought leadership, and \textit{Application End Users} providing performance feedback loops. This multi-stakeholder ecosystem necessitates staged engagement: developers discover through hands-on trials, technical decision makers validate through architecture assessments, and financial decision makers approve based on quantified outcomes. Technology SMEs and startups are prioritized as early adopters given their tolerance for emerging platforms, shorter decision cycles, and lower compliance barriers.

\paragraph{\textbf{Value Propositions.}} The canvas articulates three core value propositions validated through customer interviews, each tied to specific pain-expectation pairs identified during discovery. First, customers expressed that any solution addressing deployment complexity must deliver at least 20\% improvement in development productivity and meaningful acceleration in time-to-market; OaaS targets this expectation through unified abstraction that eliminates fragmented tooling and glue code. Second, customers indicated that reducing onboarding difficulty requires solutions that lower the skill barrier for new hires by over 25\%; OaaS addresses this expectation through declarative, object-oriented programming models that hide infrastructure details. Third, customers specified that simplifying application deployment must reduce manual maintenance efforts by approximately 30\%; OaaS targets this expectation through automated provisioning, configuration, and lifecycle management. These quantitative thresholds represent the minimum value customers require from any platform claiming to solve their core pain points, establishing clear benchmarks against which OaaS value delivery will be measured.

\begin{figure}[ht]
    \centering
    \includegraphics[width=0.8\linewidth]{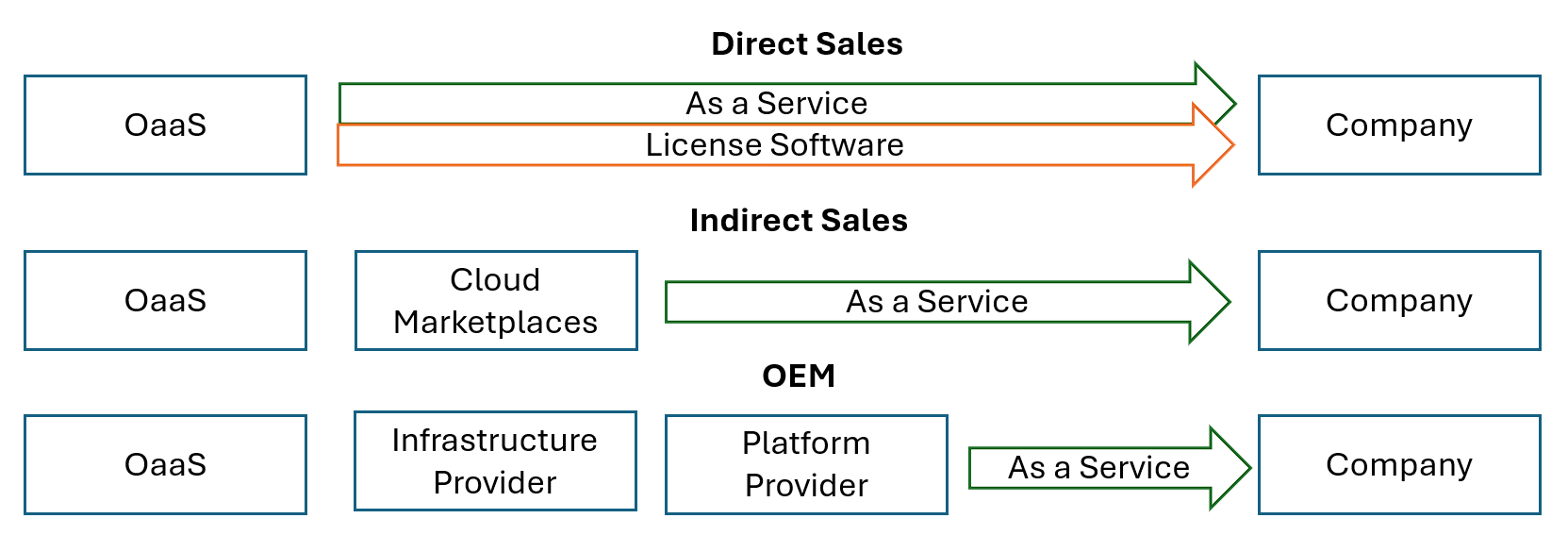}
    \caption{OaaS product channel strategy showing three validated go-to-market pathways.}
    \label{fig:ch6-product-channels}
\end{figure}

\paragraph{\textbf{Channels.}} Customer acquisition follows a multi-channel strategy aligned with validated buyer behavior, visualized in Figure~\ref{fig:ch6-product-channels}. The diagram illustrates three distinct go-to-market pathways validated through customer interviews, with cloud marketplaces as the primary focus for our SME and startup target segment.

\textit{Indirect sales through cloud marketplaces}---such as AWS Marketplace, Azure Marketplace, and Google Cloud Marketplace---serves as the primary go-to-market channel for technology SMEs and startups. Customer interviews revealed strong preference for marketplace-based discovery because it provides immediate discoverability, frictionless procurement through existing cloud billing, rapid trial-to-purchase conversion, and built-in trust from established providers. The managed service model addresses deployment complexity while preserving flexibility to migrate to self-hosted deployments.

\textit{Direct sales as licensed software} provides a complementary channel for enterprise customers requiring deep customization, on-premise deployments, and multi-year licensing agreements. This channel enables technical engagement for complex integration requirements, custom SLA negotiations, and dedicated support arrangements that enterprise buyers expect. While customers explicitly indicated preference for self-hosted deployment models to preserve architectural flexibility and avoid vendor lock-in, this channel primarily targets larger organizations with established procurement processes rather than our primary SME segment.

\textit{OEM partnerships} through infrastructure providers and platform providers represent a third channel for embedded distribution. Infrastructure providers (AWS, Azure, GCP) can bundle OaaS capabilities into their native services, while platform providers can integrate OaaS as middleware for their customers. This channel extends reach to customers who prefer turnkey solutions but requires careful partnership structuring to maintain OaaS brand identity and customer relationships.

Notably, \textit{direct sales as a fully managed platform} was explicitly rejected by customers due to concerns about vendor lock-in, limited customization, and loss of infrastructure control. This feedback shaped our channel strategy to prioritize licensed software and marketplace distribution over proprietary hosting.

\paragraph{\textbf{Customer Relationships.}} Building trust and enabling success requires ongoing engagement beyond initial sales. \textit{Content marketing}---technical blogs, architecture guides, case studies, and conference talks---establishes thought leadership and educates potential customers on OaaS capabilities. \textit{Ongoing technical support through service subscriptions} provides tier-based assistance, with premium tiers offering dedicated support engineers for enterprise customers. \textit{Online community engagement} through forums, Slack channels, and social media fosters peer-to-peer knowledge sharing, early feedback loops, and developer advocacy. \textit{Conference and trade show participation} enables face-to-face relationship building and live demonstrations that address complex technical questions. \textit{Free-tier or hands-on demos} lower adoption barriers by allowing developers to experiment with OaaS without procurement approvals, accelerating evaluation cycles.

\paragraph{\textbf{Revenue Streams.}} The business model employs a diversified revenue strategy with three complementary streams. \textit{License charges via monthly subscriptions} provide predictable recurring revenue through tiered plans based on usage volume, number of developers, or deployment scale. This aligns with enterprise software procurement cycles and enables upselling as adoption grows. \textit{Pay-per-use computational fees} offer consumption-based pricing for compute, storage, and data transfer, appealing to cost-conscious startups and variable-workload customers. This model directly addresses cost optimization expectations by ensuring customers pay only for resources consumed. \textit{Consulting and professional services fees} generate additional revenue through implementation support, migration assistance, custom integration development, and training programs, particularly for large enterprise deployments requiring hands-on expertise.

\paragraph{\textbf{Key Resources.}} Successful commercialization depends on strategic assets developed and acquired over time. The \textit{proprietary OaaS platform and intellectual property}---including patents, trade secrets, and core algorithms---form the defensible technical foundation. A \textit{skilled engineering and development team} with expertise in distributed systems, serverless computing, and developer tooling is essential for continuous innovation and production readiness. \textit{Customer data and feedback loops} captured through telemetry, support tickets, and community interactions inform product roadmap prioritization and feature development.

\paragraph{\textbf{Key Activities.}} Day-to-day operations focus on five strategic pillars. \textit{Developing community and developer relations} involves nurturing the ecosystem, organizing hackathons, publishing documentation, and engaging with early adopters to build advocacy and gather feedback. \textit{Customer support} ensures rapid issue resolution, high satisfaction, and ongoing technical assistance through tiered service levels. \textit{Developing and maintaining OaaS} includes continuous platform enhancements, bug fixes, performance optimizations, and security updates aligned with customer expectations. \textit{Fundraising} from venture capital, government grants (e.g., NSF SBIR/STTR), or strategic investors provides capital for scaling operations. \textit{Marketing and sales} drive demand generation, lead qualification, and conversion through targeted campaigns aligned with validated customer segments.

\paragraph{\textbf{Key Partners.}} Ecosystem collaboration amplifies reach and capabilities through strategic partnerships that balance mutual value creation with inherent risks. Tables~\ref{tab:ch6-partnership-value} and~\ref{tab:ch6-partnership-risks} analyze the partnership dynamics across four key partner categories, articulating value exchange and risk considerations.

\begin{table}[ht]
    \centering
    \small
    \begin{tabular}{|p{4cm}|p{5.5cm}|p{5.5cm}|}
        \hline
        \textbf{Partner Category} & \textbf{OaaS Provides} & \textbf{OaaS Receives} \\ \hline \hline
        Cloud Providers (AWS, Azure, GCP) & Drives core service usage; adds value to their platform through expanded use cases & Infrastructure access; marketplace distribution; co-marketing opportunities; technical support \\ \hline
        System Integrators \& Consultants & Innovative client solutions; partner program with training, support, and co-selling enablement & Enterprise sales channel; implementation and migration services; customer referrals \\ \hline
        Tech Vendors (IoT Devices, Edge Infrastructure) & Ecosystem expansion through complementary integrations; broader market access & Ecosystem expansion; broader use case coverage; technical partnerships \\ \hline
        Academia \& OSS Community & Research platform; OSS contributions; validation testbed; student engagement opportunities & Talent pipeline; cutting-edge innovation; developer credibility; community-driven features \\ \hline
    \end{tabular}
    \caption{Partnership value exchange: what OaaS provides and receives from key partner categories}
    \label{tab:ch6-partnership-value}
\end{table}

\begin{table}[ht]
    \centering
    \small
    \begin{tabular}{|p{3.5cm}|p{11.5cm}|}
        \hline
        \textbf{Partner Category} & \textbf{Partnership Risks and Mitigation Considerations} \\ \hline \hline
        Cloud Providers & A provider could leverage partnership insights to launch a competing native service, requiring careful IP protection and strategic positioning of unique value propositions. \\ \hline
        System Integrators \& Consultants & Natural hesitancy to recommend unproven technologies that could put their reputation at risk with clients, requiring substantial evidence of reliability, ROI, and production-readiness before securing recommendations. \\ \hline
        Tech Vendors & Poor integration quality from partners could damage OaaS user experience; IoT device manufacturers and edge infrastructure vendors showed fragmentation and proprietary protocols during discovery, requiring careful partner vetting and integration quality standards. \\ \hline
        Academia \& OSS Community & Research collaborations are speculative and may not yield direct commercial value; requires balancing open-source community engagement with proprietary competitive advantages. \\ \hline
    \end{tabular}
    \caption{Partnership risk assessment and mitigation strategies for key partner categories}
    \label{tab:ch6-partnership-risks}
\end{table}

\textit{Cloud Providers} (AWS, Azure, Google Cloud) offer infrastructure discounts, co-selling opportunities, and technical support for joint customers, creating strategic alliances that drive mutual platform value. \textit{System Integrators and Consultants} extend market reach by recommending OaaS to their enterprise clients during cloud migration or modernization projects. \textit{Tech Vendors} provide complementary technologies and integration opportunities, expanding both ecosystems, though careful partner vetting is essential to maintain user experience quality. \textit{Academia and Research Institutions} provide access to cutting-edge research, student talent, and validation testbeds for novel features, while \textit{OSS Community} contributors accelerate development through community-driven enhancements, plugins, and integrations.

\paragraph{\textbf{Cost Structure.}} Operating expenses fall into four primary categories. \textit{Infrastructure and computing costs} for hosting the platform, running CI/CD pipelines, and supporting customer workloads scale with adoption. \textit{Licensing fees for software}---including third-party tools, core components, and external integrations---add recurring overhead. \textit{Marketing and client acquisition expenses} cover digital advertising, event sponsorships, content production, and sales enablement. \textit{Personnel salaries} for engineering, sales, marketing, and operations represent the largest fixed cost.

\textit{Cloud Providers} (AWS, Azure, Google Cloud) offer infrastructure discounts, co-selling opportunities, and technical support for joint customers, creating strategic alliances that drive mutual platform value. However, the risk remains that a provider could leverage insights from the partnership to launch a competing native service. \textit{System Integrators and Consultants} extend market reach by recommending OaaS to their enterprise clients during cloud migration or modernization projects, though their natural hesitancy to recommend unproven technologies requires substantial evidence of reliability and ROI. \textit{Tech Vendors} provide complementary technologies and integration opportunities, expanding both ecosystems, though IoT device manufacturers and edge infrastructure vendors received negative feedback during discovery due to fragmentation and proprietary protocols that could compromise user experience. \textit{Academia and Research Institutions} provide access to cutting-edge research, student talent, and validation testbeds for novel features, while \textit{OSS Community} contributors accelerate development through community-driven enhancements, plugins, and integrations, though these collaborations remain speculative with uncertain commercial returns.

\paragraph{\textbf{Cost Structure.}} Operating expenses fall into four primary categories. \textit{Infrastructure and computing costs} for hosting the platform, running CI/CD pipelines, and supporting customer workloads scale with adoption. \textit{Licensing fees for software}---including third-party tools, core components, and external integrations---add recurring overhead. \textit{Marketing and client acquisition expenses} cover digital advertising, event sponsorships, content production, and sales enablement. \textit{Personnel salaries} for engineering, sales, marketing, and operations represent the largest fixed cost.

\subsection{Revenue Model and Pricing Strategy}
\label{sec:ch6-revenue-model}

The revenue model balances accessibility for startups with scalability for enterprises. The \textit{subscription-based licensing} tier offers predictable monthly or annual pricing with features segmented into Free (individual developers, limited usage), Professional (small teams), and Enterprise (large organizations, custom pricing). This model aligns with customer expectations for cost predictability while enabling land-and-expand growth strategies. The \textit{usage-based pricing} tier charges for actual compute time, storage consumption, and data transfer, appealing to customers with variable workloads who prioritize cost optimization. Hybrid plans combining base subscriptions with usage overages provide flexibility and revenue upside during traffic spikes. \textit{Consulting and professional services} are offered for implementation support, migration assistance, and custom development, generating high-margin revenue while deepening customer relationships and accelerating adoption.

Pricing strategy employs a hybrid approach combining value-based and competitive-based methodologies. Value-based pricing is determined from quantified time and cost savings on the customer side, including reduced developer hours, faster time-to-market, decreased infrastructure complexity, and lower operational overhead. These savings translate into concrete ROI metrics that justify premium pricing for customers realizing significant productivity gains. Competitive-based pricing is positioned relative to commercial serverless platforms (AWS Lambda, Azure Functions, Google Cloud Functions) to ensure market competitiveness while emphasizing differentiated value from unified abstractions, declarative SLAs, and edge-cloud capabilities that traditional FaaS offerings lack. This dual-anchor approach ensures OaaS pricing remains justified by delivered value while staying competitively positioned within the serverless market landscape. Customers expressed willingness to pay premiums for platforms that demonstrably reduce onboarding time, deployment complexity, and operational toil. 

\subsection{Go-to-Market Strategy}
\label{sec:ch6-gtm-strategy}

The go-to-market strategy prioritizes rapid adoption in validated customer segments while building ecosystem momentum. Phase 1 targets \textit{early adopters} in technology startups, research institutions, and consulting firms identified during I-Corps as most receptive to innovation and least constrained by legacy processes. Tactics include launching a free tier with generous usage limits, publishing comprehensive documentation and tutorials, and securing 5--10 lighthouse customers willing to provide testimonials and case studies. Developer evangelism through conference talks, blog posts, and open-source contributions establishes credibility and drives organic traffic.

Phase 2 expands into \textit{small-to-medium technology companies} (11--200 employees) facing acute deployment complexity and onboarding challenges. Partnerships with cloud marketplaces (AWS, Azure, Google Cloud) enable frictionless discovery and one-click deployments. Strategic partnerships with system integrators and consultants provide referral channels and professional services leverage. Customer success programs focus on reducing time-to-value, publishing ROI metrics, and identifying expansion opportunities within existing accounts.

Phase 3 pursues \textit{enterprise accounts} in financial services, healthcare, and logistics where security, compliance, and observability requirements are well-defined. Enterprise sales cycles are longer but yield higher contract values and multi-year commitments. Dedicated account teams, custom SLAs, and on-premise deployment options address enterprise buying criteria. Success metrics include 20\% month-over-month user growth in Phase 1, \$500K ARR by end of Phase 2, and \$5M ARR by end of Phase 3.

Channel strategy prioritizes direct sales for enterprise deals requiring customization and relationship building, while cloud marketplaces serve as self-service channels for SMEs and startups. Community-led growth through open-source contributions, developer advocacy, and technical content attracts bottom-up adoption that influences top-down purchasing decisions. Strategic OEM partnerships with platform providers remain exploratory but could unlock distribution leverage if properly structured to avoid channel conflict.

\vspace{2mm}
\noindent
\colorbox{blue!10}{
\parbox{0.96\linewidth}{
\underline{\textbf{Takeaway}:} \emph{
Validated market segments provide clear prioritization: technology SMEs and startups offer the optimal entry point given their acute pain points, tolerance for emerging platforms, and lower compliance barriers, while enterprise requirements define the production roadmap.
}}}
\vspace{2mm}

\section{Summary}
\label{sec:ch6-summary}

This chapter presented commercialization validation from NSF I-Corps National (Summer 2025 Cohort 3) through 101 interviews across 86 organizations. Deployment complexity (38.6\%), onboarding difficulty (35.6\%), and system complexity (24.8\%) emerged as dominant pain points, while practitioners prioritized productivity, automation, and faster onboarding—validating OaaS's focus on unified abstractions and declarative interfaces. Customer discovery validated OaaS's core capabilities while identifying security, compliance, and observability gaps critical for production readiness.

Three market segments emerged: technology SMEs and startups (33 interviews), research institutions (17 interviews), and enterprise accounts (12 interviews). Strategic analysis identified SMEs and startups as the primary early-stage target given their tolerance for emerging platforms and lower compliance barriers, with research institutions as a complementary segment. The Business Model Canvas establishes a staged commercialization strategy with diversified revenue streams (subscription licensing, usage-based pricing, professional services), strategic partnerships across four key categories (cloud providers, system integrators, tech vendors, and academia), and phased go-to-market execution targeting SME/startup organizations before enterprise expansion. The convergence of validated pain points, practitioner expectations, and demonstrated problem-solution fit establishes an evidence-based pathway from research prototype to commercial platform focused on the SME and startup market.

%%%%%%%%%%%%%%%%%%%%%%%%%%%%%%%%%%%%%%
\chapter{Conclusion and Future Research Directions}
\label{chapter7}
%%%%%%%%%%%%%%%%%%%%%%%%%%%%%%%%%%%%%%

\section{Conclusion}
\label{sec:ch7-conclusion}

Cloud computing has fundamentally transformed how applications are developed and deployed, yet a persistent gap remains between the promise of serverless computing and its practical realization. This dissertation addressed this gap through empirical validation and technical innovation, establishing the Object-as-a-Service (OaaS) paradigm as a path toward simplified, unified cloud-native platforms.

Grounded in empirical evidence from an initial practitioner interview study (Chapter~\ref{chapter1}, 21 participants) and validated through NSF I-Corps National customer discovery (Chapter~\ref{chapter6}, 101 interviews across 86 organizations), this work demonstrates that infrastructure complexity imposes substantial tax on developer productivity. Both studies consistently identified complexity in deployment, onboarding, and system management as the most pervasive challenges—particularly for small enterprises, domain experts, and resource-constrained organizations. Critically, practitioners prioritize productivity and automation over cost optimization, contradicting much academic research that emphasizes cost savings.

This dissertation makes five major contributions that collectively advance the state-of-the-art in serverless computing:

\textbf{1. Empirical Validation:} Through three complementary empirical studies—an initial practitioner interview study (Chapter~\ref{chapter1}, 21 participants), a human study evaluating developer experience (Chapter~\ref{chapter5}, 39 participants), and NSF I-Corps National customer discovery (Chapter~\ref{chapter6}, 101 interviews)—we identified and quantified critical pain points, practitioner expectations, and developer productivity gains across diverse organizational contexts. The studies reveal that practitioners consistently prioritize operational maintainability, development simplicity, and onboarding efficiency over cost optimization. Non-technical organizations face significantly higher development complexity rates, while the human study quantified 31\% faster task completion with unified abstractions, validating that simplified interfaces deliver measurable productivity improvements.

\textbf{2. Object-as-a-Service Paradigm (Chapter~\ref{chapter3}):} OaaS unifies resource, state, and workflow management into a single object-oriented abstraction, addressing fragmentation by consolidating FaaS compute, managed databases, and orchestrators into one coherent interface. The Oparaca prototype demonstrates that this unified abstraction streamlines cloud-native programming with negligible performance overhead while achieving scalability comparable to state-of-the-art systems.

\textbf{3. SLA-Driven OaaS Paradigm (Chapter~\ref{chapter4}):} Declarative Non-functional Requirement (NFR) management enables performance optimization through intuitive, high-level specifications. Developers can specify outcomes—availability, throughput, consistency, latency targets—without understanding low-level mechanisms, directly addressing practitioner demands for service quality assurance and simplified programmability.

\textbf{4. OaaS across Edge-Cloud (Chapter~\ref{chapter5}):} EdgeWeaver extends OaaS across the edge-cloud continuum with SLA-driven placement, connectivity-aware invocation, and NFR trade-off management, addressing responsiveness challenges and enabling deployment in geo-distributed, resource-constrained, and connectivity-challenged environments critical for IoT and edge computing scenarios.

\textbf{5. Commercialization Validation (Chapter~\ref{chapter6}):} NSF I-Corps National customer discovery corroborates and expands the initial practitioner study, validating market demand across diverse sectors. The findings confirmed deployment complexity, onboarding difficulty, and system complexity as dominant pain points, with industry-specific patterns revealing acute security concerns in regulated industries, elevated onboarding challenges in education, and observability gaps in healthcare. These insights establish a staged commercialization pathway with technology SMEs and startups as the primary early-stage target given their tolerance for emerging platforms and lower compliance barriers, complemented by research institutions, while deferring enterprise accounts in regulated industries until production readiness requirements (security, compliance, observability) are addressed.

\subsection{Limitations and Scope}
\label{sec:ch7-limitations}

While this dissertation makes significant contributions to serverless computing research and establishes the viability of the Object-as-a-Service paradigm, several limitations constrain the scope and generalizability of the findings:

\textbf{Dataflow Function Limitations:} While Chapter~\ref{chapter3} introduced dataflow (macro) functions as a mechanism to chain operations across objects via declarative data dependencies, the dissertation does not extend this capability with formal guarantees across the edge-cloud continuum. The complexity of ensuring end-to-end consistency, fault tolerance, and latency bounds for dataflow executions that span intermittently connected edge and cloud tiers remains unaddressed. Consequently, Chapters~\ref{chapter4} and~\ref{chapter5} focus on individual object invocations rather than guaranteed workflow orchestration.

\textbf{Runtime Implementation Trade-offs:} The Oparaca and EdgeWeaver prototypes leverage Knative as the underlying serverless runtime, simplifying implementation by reusing existing container orchestration and auto-scaling mechanisms. However, this design choice introduces performance overhead from networking, and resource management layers. The dissertation does not explore alternative serverless runtime implementations—such as WebAssembly-based execution—that could potentially reduce invocation latency and resource footprint at the cost of increased implementation complexity.

\textbf{Multi-tenancy and Network Isolation:} While the Class Runtime architecture (Chapters~\ref{chapter4} and~\ref{chapter5}) supports multi-tenancy through deployment of separate class runtimes per tenant, the dissertation does not comprehensively address network isolation mechanisms. Specifically, the integration between the Messaging Infrastructure and tenant-specific network policies, including traffic segregation, quota enforcement, and secure inter-tenant communication boundaries, remains underspecified. This gap limits deployment in environments requiring strong isolation guarantees.

\textbf{Security Scope:} Security was not a primary focus of this dissertation. Comprehensive security mechanisms, including encryption-at-rest for object state, secure key management, audit logging for compliance, and defense against common attack vectors—are not deeply explored. The dissertation prioritizes demonstrating the viability of unified abstractions and declarative NFR management rather than establishing production-grade security hardening.

\textbf{Class Runtime Updates and Migration:} The dissertation does not detail mechanisms for seamlessly migrating Class Runtimes when class definitions are updated. This includes handling live object state migration, maintaining service availability during transitions, ensuring data consistency across old and new runtime versions, and managing rollback scenarios when updates fail. Consequently, the current prototypes assume relatively static class definitions or require service interruption during updates—a limitation for continuously evolving production applications.

These limitations establish boundaries for the dissertation's claims while identifying concrete research directions addressed in Section~\ref{sec:ch7-future}. Importantly, they do not invalidate core contributions—the unified object abstraction, declarative NFR management, and edge-cloud extension demonstrate viability at research scale and establish architectural patterns for production systems. However, they highlight that transitioning OaaS from research prototype to enterprise-grade platform requires systematic work across security hardening, multi-region deployment, comprehensive tooling, and domain-specific validation.

\section{Future Research Directions}
\label{sec:ch7-future}

\begin{figure}[ht]
    \centering
    \includegraphics[width=0.5\textwidth]{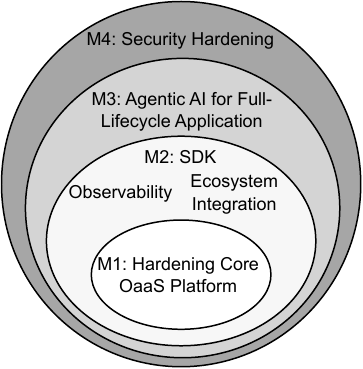}
    \caption{OaaS future research roadmap showing nested milestones from core platform hardening (M1) through ecosystem integration (M2) and agentic AI (M3) to security hardening (M4), with observability as a cross-cutting concern.}
    \label{fig:oaas-future-roadmap}
\end{figure}

The limitations identified in Section~\ref{sec:ch7-limitations} establish immediate technical challenges, while empirical findings from customer discovery (Chapter~\ref{chapter6}) reveal broader adoption barriers. Future research must address both dimensions: resolving technical gaps in the current prototypes while expanding OaaS capabilities for enterprise deployment. We organize this research agenda as nested milestones (represented as $M_x$) that build upon the dissertation's foundation (Figure~\ref{fig:oaas-future-roadmap}). Milestone M1 directly addresses the technical limitations before progressively expanding toward production readiness and transformative capabilities in M2-M4.

\subsection{M1: Addressing Technical Limitations and Core Platform Hardening}

This foundational milestone resolves the technical gaps acknowledged in Section~\ref{sec:ch7-limitations} while establishing production readiness for enterprise adoption. Research challenges span immediate technical limitations and broader platform capabilities. We first address core technical limitations before expanding to broader production readiness:

\textit{(i) Dataflow Orchestration with Guarantees}: Extending macro functions (Chapter~\ref{chapter3}) to provide end-to-end consistency, fault tolerance, and latency bounds across edge-cloud deployments. Research challenges include developing distributed transaction protocols that maintain ACID properties over intermittently connected tiers, designing compensation mechanisms for partial failures in multi-step workflows, and establishing latency prediction models that account for network variability between edge and cloud. This extension would enable developers to declaratively specify workflow-level SLAs while the platform automatically handles failure recovery and performance optimization across the continuum.

\textit{(ii) Optimized Serverless Runtime}: Exploring alternatives to Knative-based implementations to reduce invocation overhead. Research directions include custom lightweight container runtimes that minimize networking layers and resource management overhead, WebAssembly-based execution engines offering microsecond-scale cold starts and fine-grained sandboxing, and hybrid approaches that dynamically select execution strategies based on workload characteristics. Critical challenges involve maintaining auto-scaling and multi-tenancy guarantees while achieving performance improvements, developing migration paths from container-based deployments, and ensuring compatibility with existing FaaS ecosystems.

\textit{(iii) Network Isolation and Multi-tenancy}: Comprehensively addressing integration between Messaging Infrastructure (MI) and tenant-specific network policies. Research challenges include designing policy specification languages that express traffic segregation rules, quota limits, and inter-tenant communication boundaries; implementing efficient enforcement mechanisms that validate every object invocation against tenant policies with minimal latency overhead; and developing monitoring systems that detect policy violations and unauthorized access patterns in real-time. Successfully addressing these challenges enables deployment in shared infrastructure environments requiring strong isolation guarantees.

\textit{(iv) Seamless Class Runtime Migration}: Developing mechanisms for zero-downtime updates when class definitions evolve. Research directions include live migration protocols that transfer object state between runtime versions while maintaining consistency guarantees, versioning strategies supporting simultaneous operation of multiple class versions during gradual rollout, automated compatibility testing that validates new runtime versions against production workloads, and rollback mechanisms enabling instant reversion when updates cause failures. Critical challenges involve minimizing migration-induced latency spikes, preserving in-flight invocations during transitions, and managing data schema evolution across runtime versions.

Beyond addressing these immediate technical limitations, the milestone expands platform capabilities for production readiness:

\textit{(v) Enhanced SLA Support}: Expanding the NFR interface beyond current performance metrics (availability, throughput, consistency, locality) to encompass security policies as first-class requirements, cost constraints with budget-aware admission control, and regulatory compliance specifications—requiring expressive yet accessible specification languages and efficient runtime enforcement mechanisms.

\textit{(vi) Multi-region Deployment}: Enabling geo-distributed object placement with cross-region state management, consistency protocols that balance performance against data durability guarantees, and latency-aware placement algorithms that optimize for user proximity while respecting data sovereignty constraints.

\textit{(vii) Production-grade Reliability}: Implementing comprehensive fault tolerance across distributed components, automated failure detection and recovery with minimal downtime, and formal verification of correctness properties to ensure platform behavior matches specifications.

\textit{(viii) Performance Optimization at Hyperscale}: Developing efficient scheduling algorithms that minimize resource fragmentation, intelligent resource packing to maximize utilization, and cold-start elimination techniques that maintain sub-second response times even during traffic spikes.

This milestone establishes the robust technical foundation required for enterprise adoption in regulated industries, directly responding to both the dissertation's acknowledged limitations and production readiness requirements identified during customer discovery. Successfully addressing these challenges transforms OaaS from research prototype to production-capable platform while maintaining the core benefits of unified abstraction and declarative NFR management.

\subsection{M2: SDK, Ecosystem Integration, and Observability}
Building on the hardened platform, this milestone expands developer experience through comprehensive tooling and ecosystem integration. The first research thrust focuses on \textit{developer productivity tools}: idiomatic language SDKs for Python, JavaScript, Java, and Go that provide intelligent code completion and type safety; local development environments that simulate production behavior for rapid iteration; and migration tooling to simplify transitions from existing FaaS platforms. The second thrust targets \textit{operational integration}: CI/CD pipeline integration with automated testing and canary deployments; and third-party service connectors for databases, message queues, and monitoring tools. The third thrust addresses \textit{comprehensive observability}: distributed tracing across object invocations with low overhead, SLA compliance monitoring with real-time violation detection and root cause analysis, cost attribution linking resource consumption to business value, performance profiling identifying bottlenecks across the edge-cloud continuum, and actionable insights through intelligent dashboards and anomaly detection.

This milestone directly addresses empirically validated pain points: onboarding challenges (35.6\% of interviews), time-to-productivity concerns, and observability gaps (11.9\% of interviews). By providing polished SDKs and local development environments, OaaS reduces the steep learning curve identified in customer discovery. Comprehensive observability enables practitioners to understand system behavior, diagnose failures, and optimize resource usage—capabilities critical for production adoption but absent in current serverless platforms.

A particularly compelling application domain enabled by this ecosystem is \textit{Digital Twin systems}, where OaaS's unified object abstraction naturally models physical assets (sensors, actuators, machinery) as distributed objects with state synchronization across the edge-cloud continuum. OaaS's declarative SLAs provide consistency and latency guarantees essential for real-time physical-virtual synchronization, while edge-cloud placement ensures low-latency processing at the network edge. Research challenges include designing intuitive APIs for physical-virtual state mapping, ensuring synchronization under intermittent connectivity, and providing domain-specific observability for manufacturing, smart cities, and infrastructure monitoring—domains where OaaS's strengths in state management and edge deployment directly address critical requirements.

\subsection{M3: Agentic AI for Full-Lifecycle Application Management}
A transformative research frontier explores autonomous AI agents capable of managing complete application lifecycles, representing a paradigm shift from developer-driven to AI-assisted cloud-native development. Research challenges span: (i) \textit{Intent-to-implementation synthesis} using large language models to translate natural language requirements or high-level specifications into correct, optimized OaaS object definitions with appropriate SLAs and architectural patterns; (ii) \textit{Continuous optimization} through reinforcement learning agents that autonomously tune SLAs, resource allocations, and placement strategies by analyzing telemetry, predicting workload patterns, and adapting to cost-performance trade-offs; and (iii) \textit{Intelligent debugging and remediation} leveraging program analysis and causal inference to diagnose distributed failures, identify root causes across the edge-cloud continuum, and synthesize fixes through automated code patches or configuration adjustments.

This milestone addresses the ultimate developer experience challenge: enabling practitioners to express \textit{what} they want without specifying \textit{how} to achieve it, directly targeting the productivity and automation expectations prioritized by 53.5\% and 44.6\% of customer discovery participants respectively. Critical research challenges include ensuring correctness and preventing hallucination-induced errors through formal verification techniques, establishing verifiable trust boundaries between developer intent and autonomous execution, and creating interpretable decision traces that maintain developer understanding and control. Successfully addressing these challenges positions OaaS as a substrate for next-generation cloud platforms where AI agents handle infrastructure complexity while developers focus purely on business logic and application requirements.

\subsection{M4: Security Hardening}
The outermost milestone addresses comprehensive security across all layers, transforming OaaS into an enterprise-grade platform suitable for regulated industries. Research directions include: (i) \textit{Defense-in-depth architecture} implementing multi-tenant isolation through secure sandboxing, enforcing principle of least privilege via fine-grained access controls, and deploying zero-trust networking where every object invocation requires authentication and authorization; (ii) \textit{Cryptographic enforcement} ensuring data locality constraints through verifiable placement proofs, mandatory encryption at rest and in transit, and immutable audit trails for compliance verification; (iii) \textit{Compliance automation} translating regulatory requirements (PCI-DSS for payment processing, HIPAA for healthcare data, GDPR for privacy, SOC 2 for service operations) into declarative policy specifications that the platform continuously enforces and validates; (iv) \textit{Vulnerability management} integrating automated dependency scanning to detect CVEs, orchestrating security patches across distributed deployments, and implementing runtime exploit mitigation techniques; and (v) \textit{Security certification pathways} establishing clear frameworks and evidence collection mechanisms to streamline third-party audits and regulatory approvals.

This milestone directly addresses acute security concerns identified during customer discovery, where financial services and healthcare participants emphasized that security certification and compliance guarantees were prerequisites for evaluation. Research challenges span designing security abstractions that remain intuitive despite cryptographic complexity, balancing performance overhead from encryption and access control against security guarantees, and creating compliance frameworks that adapt to evolving regulatory landscapes while maintaining verifiable correctness. Successfully addressing these challenges will enable OaaS adoption in high-assurance environments where current serverless platforms face significant trust and compliance barriers.

\vspace{2mm}

Collectively, these nested research directions chart a path from production-ready platform (M1) through enhanced developer experience and observability (M2) to intelligent automation (M3) and comprehensive security (M4), addressing empirically validated pain points while progressively expanding OaaS capabilities. By systematically addressing these challenges, future work can advance cloud platforms that are simultaneously powerful, accessible, secure, and adaptive to the evolving needs of diverse applications across the computing continuum.

%%% Uncomment if there are appendices. This command sets some things up in
%%% preparation for the actual appendices.
\appendix

%%% Include appendix "chapters" here. A single appendix should start with
%%% \appchapter*{TITLE} rather than \chapter{TITLE}. This supresses the 
%%% numbering (if there is only one appendix it doesn't rate a number). If
%%% there is more than one appendix, use \appchapter to get numbered
%%% appendices (A, B, C, ...).  
%%%%%%%%%%%%%%%%%%%%%%%%%%%%%%%%%%%%%%%%%%%%%%%%%%%%%%%%%%%%%%%%%%%%%%
%%%%%%%%%%%%% Appendix
%%%%%%%%%%%%%%%%%%%%%%%%%%%%%%%%%%%%%%%%%%%%%%%%%%%%%%%%%%%%%%%%%%%%%%
% \appchapter*{A B C 1 2 3}

%%% Local Variables: 
%%% mode: latex
%%% TeX-master: "mydissertation"
%%% End: 

%%% This starts the backmatter, which so far is only the bibliography.
\backmatter

%%% This modifies some bits of the amsplain bibliography style to conform
%%% with UNT/TGS requirements.
\bibliographystyle{UNTamsplain}

%%% Include the .bib file containing the bibliographic information (in
%%% BiBTeX format...). ``bibliography.bib'' is a descriptive name for a
%%% bibliography file in BiBTeX format.
\bibliography{bibliography}

%%% That's all folks!
\end{document}